\documentclass{cernyrep}
\usepackage{lineno}
\usepackage{doi}
\usepackage{graphicx}
\usepackage{xspace}
\usepackage{url}
\usepackage{booktabs}
\usepackage{color}
\usepackage{subcaption}
\usepackage[font=small,labelfont=bf]{caption}
\usepackage{soul}
\usepackage{array}
\usepackage{multirow}
\usepackage{placeins}
\hypersetup{bookmarksnumbered,bookmarksopen,bookmarksopenlevel=1,colorlinks=false,pdfborder={0 0 0},plainpages=false,unicode=false}

\usepackage[table,svgnames]{xcolor}
\usepackage{amsmath,amssymb}
\usepackage{wrapfig}
\usepackage{floatrow}
\newfloatcommand{capbtabbox}{table}[][\FBwidth]

\usepackage[backend=biber,style=numeric-comp,sorting=none,sortcites=true]{biblatex} 
\newcommand{\Stwo}{\textit{S2}\xspace}
\newcommand{\Sthree}{\textit{S3}\xspace}
\newcommand{\kl}{$\kappa_{3}$~}

\newcommand{\tttt}{\ensuremath{t\bar{t}t\bar{t}}\xspace}

\newcommand{\ttgamma}{\ensuremath{t\bar{t}\gamma}\xspace}
\newcommand{\ttZ}{\ensuremath{t\bar{t}Z}\xspace}
\newcommand{\ttH}{\ensuremath{t\bar{t}H}\xspace}
\newcommand{\HZy}{\ensuremath{H \to Z \gamma}\xspace}
\newcommand{\ttbar}{\ensuremath{t\bar{t}}\xspace}
\newcommand{\ttbarjet}{\ensuremath{\ttbar\text{+jet}}\xspace}

\newcommand{\abinv} {\mbox{\ensuremath{\,\text{ab}^{-1}}}\xspace}

\newcommand{\MeV}{\ensuremath{\,\text{Me\hspace{-.08em}V}}\xspace}

\newcommand{\TeV}{\ensuremath{\,\text{Te\hspace{-.08em}V}}\xspace}

\usepackage{pdfpages}

\title{Highlights of the HL-LHC physics projections by ATLAS and CMS}

\author{The ATLAS and CMS Collaborations}

\begin{document}

\includepdf[pages=1]{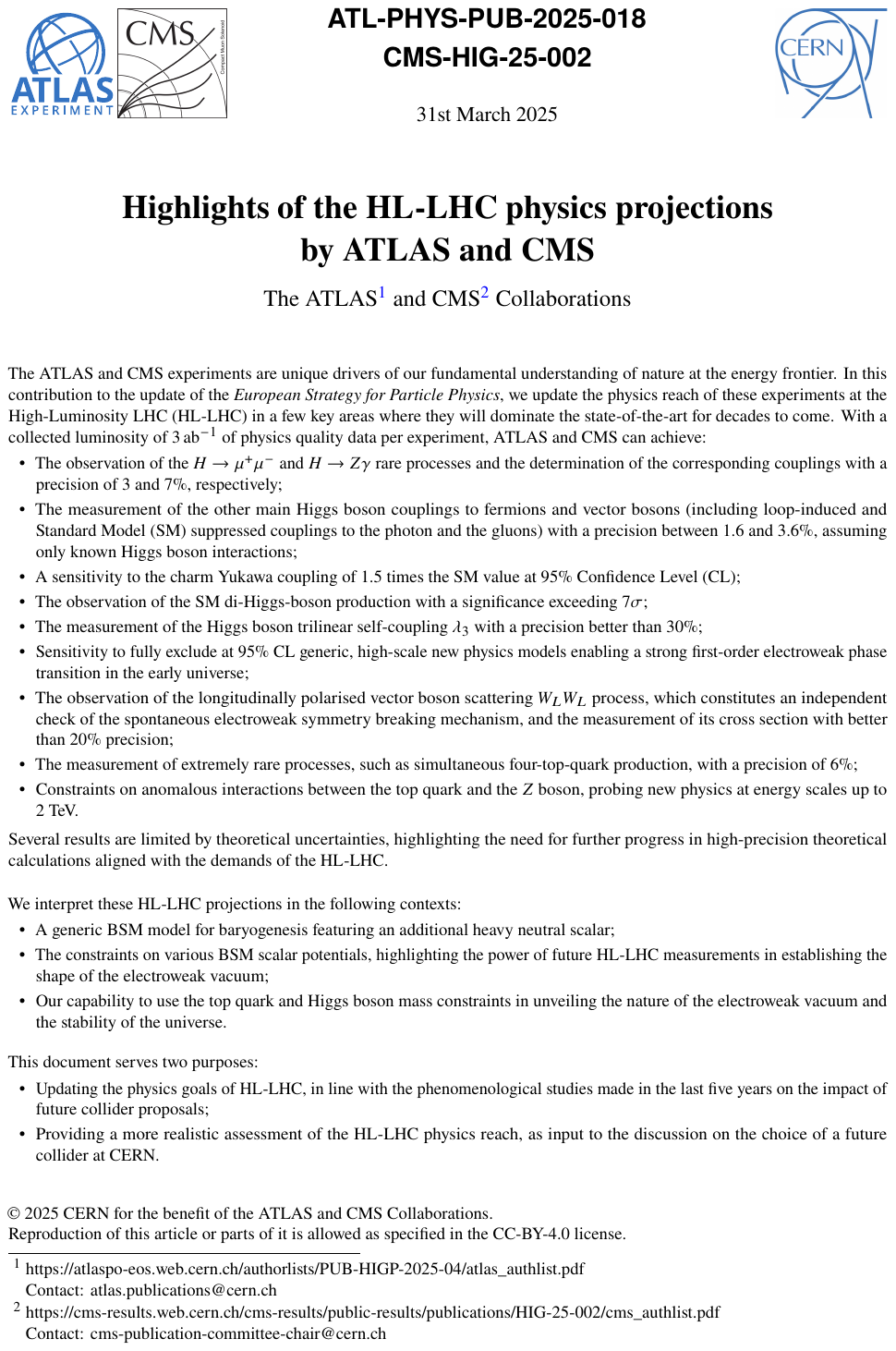}

\clearpage

\pagenumbering{arabic}

\section{Introduction} \label{sec:intro}

The High-Luminosity LHC (HL-LHC) physics programme will be crucial for deepening our understanding of fundamental physics, enabling in particular precision studies of the Higgs sector and enhancing sensitivity to rare processes and potential new physics signals. With unprecedented integrated luminosity, it will offer a unique opportunity to probe the Standard Model (SM) with extreme accuracy and explore connections to open questions in particle physics, astroparticle physics, and cosmology.
The physics reach of the upgraded ATLAS and CMS detectors at the end of their scientific programme will not only be significant in its own right but will also serve as a critical foundation for decision-making on future colliders, shaping the 2026 update of the \emph{European Strategy for Particle Physics} (ESPPU).

The HL-LHC legacy measurements of many SM parameters are expected to remain relevant for decades, even during the operation of a next future collider. Until then, our understanding of key SM sectors will rely on the ultimate precision achieved by the HL-LHC experiments. 

The HL-LHC projected precision reach of ATLAS and CMS across their physics programme has been extensively documented in detailed studies submitted to the 2020 European Strategy~\cite{Dainese:2019rgk,ATLAS:2019mfr,Cepeda:2019klc} and Snowmass 2022~\cite{CMS:2022cju}, including comprehensive investigations into their sensitivity to potential new physics scenarios. This report updates a selected set of those projections that could have groundbreaking implications for our understanding of nature. Specifically, it focuses on physics results that will remain uniquely accessible at the HL-LHC for a long time after its conclusion or that are statistically limited. The provided updates take into account the precision reached by ATLAS and CMS data analyses on the full LHC Run-2 dataset (2015-2018, $\sim 140$ fb$^{-1}$ per experiment), as well as some of the anticipated improvements for Run~3 (2022-2026, $>170$ fb$^{-1}$ per experiment).

We evaluate the precision of specific rare single Higgs boson processes such as $H \to \gamma \gamma$, $H\to \mu^+ \mu^-$, and $H\to Z \gamma$, that remain statistically limited at the LHC but will benefit from the high Higgs boson yield at the HL-LHC.
Furthermore, several physics measurements considered for this report aim to advance our understanding of electroweak symmetry breaking (EWSB) and constrain the shape of the Brout--Englert--Higgs (BEH) potential. These include di-Higgs boson ($HH$) production and constraints on the trilinear Higgs self-coupling $\lambda_3$, triple-Higgs boson production ($HHH$) to constrain the quartic Higgs coupling $\lambda_4$, the top quark mass ($m_t$) and Higgs boson mass ($m_H$), which are linked to the stability of the universe~\cite{Hiller:2024zjp}, and vector boson scattering (VBS) as a complementary approach to study EWSB.

Similarly, the HL-LHC top-quark physics reach will remain unequalled at least until a potential $e^+e^-$ collider operates at the top-quark pair production threshold. The ultimate precision on $m_t$ is imperative to significantly reduce uncertainties in many SM predictions needed at future colliders.
This document assesses the achievable precision on the top quark Yukawa coupling from both the on-shell $ttH$ channel and the off-shell four-top-quark production (\tttt), as well as the precision on the top-$Z$ and top-$\gamma$ couplings. Additionally, the precision achievable for the \tttt production cross section measurement and its potential interpretations in the context of physics beyond the SM (BSM) is discussed.
The precision which the HL-LHC will reach for many rare SM processes will not be superseded for many decades, until the next generation hadron or muon collider operation. Previous studies document this unique aspect of the HL-LHC physics programme~\cite{Dainese:2019rgk,ATLAS:2019mfr,CMS:2022cju,Cepeda:2019klc}.

The impact of the extrapolated precision is assessed in various theoretical scenarios, particularly regarding the BEH potential and the possible exclusion of a first-order electroweak phase transition in the early universe~\cite{SH2,SH3}. New interpretations beyond those previously published by ATLAS and CMS are also reported here.

Two different per-experiment integrated luminosity ($\mathcal{L}$) scenarios are considered at a centre-of-mass energy of $\sqrt{s} = $14~TeV:  a nominal  3\abinv  and an intermediate 2\abinv of data good for physics analysis,
i.e.~after taking into account detector recording inefficiencies and data quality requirements, typically accounting for a 10\% decrease in integrated luminosity.
All cross sections used in the projections considered in this report have been scaled to 14~TeV.

Whenever possible, combined ATLAS and CMS results are reported. In cases where results are available from only one collaboration, they are assumed to apply to both.
Insights gained from a decade of detector operations, along with extensive studies on the impact of the HL-LHC detector upgrades, support the assumption that the two experiments should achieve comparable sensitivity.

All projections presented in this document are based on published results from the two collaborations, considering two different systematics scenarios. 
The first scenario, referred to as \Stwo for backward compatibility with the past European strategy naming convention, assumes reduced systematic uncertainties in most of the cases, 
with theoretical systematic uncertainties halved where it appears feasible, and negligible contributions from limited Monte Carlo simulation event counts. The ``scaling'' of experimental systematic uncertainties depends on the physics object and takes into account the different origin of the uncertainties (statistics of the calibration samples, modelling of SM processes in the calibration, etc.)~\cite{ATL-PHYS-PUB-2019-005,Collaboration:2650976}. Conservatively, the same minimal systematic uncertainties were assumed for both the intermediate 2\abinv and nominal 3\abinv scenarios. Details on the specific implementations are provided in the supporting documents~\cite{ATLAS_PUBNOTE_H,ATLAS_PUBNOTE_VH,ATLAS_PUB_dihiggs,ATLAS_PUB_bbtautau,PUB_ATLAS_multi,ATLAS-PUB-NOTE-HHH,ATLAS_PUB_TOP_ttgttZ,ATLAS_PUB_TOP_fourtop,ATLAS_PUB_TOP_topmass,CMS_PAS_H,CMS_PAS_dihiggs,CMS_PAS_X_to_ZZ_tt,CMS_PAS_TOP}. An additional scenario (\Sthree) evaluates the impact of recent improvements in specific object reconstruction relative to \Stwo, focusing on analyses that have already demonstrated and documented advancements compared with the latest publications. 
The \Stwo or \Sthree projections  are  presented  only when relevant for the discussion.
The specific  improvements assumed for \Stwo and \Sthree are discussed in the relevant sections, with \Sthree relevant only in specific studies.

All scenarios for the systematic uncertainties are based on the best available knowledge. Past experience suggests that experimental and theoretical advancements will exceed expectations that are solely based on the integrated luminosity growth. Consequently, these \Stwo and \Sthree scenarios—grounded in LHC Run-2 and Run-3 knowledge—represent conservative targets.

The impact of the different running conditions, e.g.~the 140--200 average simultaneous $pp$ interactions (pileup), is not explicitly taken into account in these new extrapolations. Previous studies suggest that the HL-LHC upgrades of the  ATLAS and CMS detectors and the improvement in the reconstruction algorithms and data processing offer significantly enhanced trigger, tracking, and pileup suppression capabilities, which are expected to more than compensate for the challenges posed by the increased pileup.

The success of the HL-LHC physics programme depends on the timely and successful completion of the ambitious upgrades to the accelerator, as well as the ATLAS and CMS detectors, which are currently under construction. Equally important is addressing the significant software and computing challenges posed by the unprecedented wealth of new data. Dedicated submissions from ATLAS and CMS to the ESPPU~\cite{ATLASDetectorUpgrade,ATLAScomputing,CMScomputing} report on these efforts.

This document does not consider several key measurements expected by HL-LHC relevant for Higgs boson physics and EWSB, such as $m_W$
and the invisible Higgs boson decay rate. For these cases, previously released projections still apply. In particular, one expects an uncertainty of $\sim 5$~MeV on $m_W$~\cite{CMS:2022cju}
and an upper limit of $2.5\%$ on the $H \to \mathrm{invisible}$ branching fraction~\cite{Cepeda:2019klc}. The latter is particularly significant for dark matter searches, as many SM extensions predict that the Higgs boson could decay into invisible particles, potentially providing a portal to dark matter.
In addition, the search programme will continue exploring uncovered phase space and benefit from improved sensitivity for weakly-coupled or otherwise experimentally challenging BSM scenarios, as described in previous studies~\cite{Dainese:2019rgk,ATLAS:2019mfr,CMS:2022cju,Cepeda:2019klc}.

Additional material can be found in the supporting appendices, linked to this document.

\section{Precision Higgs boson physics and rare decays}

This section presents the update of the expected precision of several Higgs boson measurements at the HL-LHC, based on the latest analyses of the complete Run-2 dataset ($\sim$140~fb$^{-1}$ of $pp$ collisions). These projections  generally improve upon previous ones in Refs.~\cite{CMS:2022cju,ATL-PHYS-PUB-2018-054,CMS-PAS-FTR-18-011}.
The estimation on Higgs coupling determination is quoted in terms of the uncertainty on the coupling modifiers, i.e., on the ratio between a given coupling and its SM expectation.
Particular emphasis is placed on rare decay channels with clean experimental signatures, such as $H \to \gamma\gamma$, $Z\gamma$, and $\mu\mu$, which benefit significantly from the large integrated luminosity. The updated results do not account for any future optimization of the data analysis (event selection, categorization, and signal-versus-background discrimination techniques). The projected sensitivities can therefore be considered conservative.
All details on the analyses used as input and the assumptions on the systematic uncertainties can be found in Refs.~\cite{ATLAS_PUBNOTE_H} and~\cite{CMS_PAS_H} for the ATLAS and CMS projections, respectively. All the numbers quoted here and in Refs.~\cite{ATLAS_PUBNOTE_H,CMS_PAS_H} are derived in the 
\Stwo scenario.

The differential cross-section measurements for each production mode within the simplified template cross section (STXS) framework~\cite{Andersen:2016qtm,run2STXS,LHCHiggsCrossSectionWorkingGroup:2016ypw} in the $H\to\gamma\gamma$ final state performed by the ATLAS~\cite{ATLAS:2022tnm} and CMS~\cite{CMS:2021kom} Collaborations have been projected to 3\abinv per experiment.  Cross-section measurements at large transverse momentum of the Higgs boson ($p_T^H$) are still statistically limited and are particularly sensitive to BSM physics. The projected combined uncertainty in the gluon-gluon fusion (ggF) Higgs boson production cross-section times the branching fraction to diphoton $\mathcal{B}(H\to\gamma\gamma)$, evaluated in the STXS bin for $|y_H|<2.5$ and $p_T^H > 650$~GeV, is 
79\% in the \Stwo scenario at 3~ab$^{-1}$, where the predicted cross section is $0.005\pm 0.001$~fb.
Other decay channels, such as boosted $H\rightarrow \tau\tau$~\cite{CMS:2024jbe} and $H\to b\bar{b}$~\cite{CMS:2020zge,ATLAS:2021tbi}, will significantly enhance the measurement precision in this phase space. Furthermore, as shown in Ref.~\cite{Becker:2669113}, the ggF contribution to the total Higgs boson production cross-section becomes more subdominant compared to vector-boson-associated ($VH$, $V=W,Z$) production for $p_T^H > 1$~TeV. 
Reference~\cite{ATLAS_PUBNOTE_VH} presents the projections for the measurement of the $WH$ and $ZH$ production modes in $H\to b\bar{b}$ decays, based on the ATLAS full Run-2 dataset. The projected ATLAS+CMS uncertainty in the $VH$ cross section times $\mathcal{B}(H \to b\bar{b})$ with $p_T^V > 600$~GeV is 19\% for $WH$ and 22\% for $ZH$ at 3~ab$^{-1}$ in the \Stwo scenario, assuming the same sensitivity for both experiments. 

The $VH$ production mode is also crucial for probing the Higgs boson coupling to charm quarks~\cite{ATLAS_PUBNOTE_VH}. Considering the recent improvements in $b$- and $c$-jet tagging techniques~\cite{CMSbtagplots,ATLASNNbtag,ATLASGNNbtag}, a $1.6\sigma$ significance over the background-only hypothesis is expected for the $VH\to c\bar{c}$ process with 3~ab$^{-1}$ per experiment for ATLAS+CMS. Additionally, the $H$-to-$c$ quark coupling modifier $\kappa_c$ can be constrained to $|\kappa_c|<1.5$ at 95\%~CL, improving the previous projection from direct searches by 30\%~\cite{Cepeda:2019klc}.

The determination of the \HZy signal strength and the effective coupling modifier $\kappa_{Z\gamma}$ is heavily limited by statistical uncertainties.  Recently, evidence for the $H\to Z\gamma$ decay with a significance of three standard deviations~\cite{ATLAS:2023yqk} was achieved through a combination of ATLAS~\cite{ATLAS:2020qcv} and CMS~\cite{CMS:2022ahq} searches based on the full Run-2 dataset, with an observed signal strength of $\mu = 2.2 \pm 0.7$. Deviations from the SM coupling are expected in several BSM models~\cite{Low_2011,Azatov_2013}. The projections presented here are based on the latest analyses and their combination.

\begin{table}[t!]
    \centering
    \begin{tabular}{l|l|c|c}
    \toprule
    \multirow{ 2}{*}{ $\mathcal{L}$} & & \multicolumn{2}{c}{$\delta\mu~[\%]$}\\
      & &{$H\rightarrow Z\gamma$} &{$H\rightarrow \mu\mu$} \\
       \midrule
		\multirow{ 3}{*}{2~ab$^{-1}$}&ATLAS & 21  & 13 \\ 
		&CMS & 23 & 8.4 \\
        &ATLAS+CMS& 15    & 7.1 \\
       \midrule
		\multirow{ 3}{*}{3~ab$^{-1}$}&ATLAS  & 17  & 11  \\ 
		&CMS & 19  & 7.0 \\
        &ATLAS+CMS & 14    & 5.9 \\
        \bottomrule
    \end{tabular}
    \caption{Projected uncertainties in percentage on the $H\rightarrow Z\gamma$ and $H\rightarrow \mu\mu$ signal strengths ($\mu$) for the ATLAS~\cite{ATLAS_PUBNOTE_H} and CMS experiments~\cite{CMS_PAS_H}, as well as their combination, for different integrated luminosities per experiment and uncertainty scenario  \Stwo. The $H\rightarrow \mu\mu$ analyses are combined assuming no experimental correlation between the two experiments, while the theoretical uncertainty is fully correlated.}
    \label{tab:H_Zy_mumu}
\end{table}

The $H\rightarrow \mu\mu$ decay is the most promising channel for measuring Higgs boson interactions with second-generation fermions. The ATLAS~\cite{ATLAS:2020fzp} and CMS~\cite{CMS:2020xwi} analyses, based on the full Run-2 dataset, have been extrapolated to the HL-LHC~\cite{ATLAS_PUBNOTE_H,CMS-PAS-FTR-21-006}, accounting for higher signal and background production cross sections, and expected improvements in dimuon mass resolution up to 30\% due to detector upgrades~\cite{CERN-LHCC-2017-021,CERN-LHCC-2017-009}. 
Table~\ref{tab:H_Zy_mumu} summarizes the projected  precision of the signal strength measurement for $H\rightarrow Z\gamma$ and $H\to \mu\mu$, which are  expected to reach 14\% and 6\%, respectively.

In addition, the large number of produced Higgs bosons ($\sim$380 million in ATLAS+CMS for 3~ab$^{-1}$ per experiment) makes ATLAS and CMS uniquely sensitive to rare Higgs boson decays, such as $H \to J/\psi \, \gamma$ and $H \to \phi \, \gamma$, as well as exotic final states, like $H\to e\mu$ or multi-body decays, predicted by many BSM extensions of the Higgs sector. 

\begin{figure}[ht]
    \centering
    \includegraphics[width=0.45\linewidth]{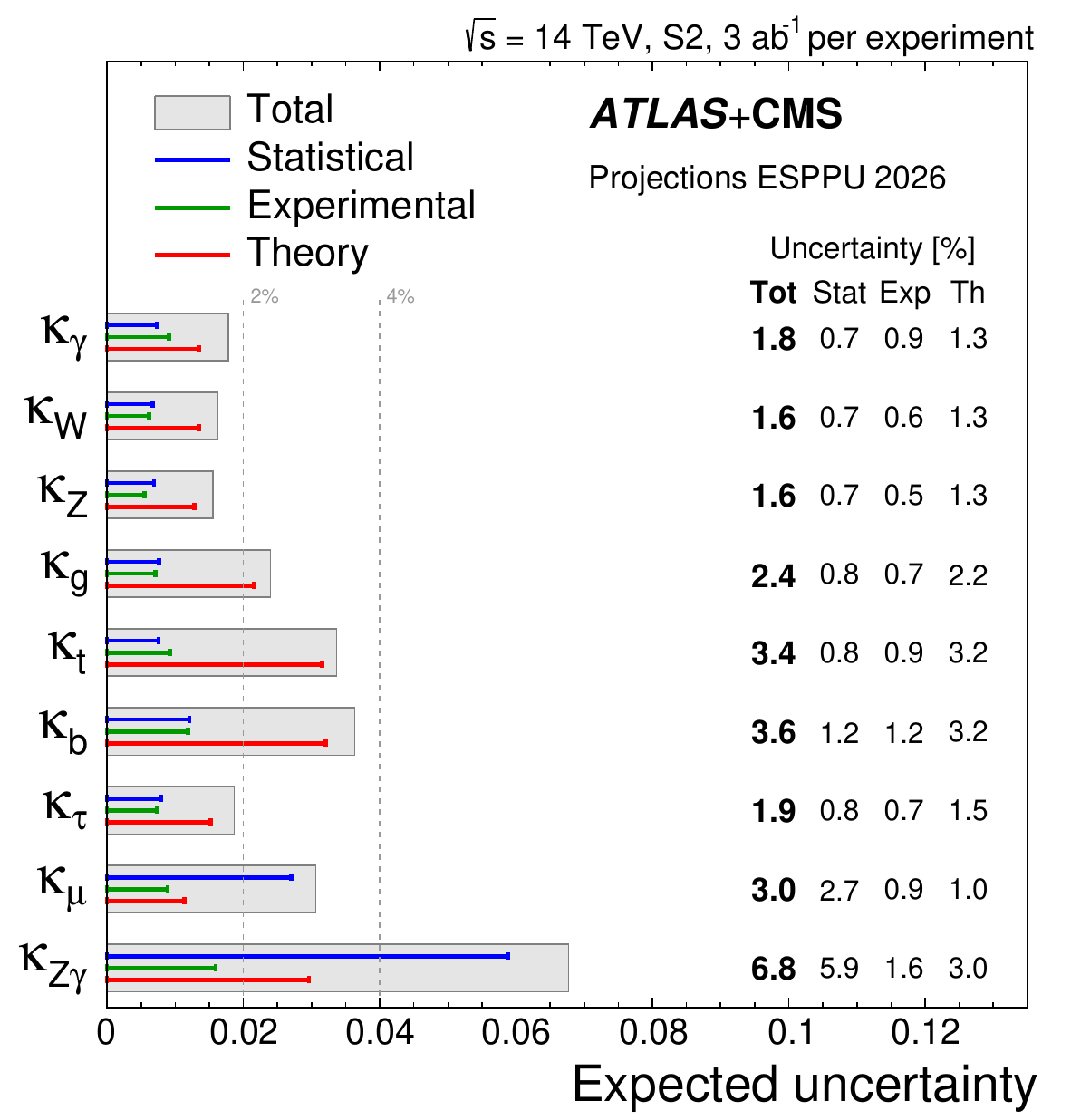}
    \includegraphics[width=0.45\linewidth]{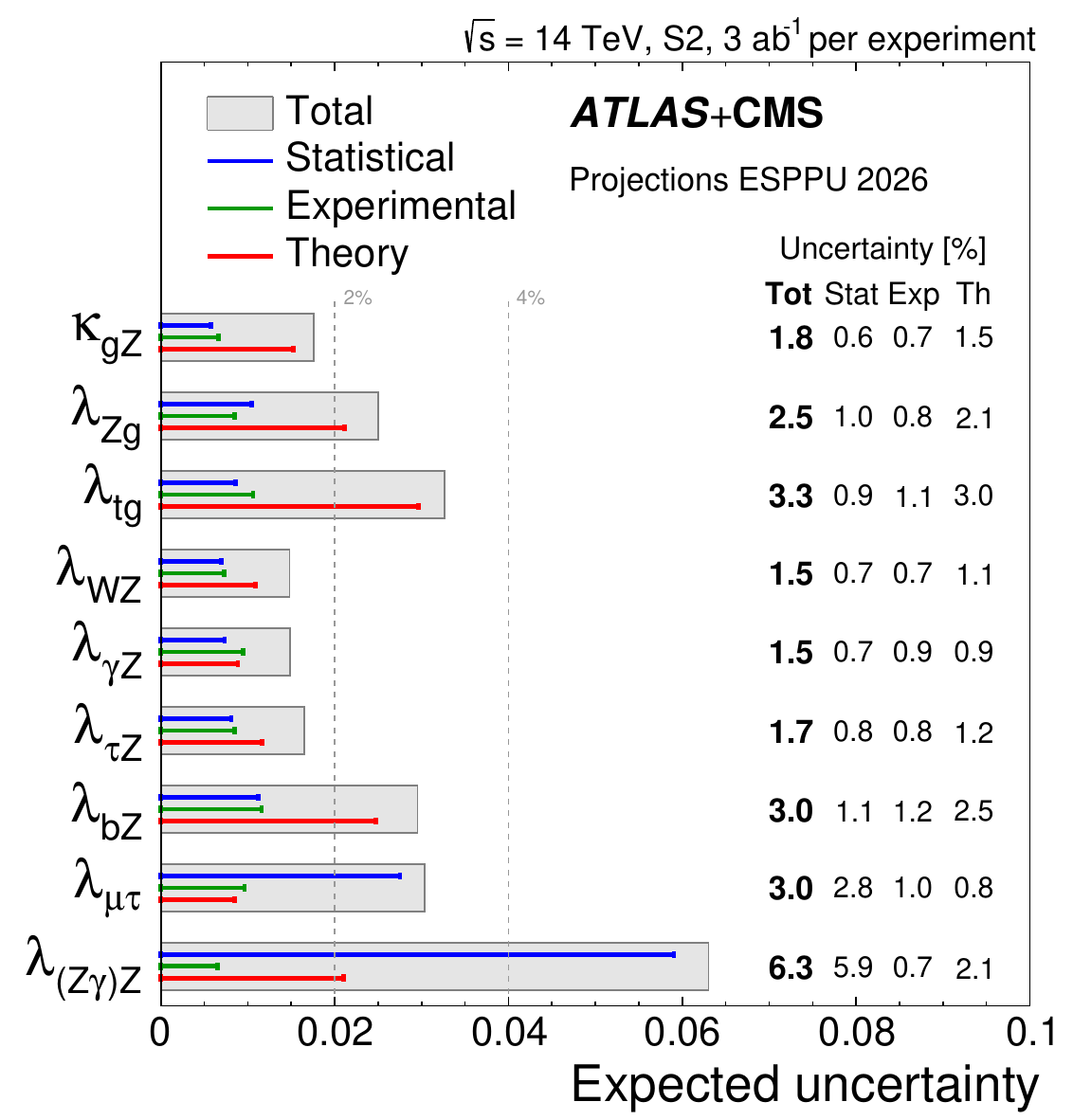}
    \caption{The projected uncertainty in the combined coupling signal strength modifiers (left) and their ratios (right) with 3~ab$^{-1}$ of $pp$ collisions under the \Stwo systematic uncertainty scenario, assuming that the Higgs boson decays only to final states predicted by the SM.}
    \label{fig:kappa_H}
\end{figure}

The most recent ATLAS projections based on full Run-2 analyses ($H\to\gamma\gamma$, $Z\gamma$, $\mu\mu$~\cite{ATLAS_PUBNOTE_H}, $VH\to b\bar{b}$~\cite{ATLAS_PUBNOTE_VH}, $H\to\tau\tau$~\cite{ATL-PHYS-PUB-2022-003}) have been combined with previous projections~\cite{ATL-PHYS-PUB-2018-054} ($H\to ZZ$, $H\to WW$, $ttH \to b\bar{b}$, $ttH \to \text{multilepton}$) to simultaneously extract the Higgs boson coupling modifiers~\cite{ATLAS_PUBNOTE_H}: $\kappa_\gamma$ (Higgs-photon), $\kappa_W$ (Higgs-W boson), $\kappa_Z$ (Higgs-Z boson) , $\kappa_g$ (Higgs-gluon), $\kappa_t$ (Higgs-top),  $\kappa_b$ (Higgs-bottom), $\kappa_\tau$ (Higgs-tau), $\kappa_\mu$ (Higgs-muon), and $\kappa_{Z\gamma}$ (Higgs-Z-photon). In the results for the $\kappa_j$ parameters presented here the Higgs boson branching fraction to BSM final states is fixed to zero. These projections have been further combined with CMS coupling projections presented in Ref.~\cite{CMS-PAS-FTR-18-011}, except for $\kappa_\mu$ and $\kappa_{Z\gamma}$, which are taken from the above-described most recent full Run-2 CMS projections. Following the procedure adopted in Refs.~\cite{Dainese:2019rgk,ATLAS:2019mfr},
the combination is performed separately for each $\kappa$ using the best linear unbiased estimator (BLUE) methodology, described in Ref.~\cite{Valassi:2003mu}, considering the theory uncertainties associated with each $\kappa$ measurement to be fully correlated among experiments. It should be noted that the extraction of the $\kappa$ signal modifiers is done under the assumption of a SM value for the Higgs boson width. 
This approach does not account for potential correlations between parameters. The results at 3~ab$^{-1}$ for the \Stwo scenario are summarized in Fig.~\ref{fig:kappa_H} (left). The comparison between this and the previous HL-LHC projection is shown in Fig.~\ref{fig:kappa_H_imp} of Appendix~\ref{app:app}. 
The precision of $\kappa_\mu$ and $\kappa_{Z\gamma}$ improves by 30\%, while the other couplings show small changes. 
It is important to note that theory uncertainties are still dominant for most couplings. 
Reducing theoretical uncertainties below the factor-of-two benchmark assumed in \Stwo would significantly enhance the precision of most of the coupling modifiers.
To eliminate the dependence on the Higgs boson’s total width and mitigate common systematic uncertainties, the projected measurement of ratios of Higgs boson coupling modifiers is also provided. A key reference parameter, $\kappa_{gZ} = \kappa_g\kappa_Z/\kappa_H$, is introduced to represent the high-precision measurement of the $gg \to H \to ZZ$ production rate. The cross-sections in other channels are then reformulated in terms of $\kappa_{gZ}$ and a set of coupling strength scale factor ratios, defined as $\lambda_{XY} = \kappa_X/\kappa_Y$. The results are presented in Fig.~\ref{fig:kappa_H} (right). The properties of the Higgs boson will be also probed through other measurements such as cross-section times branching fractions. Unlike the measurements of the coupling modifiers $\kappa$, these measurements will not be limited by large theory uncertainties.

Finally, assuming the SM, the ATLAS+CMS combination of the $H \to ZZ$ off-shell cross section can constrain the Higgs width with an uncertainty of 0.7 MeV with 3\abinv~\cite{CMS-PAS-FTR-18-011}, projecting the current measurements under the \Stwo scenario.

\section{Di-Higgs boson physics}
\label{sec:non-res}

\begin{figure}[tb!]
    \centering
        \includegraphics[width=0.49\textwidth]{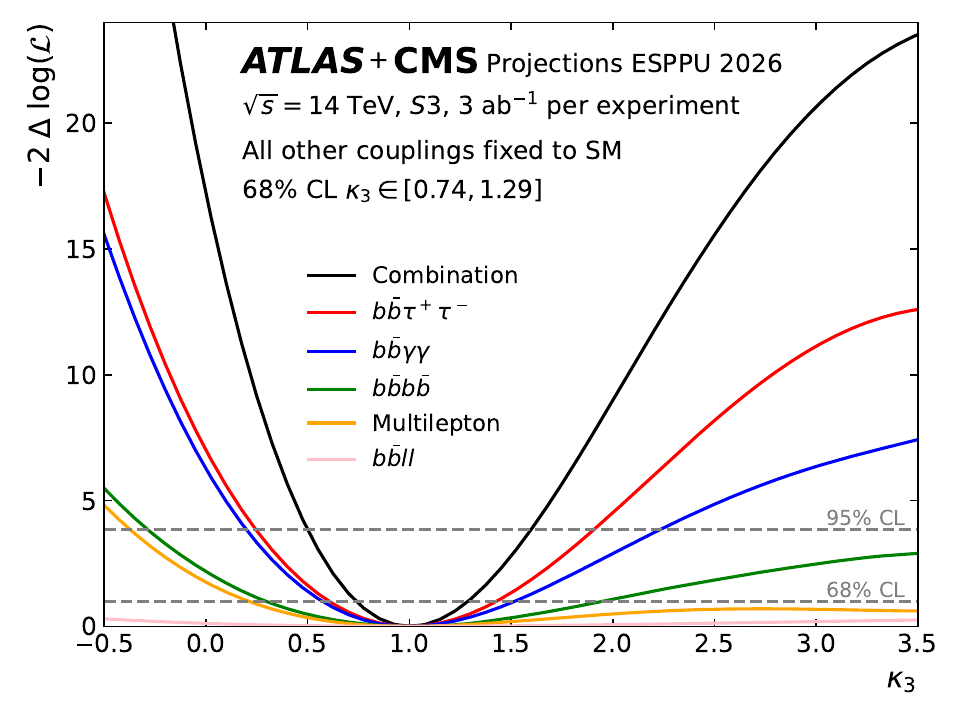}~
        \includegraphics[width=0.49\textwidth]{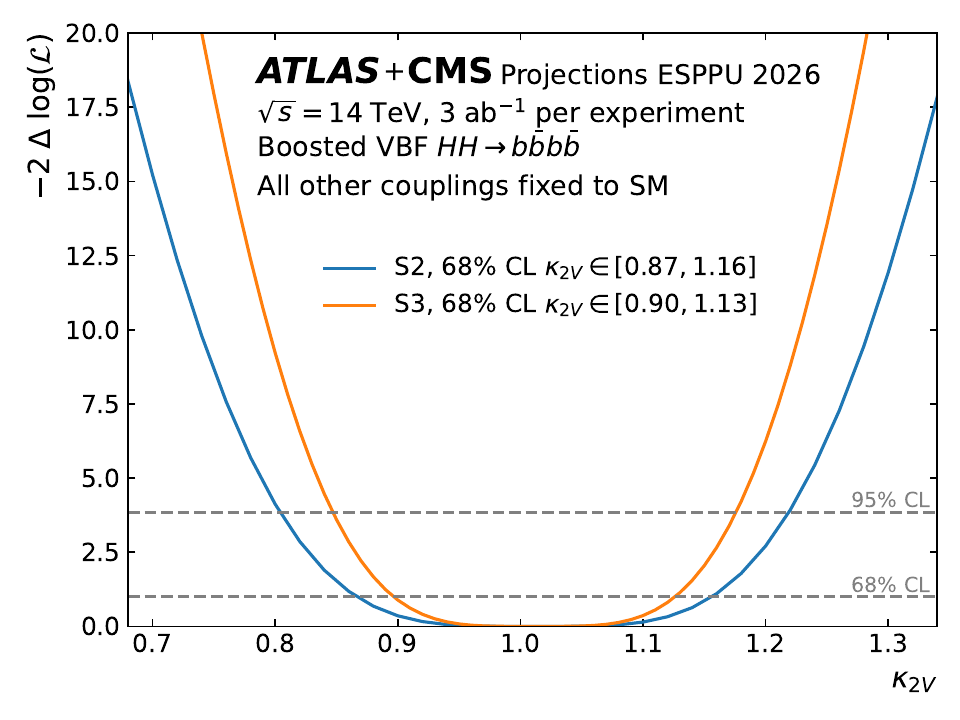}\\
     \caption{Left: Expected ATLAS+CMS \kl likelihood scans for single decay channels and the combination for 3~ab$^{-1}$  for the \Sthree scenario, obtained fixing $\kappa_{3}^\mathrm{true}=1$. Right: The
     ATLAS+CMS projections for $\kappa_{2V}$ in the \Stwo and \Sthree scenarios, fixing $\kappa_{2V}^\mathrm{true}=1$.\label{fig:HHmain_figure}}
\end{figure}

The measurement of $HH$ production is sensitive to modifications to the BEH potential, since it directly probes the deviation of the trilinear coupling parameter $\lambda_{3}$ from the SM, parametrized via the coupling strength modifier $\kappa_{3} = \lambda_{3}/ \lambda_{3}^{\text{SM}}$.
In recent years, major improvements to the analysis techniques have increased the sensitivity of ATLAS and CMS to this rare process. In particular, the use of graph-based architectures for deep-learning based jet tagging has enhanced the sensitivity to various decay modes (e.g.~$H \to b \bar b$ and $H \to \tau^+ \tau^-$) at both small and large Higgs boson $p_T$, where the Higgs boson decay products are boosted and overlap.
Figure~\ref{fig:HHmain_figure} (left) shows the combination of the ATLAS and CMS extrapolations for the decay modes listed in Table~\ref{tab:HHresult}.
The details of the assumptions used in the \Stwo and \Sthree scenarios can be found in Refs.~\cite{ATLAS_PUB_dihiggs,CMS_PAS_dihiggs}. In this study, the \Sthree scenario is defined considering a 5\% improvement both in $b$-jet tagging~\cite{CMSbtagplots,ATLASNNbtag,ATLASGNNbtag}  and hadronic tau reconstruction efficiencies~\cite{CMStau}, expected already for the incoming Run-3 results.
Beyond Run~3, additional improvements are expected, in terms of trigger, detector, and analysis techniques. 
In the combination,
the ATLAS $b\bar{b}\tau^+\tau^-$  projection~{\cite{ATLAS_PUB_bbtautau}}, the CMS resolved and boosted projections $b\bar{b}b\bar{b}$~\cite{CMS_PAS_dihiggs}, the ATLAS multilepton~\cite{PUB_ATLAS_multi} and ATLAS $b\bar{b}\ell^+\ell^-$ \cite{ATLAS_PUB_dihiggs} projections have been adopted for both experiments, since they can reach similar sensitivity by using the same experimental techniques. In the case of the $b\bar{b}\tau^+\tau^-$ channel, the CMS Run-2 sensitivity was limited by the trigger. An improved trigger has already been deployed by CMS for Run~3~\cite{PUB_CMS_bbtautautrigger}, achieving similar performance as the ATLAS trigger.
The $b\bar{b}\gamma\gamma$ projection is based on independent ATLAS~\cite{PUB_ATLAS_bbyy} and CMS~\cite{CMS_PAS_dihiggs} projections.

Table~\ref{tab:HHresult} shows the expected significance on the $HH$ signal yield and the corresponding 68\% confidence intervals (CIs) on $\kappa_3$. Values are quoted per decay channel and per experiment, for the two scenarios. The combination of ATLAS and CMS projections result in an expected $> 5\sigma$ observation of $HH$ production already with 2 \abinv in the \Stwo scenario, increasing to more than $7\sigma$ with 3\abinv (to be compared to the $4\sigma$ projection reported for the previous European Strategy~\cite{Dainese:2019rgk,ATLAS:2019mfr}). 
Going from \Stwo to \Sthree brings 5\% gain in precision, while the increase in luminosity from 2 to 3~\abinv brings a gain of 20\% on the signal significance. While a single-experiment observation is  unlikely at 2\abinv, it's possibly in reach at 3\abinv, also in view of further analysis optimization. A precision on \kl below 30\%, namely -26\% $/$ +29\%, can be obtained in the  \Sthree scenario with 3\abinv.

\begin{table}[t]
    \centering
\resizebox{\textwidth}{!}{
    \begin{tabular}{lcc|cc|cc}
    \hline
         \rule{0pt}{11pt} & \multicolumn{2}{c|}{2~ab$^{-1}$ (\Stwo)} & \multicolumn{2}{c|}{3~ab$^{-1}$ (\Stwo)} & \multicolumn{2}{c}{3~ab$^{-1}$ (\Sthree)} \\
         & ATLAS & CMS & ATLAS & CMS & ATLAS & CMS \\
    \hline
         \multicolumn{7}{c}{$HH$ statistical significance}\\
    \hline
        \rule{0pt}{11pt}$b\bar{b}\tau^+\tau^-$ & 3.0$^\dagger$ & 1.9 & 3.5$^\dagger$ & 2.4 & \bf{3.8$^\dagger$} & \bf{2.7} \\
        $b\bar{b}\gamma\gamma$ & 2.1$^\dagger$ & 2.0$^\dagger$ & 2.4$^\dagger$ & 2.4$^\dagger$ & \bf{2.6$^\dagger$ }& \bf{2.6$^\dagger$} \\
        $b\bar{b}b\bar{b}$ resolved & 0.9 & 1.0$^\dagger$ & 1.0 & 1.2$^\dagger$ & \bf{1.0} & \bf{1.3$^\dagger$} \\
        $b\bar{b}b\bar{b}$ boosted & $-$ & 1.8$^\dagger$ & $-$ & 2.2$^\dagger$ & $-$ & \bf{2.2$^\dagger$} \\
        Multilepton & 0.8$^\dagger$ & $-$ & 1.0$^\dagger$ & $-$ & \bf{1.0$^\dagger$} & $-$ \\
        $b\bar{b}\ell^+\ell^-$ & 0.4$^\dagger$ & $-$ & 0.5$^\dagger$ & $-$ & \bf{0.5$^\dagger$} & $-$ \\
        Combination & 3.7 & 3.5 & 4.3 & 4.2 & \bf{4.5} & \bf{4.5} \\
    \hline
        ATLAS+CMS & \multicolumn{2}{c}{6.0} & \multicolumn{2}{c}{7.2} & \multicolumn{2}{c}{\bf{7.6}} \\
    \hline
        \multicolumn{7}{c}{\kl 68\% confidence interval}\\
     \hline
         \rule{0pt}{11pt}$b\bar{b}\tau^+\tau^-$ & $[0.3,~1.8]^\dagger$ & $[0.1,~3.0]$ & $[0.4,~1.7]^\dagger$ & $[0.2,~2.2]$ & \bf{$[0.5,~1.6]^\dagger$} & \bf{$[0.3,~2.0]$} \\
         $b\bar{b}\gamma\gamma$ & $[0.3,~2.0]$$^\dagger$ & $[0.2,~2.3]$$^\dagger$ & $[0.4,~1.8]$$^\dagger$ & $[0.3,~2.0]$$^\dagger$ & \bf{$[0.5,~1.7]$$^\dagger$} & \bf{$[0.4,~1.9]$$^\dagger$} \\
         $b\bar{b}b\bar{b}$ resolved & $[-0.7,~6.3]$ & $[-0.6,~7.6]^\dagger$ & $[-0.5,~6.1]$ & $[-0.3,~7.3]^\dagger$ & \bf{$[-0.5,~6.1]$} & \bf{$[-0.3,~7.2]^\dagger$} \\
         $b\bar{b}b\bar{b}$ boosted & $-$ & $[-0.6,~8.5]^\dagger$ & $-$ & $[-0.4,~8.2]^\dagger$ & $-$ & \bf{$[-0.4,~8.2]^\dagger$ }\\
         Multilepton & $[-0.2,~4.9]^\dagger$ & $-$ & $[-0.1,~4.7]^\dagger$ & $-$ & \bf{$[-0.1,~4.7]^\dagger$} & $-$ \\
         $b\bar{b}\ell^+\ell^-$ & $[-2.4,~9.3]^\dagger$ & $-$ & $[-2.2,~9.2]^\dagger$ & $-$ & \bf{$[-2.1,~9.1]^\dagger$} & $-$ \\
         Combination & $[0.6,~1.5]$ & $[0.4,~1.7]$ & $[0.6,~1.5]$ & $[0.5,~1.6]$ & \bf{[$0.6,~1.4]$} & \bf{$[0.6,~1.5]$} \\

     \hline
       ATLAS+CMS & 
       \multicolumn{2}{c}{\multirow{2}{*}{$-32 \% \; /+\!37 \%$}} & 
       \multicolumn{2}{c}{\multirow{2}{*}{$-27 \% \; /+\!31 \%$}} & 
       \multicolumn{2}{c}{\multirow{2}{*}{\bf{$-26 \% \; /+\!29 \%$}}} \\
        uncertainty  \\
    \hline
    \multicolumn{7}{l}{{\it $\dagger$ used in the ATLAS+CMS combination}}\\
    \end{tabular}
}
    \caption{Combined ATLAS and CMS expected statistical significance for $HH$ production and the corresponding 68\% confidence interval on \kl at 3~ab$^{-1}$, derived assuming $\kappa_{3}^\mathrm{true}=1$. The last row reports the projected ATLAS+CMS percentage uncertainty on \kl in the various scenarios. The measurement labelled by the $\dagger$ symbol have been used in the ATLAS+CMS combination. When the~$\dagger$ symbol is present on only one of the two experiments, this measurement has been extrapolated to 6\abinv assuming the same sensitivity on that channel for the two experiments.
    \label{tab:HHresult}}
\end{table}

\begin{wrapfigure}{r}{0.55\textwidth}
\vspace{-3mm}
    \includegraphics[width=\textwidth]{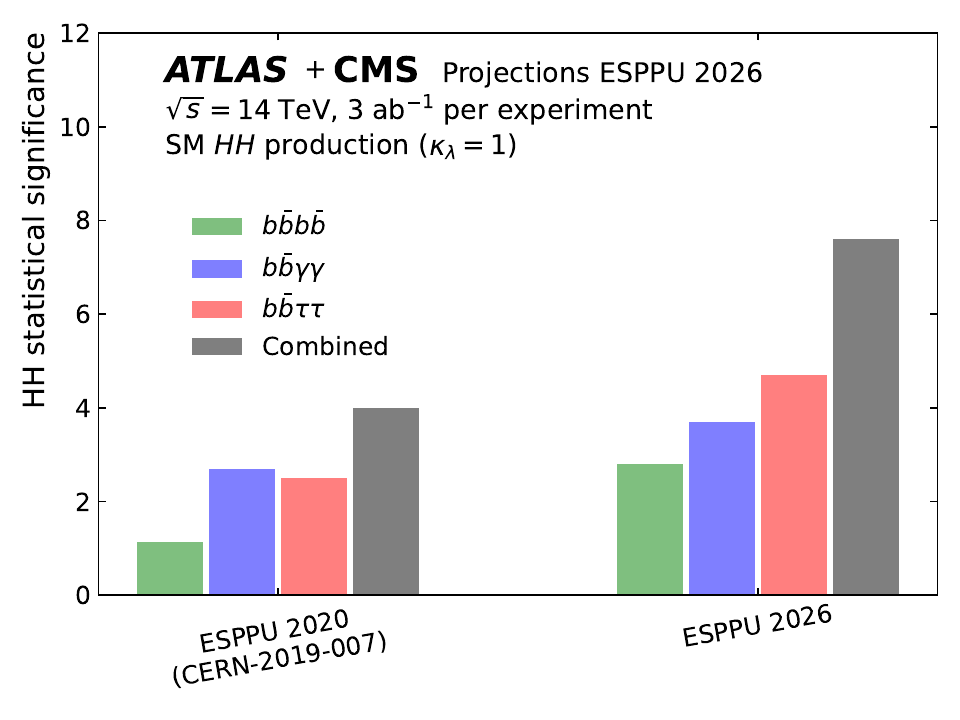}
    \caption{Comparison of the ESPPU 2020 and ESPPU 2026 projected 3\abinv $HH$ sensitivities from various final states, and their combinations.
    \label{fig:HHproj_comparison}}
\vspace{-3mm}
\end{wrapfigure}

A comparison of this projection to that of the previous European Strategy (see Fig.~\ref{fig:HHproj_comparison}) shows how various improvements in the  analysis technique translate into a much stronger projection. Even neglecting the improvements coming from the detector upgrades, it is expected that further optimization in the ongoing LHC run and during the HL-LHC phase would follow this trend and the uncertainty on \kl will then be reduced well below 30\%.

Modifications of the VVHH coupling are parametrized by the $\kappa_{2V}$ coupling modifier, which can be accessed via the measurement of vector boson fusion (VBF) $HH$ production. Figure~\ref{fig:HHmain_figure} (right) shows the sensitivity expected from ATLAS+CMS on $\kappa_{2V}$ with 2 and 3 \abinv, assuming $\kappa_{2V}^\mathrm{true} = \kappa_{2V}^\mathrm{SM} = 1$. 
The combined projection is based on the ATLAS boosted VBF $HH\rightarrow b\bar{b} b \bar{b}$   search~\cite{ATLAS:2024lsk,ATL-PHYS-PUB-2025-005}, as the sensitivity of this process is dominant. In the \Stwo scenario, a $\kappa_{2V}$ precision of $\sim 13\%$ is expected with 3~\abinv.

\begin{figure}[th]
    \centering
    \hspace{0.010\textwidth}
    \includegraphics[width=.4\textwidth,trim=3mm 8mm 17mm 18mm,clip]{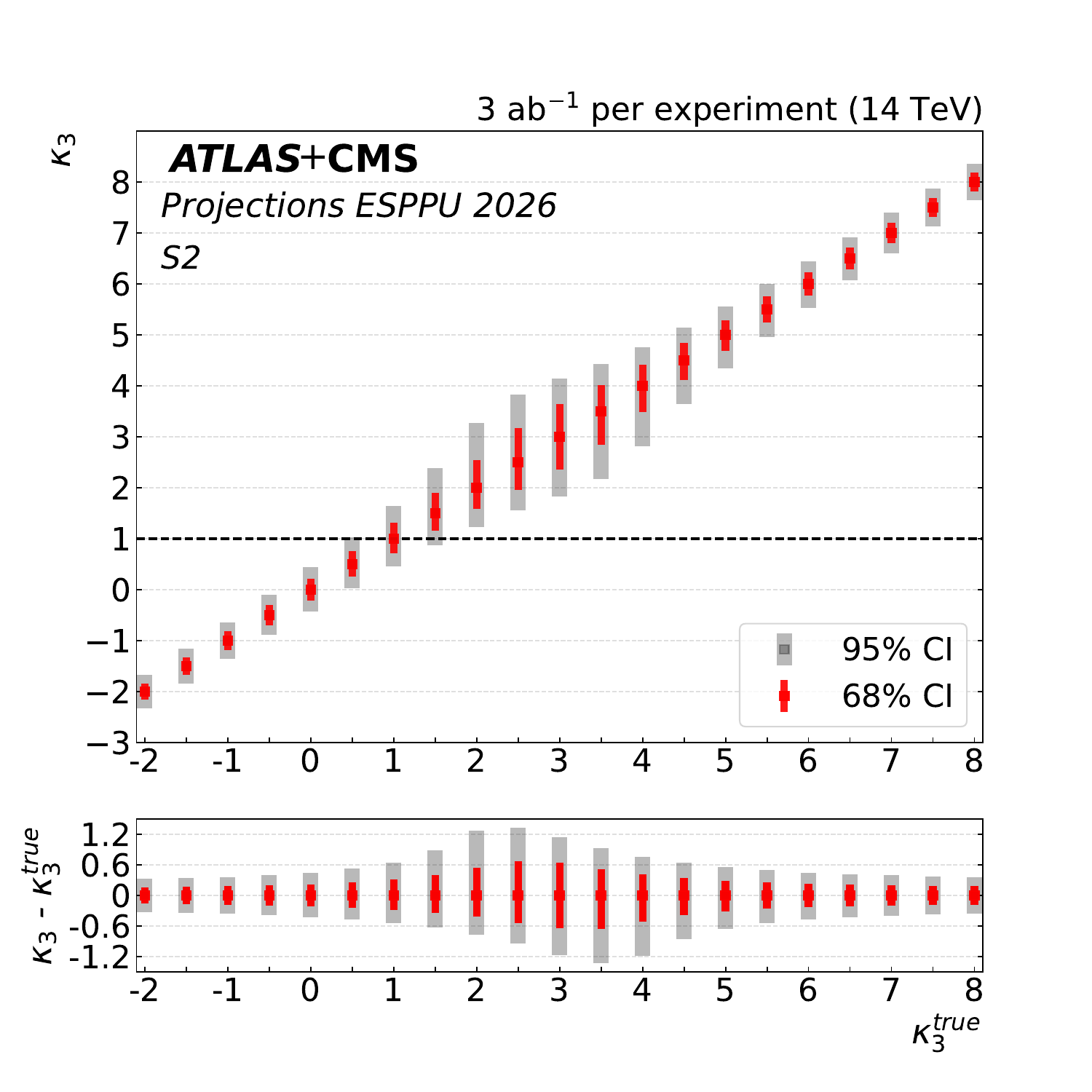}
    \hspace{0.010\textwidth}
    \includegraphics[width=0.46\textwidth,trim=0mm 2mm 12mm 8mm,clip]{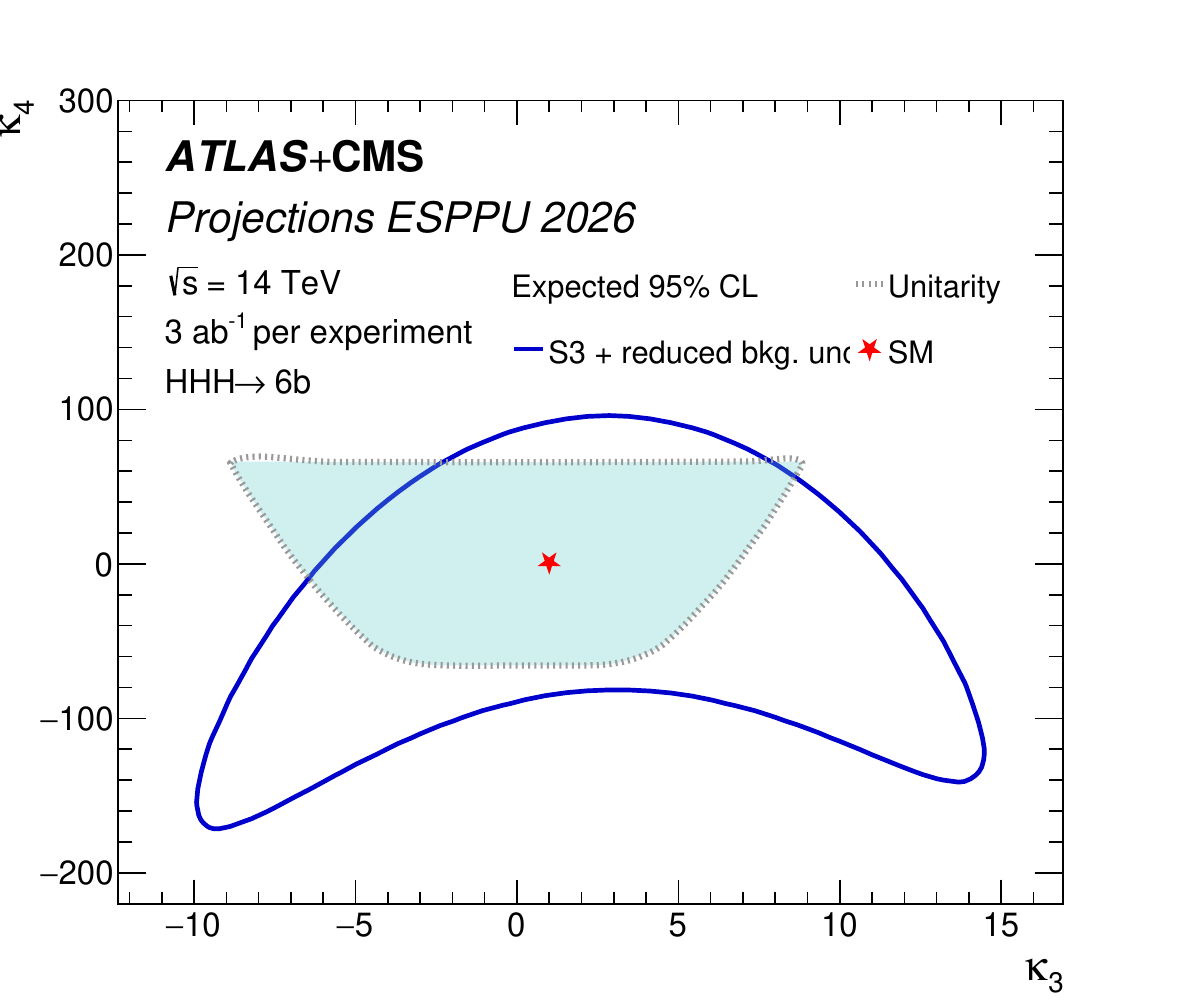}
    \caption{
      Left: The ATLAS+CMS projection on the precision of the determination of \kl as a function of $\kappa_{3}^\mathrm{true}$.  The 68\% and 95\% confidence intervals are shown in the upper plot, while the lower plot shows the \kl deviation from the simulated $\kappa_{3}^\mathrm{true}$ value and its 68\% and 95\% confidence intervals.
      Right: 95\% CL constraints from the $HHH$ search projection on $\kappa_3$ and $\kappa_4$.
      Results are shown for 3~ab$^{-1}$ per experiment at $\sqrt{s}=14$~TeV in scenario
      %s \Sone, and separately 
      \Sthree with data-driven background uncertainties. Unitarity limits, as calculated in Ref.~\cite{Stylianou:2023xit}, are overlaid in the region bounded by the grey dashed line. 
    }
    \label{fig:klerr_vs_kl_and_HHH}
\end{figure}

In Fig.~\ref{fig:klerr_vs_kl_and_HHH} (left), the ATLAS+CMS projections on the precision of the \kl determination  for different possible values of the Higgs trilinear coupling are displayed for the \Stwo scenario and 3 \abinv. 
Because of a destructive interference between the box diagram and the diagram with the trilinear coupling in ggF production, the $HH$ production cross section at the HL-LHC is minimal when the true  value of \kl ($\kappa_{3}^\mathrm{true}$) is about 2.5.
Regardless of the consequent drop in signal yield, 
the ATLAS+CMS  \kl combination can exclude the SM value of \kl
at 95\% CL for $\kappa_{3}^\mathrm{true} \gtrsim 1.7$ and $\kappa_{3}^\mathrm{true} \lesssim 0.5$.
In addition, thanks to the improvement in precision, the shape of the expected likelihood as a function of \kl has a unique minimum, regardless of the true value of \kl,  i.e.~the second local minimum observed in the previous projection~\cite{Cepeda:2019klc} is removed.

\section{Triple Higgs boson production and quartic Higgs couplings}

While Higgs boson pair production provides the most precise determination of the Higgs trilinear self-coupling,  the search for $HHH$ production represents the only direct access to the Higgs quartic self-coupling $\lambda_{4}$.  Similarly to $\lambda_{3}$, deviations from the SM quartic coupling $\lambda_{4}$ are parametrized via the coupling strength modifier $\kappa_{4} = \lambda_{4} / \lambda_{4}^{SM}$. 

Projections for the HL-LHC reach discussed here are based on the available experimental search for $HHH$ production in the $6b$ final state by the ATLAS Collaboration~\cite{ATLAS:2024xcs} with LHC Run-2 data, and the corresponding projection study for the HL-LHC~{\cite{ATLAS-PUB-NOTE-HHH}. More search channels with comparable sensitivity will be added, improving the experimental bound through a global combination.
Projections to 3 ab$^{-1}$ for both experiments are derived from this study assuming the ATLAS sensitivity in the 
\Sthree scenario.
The two-dimensional 95\% CL contour in $\kappa_3$ and $\kappa_4$ is shown in Fig.~\ref{fig:klerr_vs_kl_and_HHH} (right) for 3~ab$^{-1}$. In the figure, unitarity limits~\cite{Stylianou:2023xit} are overlaid in the region bounded by the grey dashed line. The expected limit on the $HHH$ cross section for the  \Sthree scenarios is 86 times the SM expectation. With this precision, ATLAS and CMS will start excluding portions of the region bounded by the unitarity limit.

\section{Searches for a heavy scalar}
\label{sec:resonant}

Multiple SM extensions predict measurable effects on Higgs-related quantities, such as $H$ and $HH$ couplings. At the same time, they typically have further implications for LHC physics, e.g.~predicting the existence of new particles that can be directly searched for. In this study, we consider a model with a heavy scalar $S$ extending the Higgs sector. Such a new resonance could be probed in various signatures, including $S\rightarrow HH$, $S\rightarrow ZZ$, and $t\bar t S \to t \bar t t \bar t$.

The projection of the resonant searches is based on the CMS Run-2 results for the $HH$~\cite{CMS:2024phk} and $ZZ$~\cite{CMS-PAS-HIG-24-002,CMS_PAS_X_to_ZZ_tt} signatures, extrapolated to 2 and 3~ab$^{-1}$ assuming the same performance for both experiments. More details can be found in Ref.~\cite{CMS:2024phk}.
The expected upper limit on the cross section for the $S \to HH$ and $S \to ZZ$ processes is shown as a function of the resonance mass in Fig.~\ref{fig:Resonant_limits} and Fig.~\ref{fig:Resonant_limitsapp} (left)  of Appendix~\ref{app:app}, respectively. In both cases a potential signal is considered in the narrow-width approximation, where the decay width is assumed to be negligible compared to the detector resolution. 
The $S\to H H$ projection combines the currently available prospects in the $b\bar b \tau^+ \tau^-$, $b\bar b \gamma \gamma$, and $b\bar b b\bar b$ (boosted) final states, which is conservative as also other channels are expected to contribute at masses up to $\sim$1~TeV~\cite{CMS-PAS-HIG-20-012,CMS-PAS-HIG-22-012}.

\begin{wrapfigure}[16]{r}{0.55\textwidth}
\vspace{-5mm}
    \includegraphics[width=\textwidth]{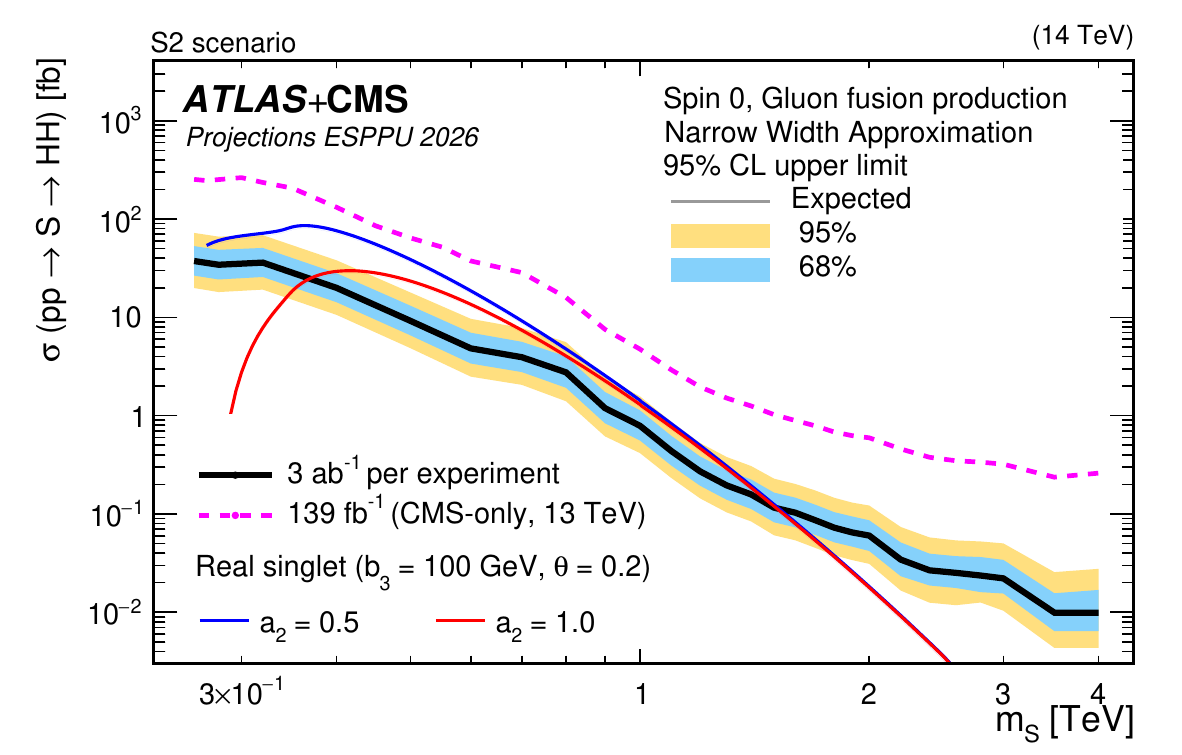}
    \caption{Expected 95\% CL upper limits on the $\sigma(pp \to S \to HH)$ cross section as a function of the scalar mass $m_S$, produced via gluon fusion using the narrow width approximation. The projection is derived assuming 3 \abinv\ per experiment for the \Stwo\ scenario at 14~TeV. For comparison, production cross section curves for the model described in Section~\ref{sec:precisionheavyscalarsinglet} are shown, for two values of the scalar portal coupling $a_2$.
    \label{fig:Resonant_limits}}
\vspace{-1mm}
\end{wrapfigure}

A scalar coupling to top quarks may alter the \tttt process.
In this context, we consider $S$ production in association with a top-quark pair ($t\bar{t}S$), with the $S$ decaying into a $t\bar{t}$ pair. Starting from ATLAS searches with full Run-2 data~\cite{EXOT-2022-13,EXOT-2019-26}, we project the combined ATLAS+CMS HL-LHC reach. 
Figure~\ref{fig:Resonant_limitsapp} right illustrates the expected upper limit on $\sigma(pp\rightarrow t\bar{t}S) \times \mathcal{B}(S \rightarrow t\bar{t})$, at 95\% CL.
Further constraints can be derived from searches for inclusive heavy scalar production with subsequent $S \to t \bar t$ decay.  While this channel was found subdominant with respect to the searches listed above, the overall interplay between all these searches and precision Higgs physics can probe large portions of the parameter space of specific BSM models.

\vspace{2cm}

\section{Constraining the shape of the BEH potential and a first-order phase transition}

Measurements of the Higgs boson self-coupling are crucial to inform on the shape of the EWSB potential, which can be expressed as
\begin{eqnarray}
   V^{\rm SM}(\phi) =  \frac18 \left(\phi^2-1\right)^2 ,~\mathrm{and}~ 
       \phi = \frac{H}{\sqrt{2}}+ 1\,,\, 
    m_H  = \frac{1}{\sqrt{2}}   \,. 
    \label{eq:SMH3main}
\end{eqnarray}
where  $\phi$ is the Higgs doublet given in terms of the physical Higgs scalar ($H$). The notation has been simplified taking into account that $m_t = y_t v/\sqrt{2}$ and $y_t \sim 1$, so that $v = \sqrt{2}$ when expressed in units of $m_t$  (see Appendix~\ref{app:theory}).

At tree level, the Higgs boson pair production is sensitive to the trilinear coupling \kl that mediates the $HHH$ interaction. Assuming that all other couplings involved in a given Higgs boson production process are known with sufficient precision, information about the triple self-coupling can be extracted from the measured event rates of $HH$ production. However, measuring the Higgs self-coupling alone is insufficient to fully determine or constrain the shape of the EWSB potential. In order to provide model-independent conclusions on fundamental questions, such as whether the EWSB transition is first-order (a key element for EW baryogenesis), an assumption about the form of the potential is generally required.

In the following, the projected HL-LHC constraints on \kl is used to constrain BSM scenarios where new heavy particles lie beyond a large energy scale cut-off $\Lambda$ and hence cannot be produced at the LHC. In such cases, the scalar potential can always be expressed as a deformation of the SM EWSB potential given in Eq.~(\ref{eq:SMH3main}).
Four EWSB potentials that may arise in realistic BSM scenarios are considered in this study.
In the context of a SM effective field theory (SMEFT)~\cite{SMEFT,Brivio:2017btx}, a dimension-6 EFT (SMEFT~6, Eq.~(\ref{eq:SMEFT8simp_app2}) of Appendix~\ref{app:theory}) and dimension-8 EFT (SMEFT~8, Eq.~(\ref{eq:SMEFT8simp_app3}) of Appendix~\ref{app:theory}) potentials are considered.
In addition, modifications of the low-energy SM BEH potential (see Eq.~(\ref{eq:SMH3main})) by a small term with a logarithmic (Eq.~(\ref{eq:log})) or an exponential (Eq.~(\ref{eq:exp})) dependence on the scalar doublet inner product $\Phi^\dagger \Phi$ are also discussed.

These four alternative scenarios predict a strong first-order phase transition (FOPT)
in the early universe~\cite{Huang:2016cjm,Ramsey-Musolf:2024ykk}, for $\kappa_3>\kappa_3^{\mathrm{min}}$ and 
$\kappa_4>\kappa_4^{\mathrm{min}}$, which is not realized in the SM~\cite{Aoki:2006we}.

\begin{wrapfigure}{R}{0.55\textwidth}
  \vspace{-3mm}
    \includegraphics[width=0.92\textwidth]{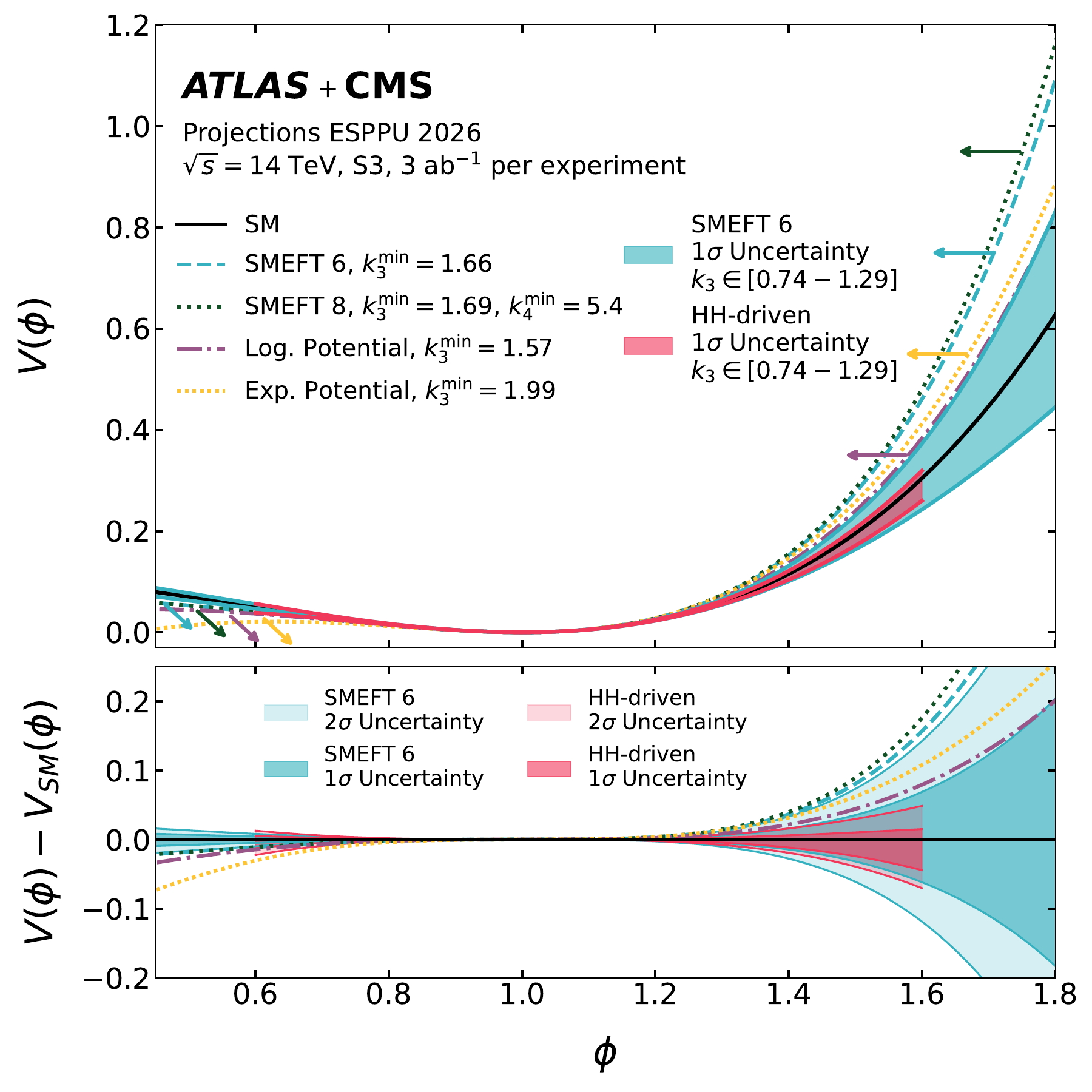}\\
    \hspace{3mm}\includegraphics[width=0.91\textwidth]{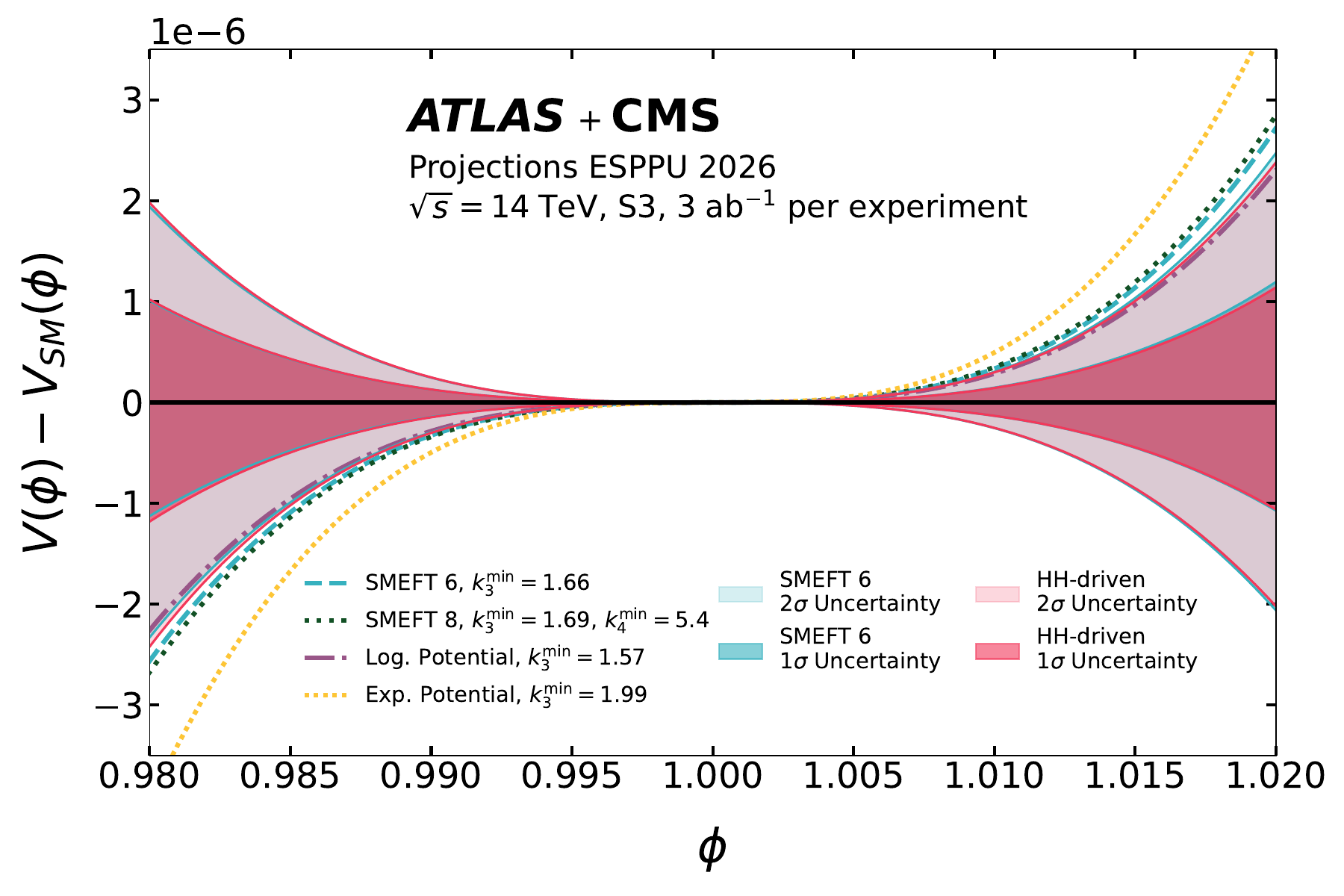} 
    \caption{(Top) BEH potentials in various models which predict a first-order phase transition~\cite{ADD}. The models are compared with the SM BEH potential. Two approaches (SMEFT~6 and HH-driven) are used to show the expected uncertainties on the Higgs self-coupling achieved by combining ATLAS and CMS at 3\abinv in the \Sthree scenario. The dashed lines show the boundary of the regions for which the alternative models predict a strong first-order phase transition. The arrows indicate the region where the strong first-order phase transition happens. Further details can be found in the text. The bottom panel shows the difference between the potential $V(\phi)$ and its SM expectation $V_{SM}(\phi)$. Here, the 68\% and 95\% CL uncertainty bands on the shape of $V(\phi)$ are shown, for the HH-driven and SMEFT~6 potentials (see text). (Bottom)  A zoom into the $V(\phi)-V_{SM}(\phi)$ difference around the minimum of $V(\phi)$, corresponding to the validity range of the HH-driven band.\label{fig:Potentials_withExclusion}}
\end{wrapfigure}

In Fig.~\ref{fig:Potentials_withExclusion} (top), the SM BEH potential and the 
allowed band from the expected experimental precision on \kl are compared to these considered alternative scenarios of EWSB.
The four non-solid lines correspond to the four alternative potentials and display the limit of the variation of the shapes that could still imply a strong FOPT in the early universe. 
These lines are obtained for values of the parameters as indicated on the legend of the figure.
Only variations of the shapes in the direction indicated by the arrows on the four curves would allow a strong FOPT.  In other words, the arrows on each of these curves indicate the direction in which the modified potentials implying a strong FOPT lie on the plot.

The darker blue and red shaded areas show the allowed band for the BEH potential at 68\% CL, derived from the ATLAS and CMS projections.
The red area illustrates the allowed range on the third derivative of the potential (which is sensitive to $\lambda_3$), ignoring higher-order terms  ($\kappa_i=0$ with $i\ge4$, see Eq.~(\ref{eq:SMEFT8simp_app2}) and Eq.~(\ref{eq:SMEFT8simp_app3}) in Appendix~\ref{app:theory}). Its finite range around the minimum aims to represent the fact that LHC constraints coming from $HH$ production can only determine $\kappa_3$ (at LO) in a model-independent way close to the minimum. It is therefore labelled as ``HH-driven'', as it represents what the experiments are directly sensitive to. 
This band is shown in the region $0.6<\phi<1.6$ for illustrative purpose, while its range of validity is highlighted in the bottom plot, as explained below. When the \kl constraints are interpreted in the context of a specific EFT, the overall shape of the potential can be described by one parameter (as for SMEFT~6) or two (as for SMEFT~8). 
This limited set of degrees of freedom determines all higher derivatives up to $\kappa_6$ ($\kappa_i=0$ for $i\ge7$), hence the shape of the potential across a much wider range of $\phi$. This is illustrated by the blue area of the plot, which represents the allowed range for the potential in a SMEFT~6. 
The bottom panel of Fig.~\ref{fig:Potentials_withExclusion} shows the difference with respect to the SM potential in a narrow range around $\phi=1$. The light blue and light red contours represent the SM potential variations corresponding to the 68\% and 95\% CL uncertainty bands on $\kappa_{3}$.
The blue and the red bands are identical around the minimum  of the potential (see  Fig.~\ref{fig:Potentials_withExclusion} (bottom)), where the effect of higher-order terms  ($\kappa_i=0$ with $i\ge4$) is negligible.
The same plots for different scenarios and integrated luminosities are given in Fig.~\ref{fig:Potentials_withExclusionAt2000_S2} of Appendix~\ref{app:app}. In the same appendix, Fig.~\ref{fig:Potentials_withExclusionAt2000_S3_zoomOnly} shows the comparison between the 68\% and 95\% CL uncertainty bands and the boundary of the region where a strong FOPT occurs, for each of the four considered scenarios: SMEFT~6, SMEFT~8, exponential, and logarithmic potentials.

Figure~\ref{fig:Potentials_withExclusion} (bottom) shows that at 3\abinv the 95\% CL red band  excludes as good as all possible strong-FOPT scenarios across the four alternative hypotheses, while some region of the parameter space cannot be excluded at 2\abinv. 
Being able to make such a statement at the end of the HL-LHC operation would be of extreme importance to our understanding of the origin of the universe and would be a unique result for decades to come.

\section{Interplay between precision Higgs physics and searches for a heavy scalar singlet} \label{sec:precisionheavyscalarsinglet}
A minimal extension of the SM involves the inclusion of a new real scalar singlet, $S$, which couples to the SM via the Higgs portal. The scalar potential takes the form:
\begin{equation}\label{eq:UVLagrlin}
V(\Phi,S) = - {\mu}^2_H |\Phi|^2 + \lambda_H |\Phi|^4 +   b_1 S - \frac{\mu_S^2}{2} S^2  +     \frac{ {b_4}}{4} S^4 +\frac{ {a_2}}{2}  |\Phi|^2 S^2 + \frac{ {b_3}}{3}S^3  + \frac{a_1}{2} |\Phi|^2 S.
\end{equation}
While minimal, this extension presents several interesting features, including universal modifications of the Higgs boson couplings to SM particles, the presence of an additional scalar state that, if sufficiently heavy, can decay into a pair of Higgs bosons, and modified $\kappa_3$ and $\kappa_4$. The enriched scalar potential dynamics enable the possibility of a strong FOPT, as explored in~\cite{Kotwal:2016tex,Huang:2016cjm,Niemi:2024axp,Zhang:2023jvh}. 
This phase transition could also be probed in a complementary way through gravitational wave observations~\cite{Ramsey-Musolf:2024ykk}. 
Additionally, the stability of the vacuum for large field configurations can be affected~\cite{Hiller:2024zjp,SH3}, potentially leading to further constraints.

Given the broad phenomenology of the singlet extension, a comprehensive set of precision measurements and searches conducted by the ATLAS and CMS Collaborations can be utilized to constrain the model. These searches are very powerful in excluding effectively the parameter space configurations that would otherwise allow a strong FOPT as shown in the following.

The first category of searches involves resonant scalar decays into vector bosons ($VV$) and $HH$. These constraints can be further strengthened by limits on \kl obtained from non-resonant %di-Higg
$HH$ searches, as discussed in Section~\ref{sec:non-res}. Additionally, upper bounds on universal modifications to Higgs boson couplings with SM particles provide further restrictions on the model’s viability.

A summary of the expected ATLAS+CMS exclusions with 3\abinv is presented in Fig.~\ref{fig:scalarexclusion_FOTPfcc-ee} (top-left) in the plane of the scalar portal coupling, $a_2$, versus the scalar singlet mixing angle $\theta$~\cite{ADD}, for a given exemplary choice of the theory parameters $m_S$, $b_3$, and $b_4$~\cite{ADD} for which a strong FOPT is possible. 
A significant portion of the viable parameter space is excluded, thanks to the interplay between measurements and searches. 
The dark blue hatched region identifies the parameter space where a strong FOPT is possible, leading to an explanation of the universe matter-antimatter asymmetry. Note that the exclusion by the searches is different here than in the case of a strong FOPT within the EFT framework~(see Fig.~\ref{fig:Potentials_withExclusion}), as here the symmetry breaking dynamics can be richer with a two step transition with different low energy scales. The EFT approach, on the other hand, is valid only in the assumption that the scale of new physics is sufficiently high.

Figure~\ref{fig:scalarexclusion_FOTPfcc-ee} (top-right) shows the 68\% and 95\%  CL HL-LHC  exclusion reach in the plane of the Higgs boson coupling to $ZZ$ relative to the SM one versus \kl, as discussed for example in Ref.~\cite{Huang_2016}. The projected bounds on \kl and the Higgs boson coupling to $ZZ$ are compared to the exclusion regions from the direct searches for $S \to HH$ and $S \to ZZ$, for the same choice of  $m_S$, $b_3$, and $b_4$ as in the top-left plot of Fig.~\ref{fig:scalarexclusion_FOTPfcc-ee}. Most of the strong FOPT phase space is excluded for this choice of parameters by the complementarity of measurements and searches.
Similar or stronger conclusions can be reached in 
Figs.~\ref{fig:Singletmasses} and~\ref{fig:Singletmassesprojections} for a benchmark of few representative choices of the $m_S$, $b_3$, and $b_4$ parameters. While the shape of the exclusion bounds derived from the $S \to HH$ and $S \to ZZ$ searches depend on the choice  of the theory parameters, direct searches always contribute to strongly limit the allowed parameter space. 

Figure~\ref{fig:scalarexclusion_FOTPfcc-ee} (bottom-left) shows the same information across all possible choices of the theory parameters $m_S$, $b_3$, and $b_4$. In this case, only the exclusion bounds from 
\kl and the Higgs boson coupling to $ZZ$ are shown, since they do not depend on the choice of the theory parameters, unlike the bounds from the direct searches. These search bounds strongly exclude the parameter space further as can be inferred from Figs.~\ref{fig:scalarexclusion_FOTPfcc-ee} (top-right) and~\ref{fig:Singletmassesprojections}. In conclusion, the exclusion of a large portion of the phase space with a strong FOPT is achieved in scenario \Sthree with 3 \abinv, thanks to the precision that the HL-LHC experiments will provide on the $HH$ measurements.

\begin{figure}[tbp]
    \centering
    \includegraphics[width=0.45\textwidth,trim=1mm 1mm 11mm 8mm,clip]{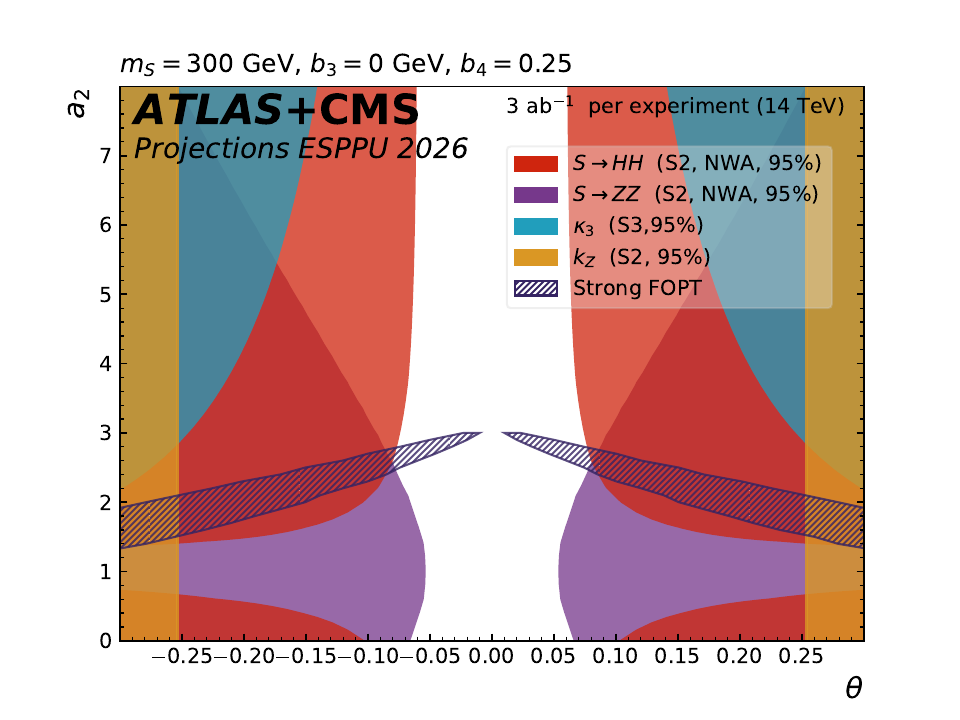}
    \includegraphics[width=0.45\textwidth,trim=0mm 1mm 11mm 8mm,clip]{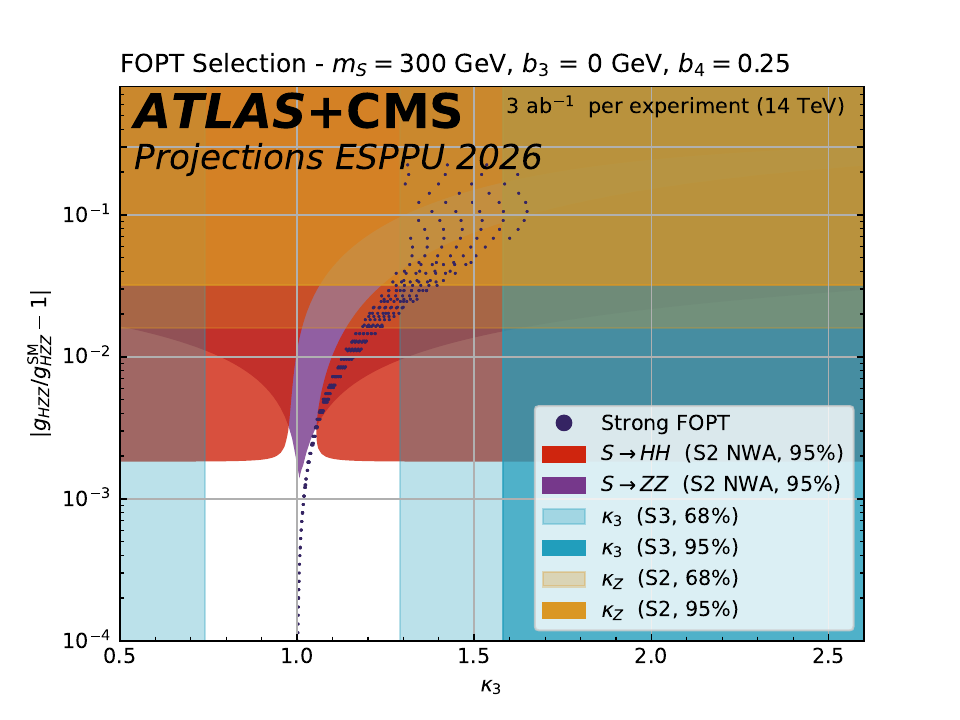}\\
    \includegraphics[width=0.45\textwidth,trim=3mm 1mm 11mm 8mm,clip]{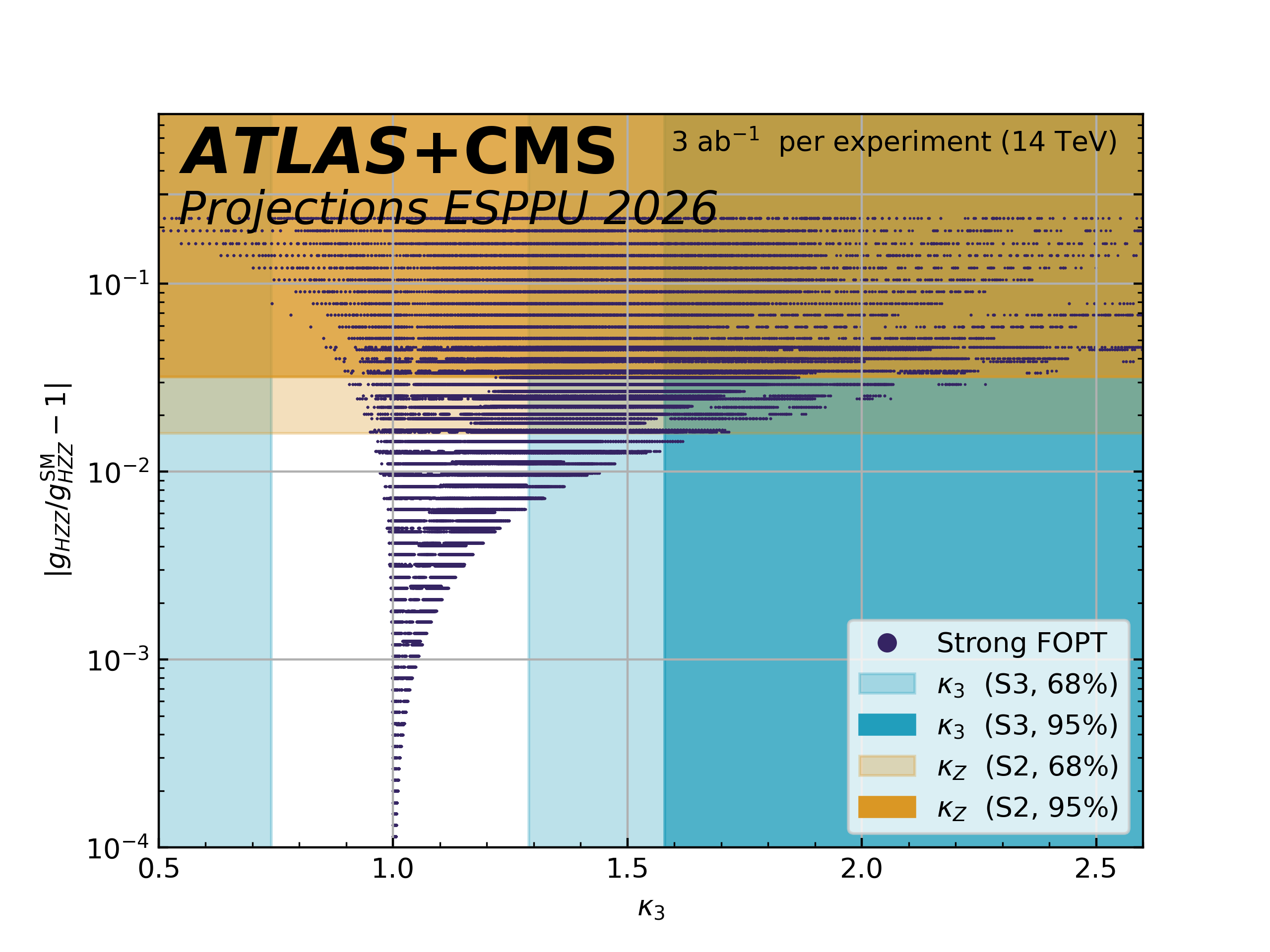}
    \includegraphics[width=.45\textwidth,trim=1mm 1mm 11mm 8mm,clip]{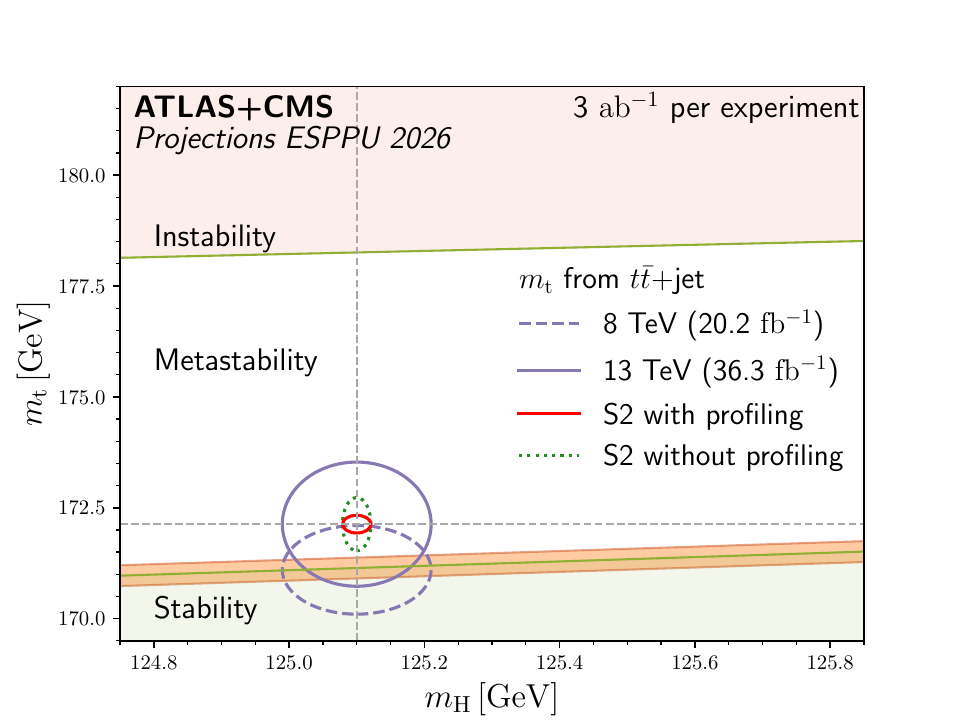}
        \caption{
        Top left: Bounds on the heavy scalar model, in the plane of the scalar portal coupling, $a_2$, versus the scalar singlet mixing angle $\theta$~\cite{ADD}. The dark blue hatched contours show the regions of the $a_2$ versus $\theta$ parameter space in the scalar singlet model where a strong first-order phase transition is possible for $m_S=300$ GeV, $b_3=0$ GeV and $b_4=0.25$. The other contours show the 95\% CL exclusion in this plane from the resonant searches into $S\rightarrow HH /ZZ$ signatures, from the $H$ coupling to $Z$ and from $\kappa_3$ constraints. 
        Top right: The same contours are shown in the plane of the deviation of the Higgs boson coupling to the $Z$ with respect to the SM one, versus \kl. The dark blue points show the area where a strong first-order phase transition in the early universe is possible within the scalar singlet model discussed in the text for $m_S=300$ GeV, $b_3=0$, and $b_4=0.25$.
        Bottom left: Exclusion bounds in the plane of the Higgs boson to $ZZ$ coupling with respect to the SM one versus \kl;  68\% and 95\% exclusion bounds are displayed. The dark blue points populate the area where a strong first-order phase transition in the early universe is possible within the scalar singlet model discussed in the text for all choices of $m_S$, $b_3$, and $b_4$.
        Bottom right: Projections for the HL-LHC measurements of the Higgs boson and top quark mass. The top quark mass measurement in \ttbarjet is shown from ATLAS at 8\TeV~\cite{ATLAS:2019guf} and CMS at 13\TeV~\cite{CMS:2022emx}. The ATLAS+CMS projection is shown with profiling of the systematic uncertainties in the extraction of the top quark mass, based on the \Stwo scenario. Figure adapted from Ref.~\cite{Bednyakov:2015sca} with unchanged value and uncertainty in the strong coupling $\alpha_\mathrm{S}$. The band between the stable and metastable region represents the uncertainty in $\alpha_\mathrm{S}$.
    \label{fig:scalarexclusion_FOTPfcc-ee}}
\end{figure}

\section{\texorpdfstring{Precision measurement of $m_t$ and $m_H$ and their impact on the EW vacuum stability}{Precision measurement of m(t) and m(H) and their impact on the EW vacuum stability}}\label{sec:topmass}

In the hypothesis that the SM is valid up to the Planck scale, the values of $m_H$ and $m_t$  provide information on the stability or metastability of our universe, as discussed, e.g., in Ref.~\cite{Andreassen_2018}. 
They provide a fundamental test of the SM and of potential tensions therein.

The $m_t$ measurement programme at ATLAS and CMS comprises numerous measurements employing different techniques~\cite{CMS:2024irj,atlascollaboration2024climbingatlas13tev}. These measurements have varying sensitivities to theoretical uncertainties, which are projected to be the limiting factor in most cases. In the long term, the precision of $m_t$ will be driven by the technique with the greatest theoretical improvement. In this Section, we project the expected uncertainty for two of these techniques, both extrapolated to 3\abinv:
a measurement of $m_t$  in \ttbarjet  from CMS using Run-2 2016 data~\cite{CMS:2022emx}, and  from ATLAS using  Run-1 data~\cite{ATLAS:2019guf}, and two measurements using \ttbar events with boosted top quarks from ATLAS~\cite{ATLAS:2022kqg} and CMS~\cite{CMS:2022kqg}  using %138\fbinv of 
the LHC Run-2 dataset. 
The CMS analyses are performed via a profile likelihood fit in which  the systematic uncertainties are simultaneously constrained with data in the likelihood minimization. Instead, the \ttbarjet resolved ATLAS  measurement externalizes all uncertainties, i.e.~the systematic uncertainties are kept fixed in the fit. The results are referred to in the following as ``with profiling'' and ``without profiling'', respectively.

The extrapolations are derived  
in the \Stwo scenario for the evolution of systematic uncertainties~\cite{ATLAS_PUB_TOP_topmass,CMS_PAS_TOP}. For the \ttbarjet measurements, the theory uncertainties reduction can be achieved utilizing the outcome of ongoing NNLO calculations (see, for example, Ref.~\cite{Badger:2024fgb}). Important improvements in  Monte Carlo generators and parton shower and hadronization modeling are required to realize this scenario.
While the current precision of the measurements with and without profiling is similar, the profiled analysis technique is expected to reach significantly higher precision with the large HL-LHC data sample, since the large available statistics might lead to strong constraints on systematic uncertainties. This requires additional theoretical work on improving the uncertainty model, in particular for Monte Carlo modeling uncertainties. With these enhancements, the method is projected to reach a precision of 200\MeV.
Instead, the method without profiling  relies less crucially on such improvements, and it would achieve a 600\MeV precision on $m_t$.
These two projections therefore represent a lower and upper bound on the achievable uncertainty in $m_t$ from the study of \ttbarjet events, respectively.
The  extrapolations using boosted top quarks and profiling project an uncertainty of 200-300 MeV for ATLAS~\cite{ATLAS_PUBNOTE_topboostedmass} and $\sim$400~MeV for CMS. 
The interpretation of these measurements as a determination of $m_t$ requires further theoretical work.

The $m_H$ measurement projections based on 3~\abinv of integrated luminosity per experiment are presented in Refs.~\cite{ATL-PHYS-PUB-2018-054} and~\cite{CMS-PAS-FTR-21-007}. These projections result in a total uncertainty on $m_H$ from $H \to ZZ^* \to 4\ell$ ($\ell=e$ or $\mu)$ events of 38~MeV for ATLAS and 26~MeV for CMS, leading to a combined uncertainty of 21~MeV.  

These projections are shown in Fig.~\ref{fig:scalarexclusion_FOTPfcc-ee} (bottom-right), where the constraints at the end of the  HL-LHC phase are depicted assuming that the central values of the two masses remain the same as measured at present. This result is compared to the regions of stability of the EW vacuum. With the full HL-LHC dataset, depending on the central value of $m_t$, ATLAS and CMS may be able to make a conclusive statement on the stability of the vacuum assuming the SM is valid up to the Planck scale.

It is important to stress the fact that a precise determination of $m_t$  and $m_H$ has a much broader reach. For instance, with an uncertainty of $\sim$100~MeV on  $m_t$ combined with an ${\cal O}(1)$~MeV precision on $m_W$ one could probe ${\cal O}(10^{-3})$ BSM effects on the electroweak observables.
This would be an important input for the physics programme of a future collider and should be considered as a major priority for the HL-LHC.

\begin{wrapfigure}[21]{R}{0.5\textwidth}
\vspace{-4mm}
    \includegraphics[width=.95\textwidth]{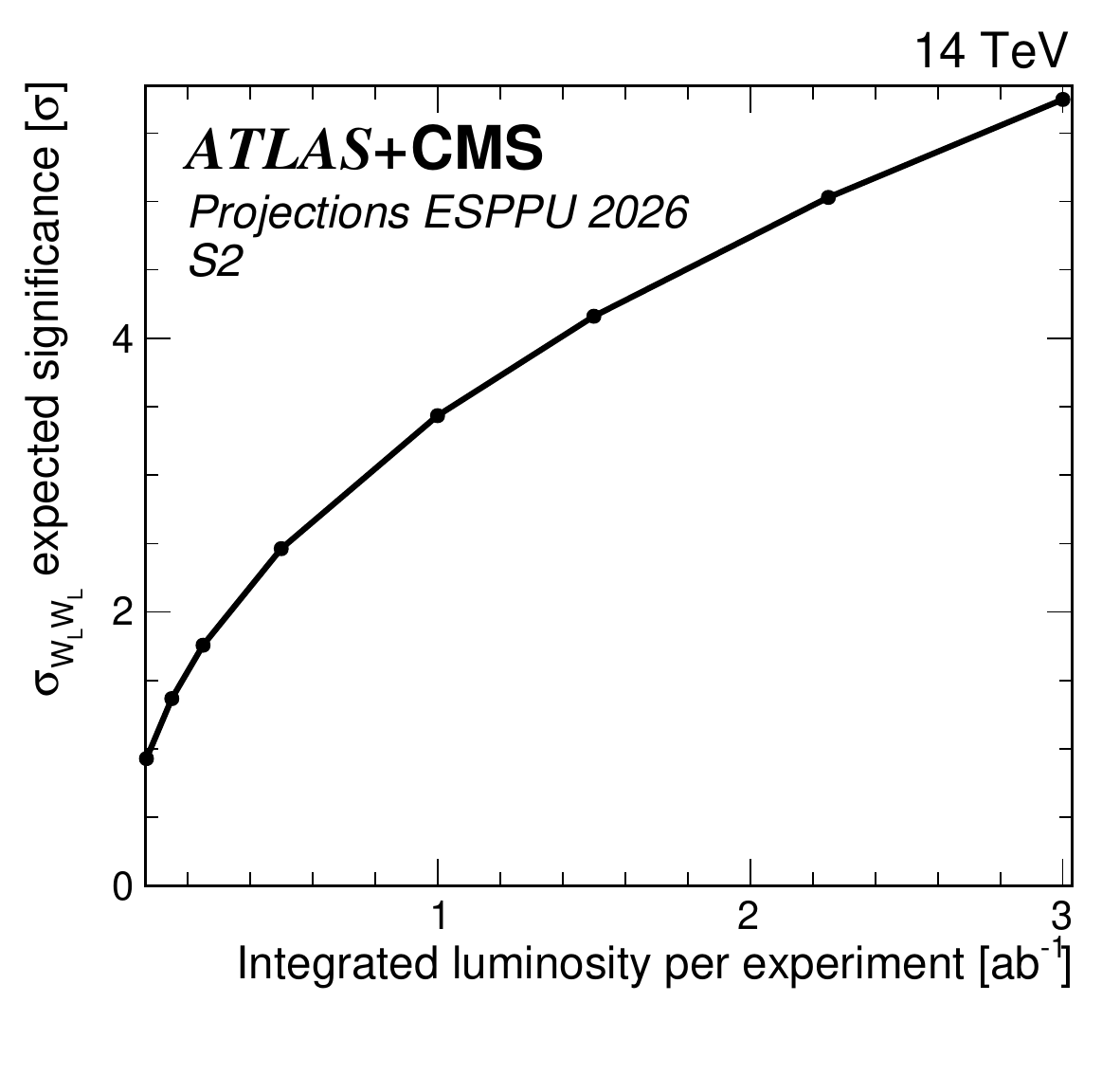}
    \caption{Expected significance of the longitudinally polarized $WW$ scattering as a function of the luminosity in the \Stwo scenario.}
    \label{fig:VVcombo}
\vspace{-3mm}
\end{wrapfigure}

\section{Diboson scattering}
The identification and measurement of the scattering of vector bosons (VBS) bears great relevance at hadron colliders, as these processes are tightly connected with the EWSB. In the absence of a Higgs boson in the SM, the scattering cross section of longitudinally polarized vector bosons would diverge as the centre-of-mass energy increases. Since the Higgs boson couples to massive particles, its presence in the couplings implies that longitudinally polarized vector bosons scatter. 
The measurement of longitudinally polarized VBS is thus a key target at the HL-LHC, as a test of the EWSB mechanism complementary to those derived from Higgs-related measurements.

A sensitivity extrapolation of $W^\pm W^\pm$ and $WZ$ scattering is carried out in the \Stwo scenario, based on the CMS study described in Ref.~\cite{CMS-PAS-FTR-21-001} and considering 3\abinv of data each for ATLAS and CMS with equal experimental sensitivity for the two experiments. 
The  ATLAS+CMS combined uncertainties on the $W^\pm W^\pm$, $W_L^\pm W_L^\pm$, and $W^\pm Z$ production cross sections are projected to 2.9\%, 17.5\%, and 5.9\%, respectively. These uncertainty projections are dominated by statistical uncertainty.
The expected significance on $W_L^\pm W_L^\pm$ production is shown in Fig.~\ref{fig:VVcombo}. The ATLAS+CMS combination is expected to cross the $5 \sigma$ observation threshold~\cite{CMS-PAS-FTR-21-001} between 2~\abinv and 3~\abinv of luminosity per experiment. This is one of the strong motivations for a full exploitation of the target HL-LHC integrated luminosity.

\begin{figure}[t!]
  \centering
    \includegraphics[width=0.48\textwidth,trim=10mm 8mm 3mm 5mm,clip]{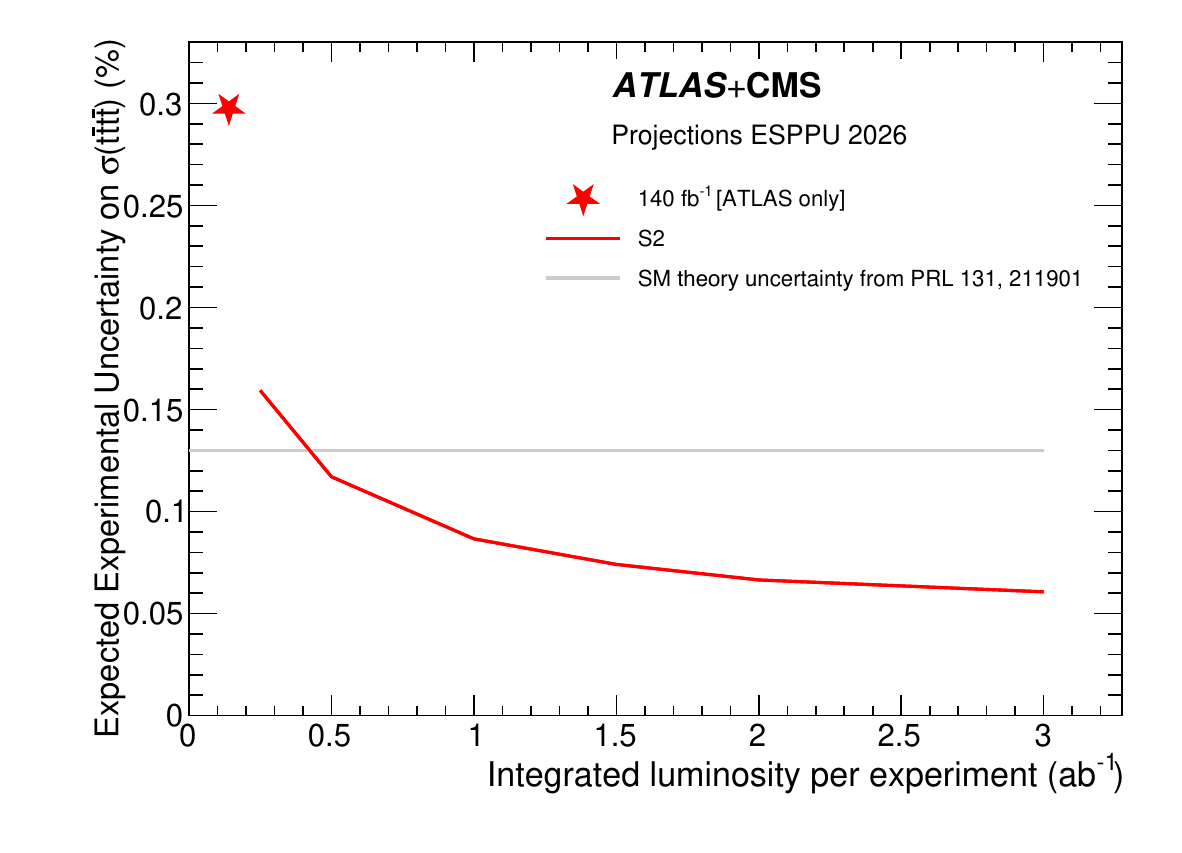}
   \includegraphics[width=0.47\textwidth]{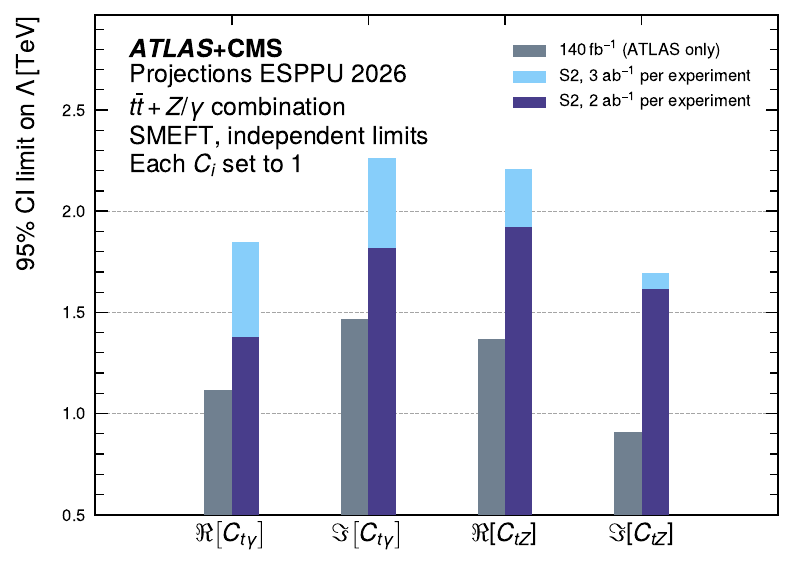}
   \caption{Left: Expected experimental uncertainty in the $\tttt$ cross section measurement.
  Right:  Projected 95\% confidence intervals on the new physics energy scale $\Lambda$ from the combined constraints of $t \bar t Z$ and $t \bar t \gamma$ cross section measurements on EFT operators related to the electroweak dipole moments of the top quark, derived for EFT operator couplings $C_i=1$. 
  \label{fig:4top}}
\end{figure}

\section{Rare top quark processes and EFT bounds}

The production of four top quarks (\tttt) is a rare process that was recently observed for the first time by both ATLAS and CMS~\cite{ATLAS:2023ajo,CMS:2023ftu} using data from LHC Run~2. As a very rare process in top-quark physics, \tttt\ production is highly sensitive to new physics effects.
The prospects for measuring the  \tttt\ cross-section at the HL-LHC at $\sqrt{s}=14$~TeV presented here are based on the full  Run-2 results by ATLAS~\cite{ATLAS:2023ajo,ATLAS_PUB_TOP_fourtop}, extrapolated to 3~ab$^{-1}$ of combined ATLAS and CMS data, assuming the same detector performance. The final states with two same-charge leptons or at least three leptons are considered for this study. 
The projected experimental precision in the \Stwo scenario is significantly smaller than the current uncertainty in the SM theoretical cross-section~\cite{vanBeekveld:2022hty}, as can be seen from Fig.~\ref{fig:4top} (left).  With 3~ab$^{-1}$, the expected uncertainty in the four top-quark production cross-section is 6\%, with the systematic uncertainty improvements of the \Stwo scenario.

The top-quark Yukawa coupling $y_{t}$ enters the Feynman diagrams for processes where a pair of top quarks is mediated by a Higgs boson. This makes \tttt production inherently sensitive to the top-quark Yukawa coupling. Additionally, the top-quark Yukawa coupling also affects the \ttH process, which is a background for \tttt production.
The cross section for \tttt production can be expressed as a function of the top-quark Yukawa coupling strength modifier, $\kappa_{t}$, assuming a 
$C\!P$-even coupling as in the SM. 
The \ttH cross section also depends on this parameter. However, unlike \tttt production, the kinematic distributions in \ttH events remain unchanged unless a $C\!P$-odd term is non-zero. 
The expected 95\% CL upper limit on $y_{t}$ as a function of the integrated luminosity, provides complementary information on $y_{t}$ when the Higgs boson is off-shell.  The extrapolation is performed  leaving the \ttH yield floating in the fit.  As shown in Appendix~\ref{app:app}, a 95\% CL limit on $y_{t} $ $\lesssim 1.5$ (see Figure~\ref{fig:top_SMtttt_yukawa} (left)) is obtained in the \Stwo scenario.  Figure~\ref{fig:top_SMtttt_yukawa} (right) shows the  95\% CL $y_{t}$ limit when of the \ttH yield is fixed to its SM value.

Within the EFT framework, the \tttt production process is particularly sensitive to the following heavy-flavour fermion operators: $\mathcal{O}_{tt}^1$, $\mathcal{O}_{QQ}^1$, $\mathcal{O}_{Qt}^1$, and $\mathcal{O}_{Qt}^8$ ~\cite{Zhang:2017mls}. These operators provide a powerful probe of new physics scenarios that enhance interactions between third-generation quarks. 
The expected 95\% CL intervals on the EFT coupling parameters, assuming a single parameter variation in the fit, are summarized in Table~\ref{tab:EFTresults} of Appendix~\ref{app:app}. 
The limits show a substantial improvement over the current ATLAS-only results. 

The projection of the inclusive $\tttt$ cross-section derived in the \Stwo scenario can be used to set limits on the $O^8_{Qt}$ EFT operator of $-1.9 < C^8_{Qt}< 2.7$. The inclusive cross-section measurement is systematically limited already at 2~ab$^{-1}$ (see Fig.~\ref{fig:top_topX_HL-LHC}), whilst differential measurements of this process will be possible with 3\abinv of data. As such, the sensitivity to the relevant EFT operators will significantly improve.

The associated production of a top-quark pair with a photon (\ttgamma) or a $Z$ boson (\ttZ) provide a powerful probe for potential new physics effects within the framework of the SMEFT. The photon transverse momentum differential distribution, measured in \ttgamma production, is particularly sensitive to EFT parameters related to the electroweak dipole moments of the top quark. Similarly, the transverse momentum of the $Z$ boson, measured in \ttZ production, adds complementary information.

The HL-LHC ATLAS+CMS projections to 2 and 3 \abinv of the sensitivity of these measurements are based on an ATLAS study~\cite{ATLAS:2024hmk,ATLAS_PUB_TOP_ttgttZ}. These projections, shown in Fig.~\ref{fig:4top} (right), illustrate the sensitivity of the HL-LHC on probing the EFT operators related to the electroweak dipole moments of the top quark, in particular when the full 3\abinv dataset is utilized, assuming the same sensitivity for both experiments.

\section{Summary}

The upgraded ATLAS and CMS detectors are unique drivers of our fundamental understanding of nature at the energy and intensity frontiers, and the HL-LHC is expected to provide crucial answers. The physics reach of the HL-LHC measurements will dominate the state-of-the-art for decades to come.

We update the prospects for rare Higgs boson decay modes for $H\to \gamma\gamma$, $H\to \mu^+ \mu^-$, and $H\to Z \gamma$. The latter two reach a precision of 3\% and 7\%, respectively. The other main Higgs boson couplings are measured with a precision between  1.6 and 3.6\%, limited by theoretical uncertainties. In addition, differential cross sections can be measured with good sensitivity in the high $p_T^H$ region, potentially sensitive to new physics effects.
The projections for $HH$ measurements in various final states and their combination are also presented. A significance of more than 7$\sigma$ can be reached for a combined ATLAS+CMS measurement at 3\abinv, with nearly single-experiment observation. Expressed in relation to the trilinear self-coupling $\lambda_3$, {\bf a precision better than 30\% will be achieved} in absence of BSM effects. Additionally, we report on the expected reach of the $HHH$ production search. 
The Higgs self-coupling results are utilised in the broader context of a generic BSM model for baryogenesis featuring an additional heavy scalar. By combining projected constraints from precision Higgs boson measurements, di-Higgs boson results, and relevant resonant searches, we derive projections on how the HL-LHC could constrain various BSM scalar potentials, highlighting the power of future measurements in {\bf constraining the shape of the electroweak vacuum potential}.
We show that we can constrain the vast majority of the parameter space of these models. Additionally, we exclude as good as fully generic BSM potentials assuming that new heavy particles lie beyond a large energy cutoff scale, for which a strong first-order phase transition can take place and potentially explain the matter-anti-matter asymmetry in the universe.
We demonstrate that the expected precision on the top-quark mass could reach as low as 200 MeV. Combined with a precision on $m_H$ of 21 MeV, this will enhance our ability to explore the nature of the electroweak vacuum and assess the stability of the universe.

Additionally, we review the prospects for measuring vector-boson scattering cross section, especially with longitudinally polarized $W$ bosons, which will be measured with better than a 18\% precision.

Finally, we provide prospects for several rare and high-mass processes involving top quarks, where the HL-LHC will achieve long-term world's best sensitivity: rare top quark processes like the four-top-quark production can be measured with a 6\% precision.

It is important to emphasize that most of these projections are limited by theoretical uncertainties. The ultimate exploitation of the full 3\abinv HL-LHC dataset will require a strong investment to advancing the LHC-related precision theoretical physics programme. With these improvements, the final reach of the HL-LHC might exceed the current expectations presented in this document.

In the coming years, the ATLAS and CMS detectors will be equipped with exceptional experimental capabilities to explore both the energy and the luminosity frontiers unlocked by the HL-LHC. Even before precisely quantifying the innovations and resulting improvements enabled by these upgrades, we have demonstrated that the planned HL-LHC programme—aiming for an integrated luminosity of 3\abinv for both ATLAS and CMS—will undoubtedly leave a lasting impact on the future directions of high energy physics.

\clearpage

\appendix

\section{Supporting material}
\label{app:app}

\begin{figure}[ht]
         \centering
           \includegraphics[width=0.44\linewidth]{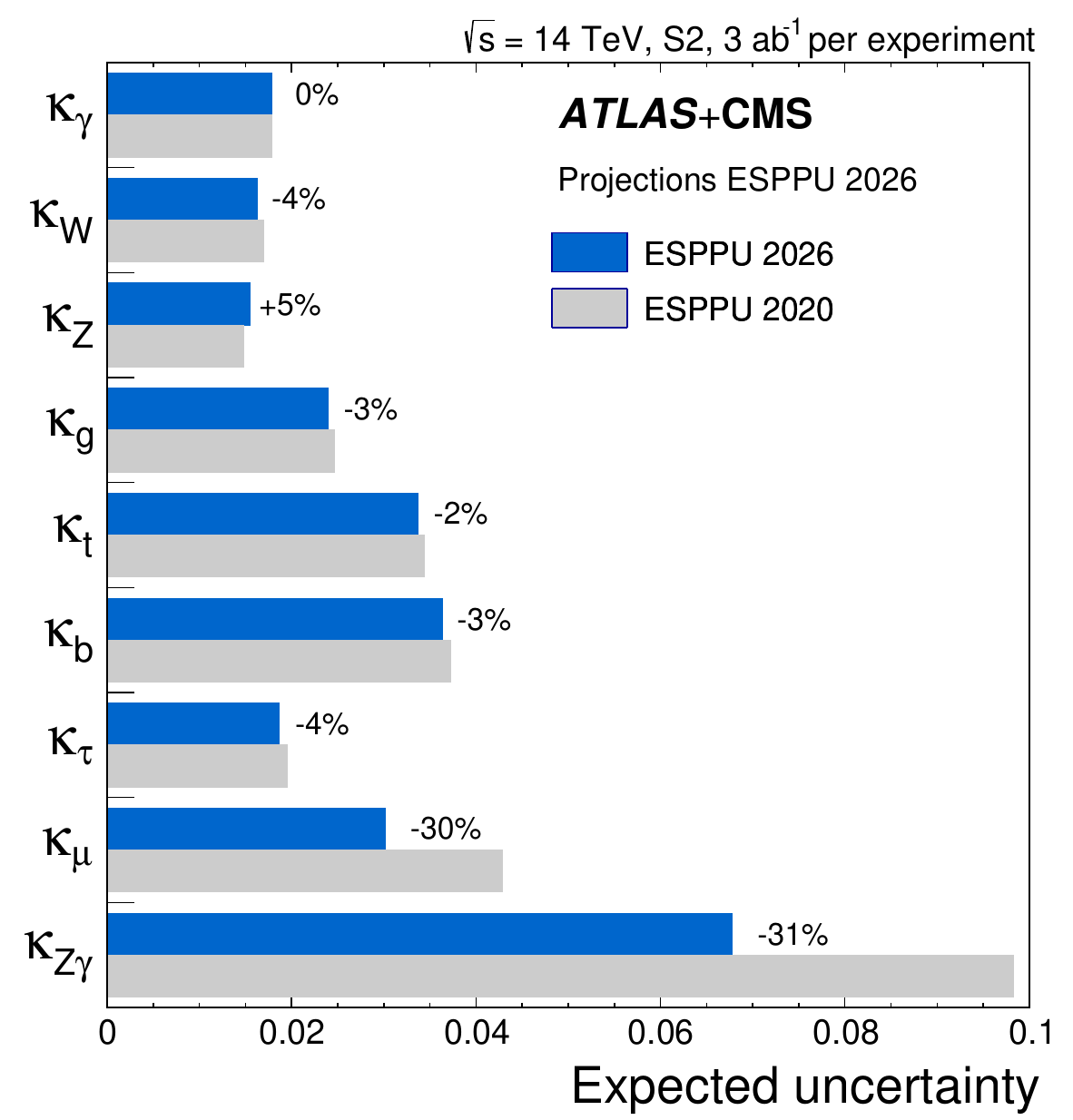}
           \includegraphics[width=0.44\linewidth]{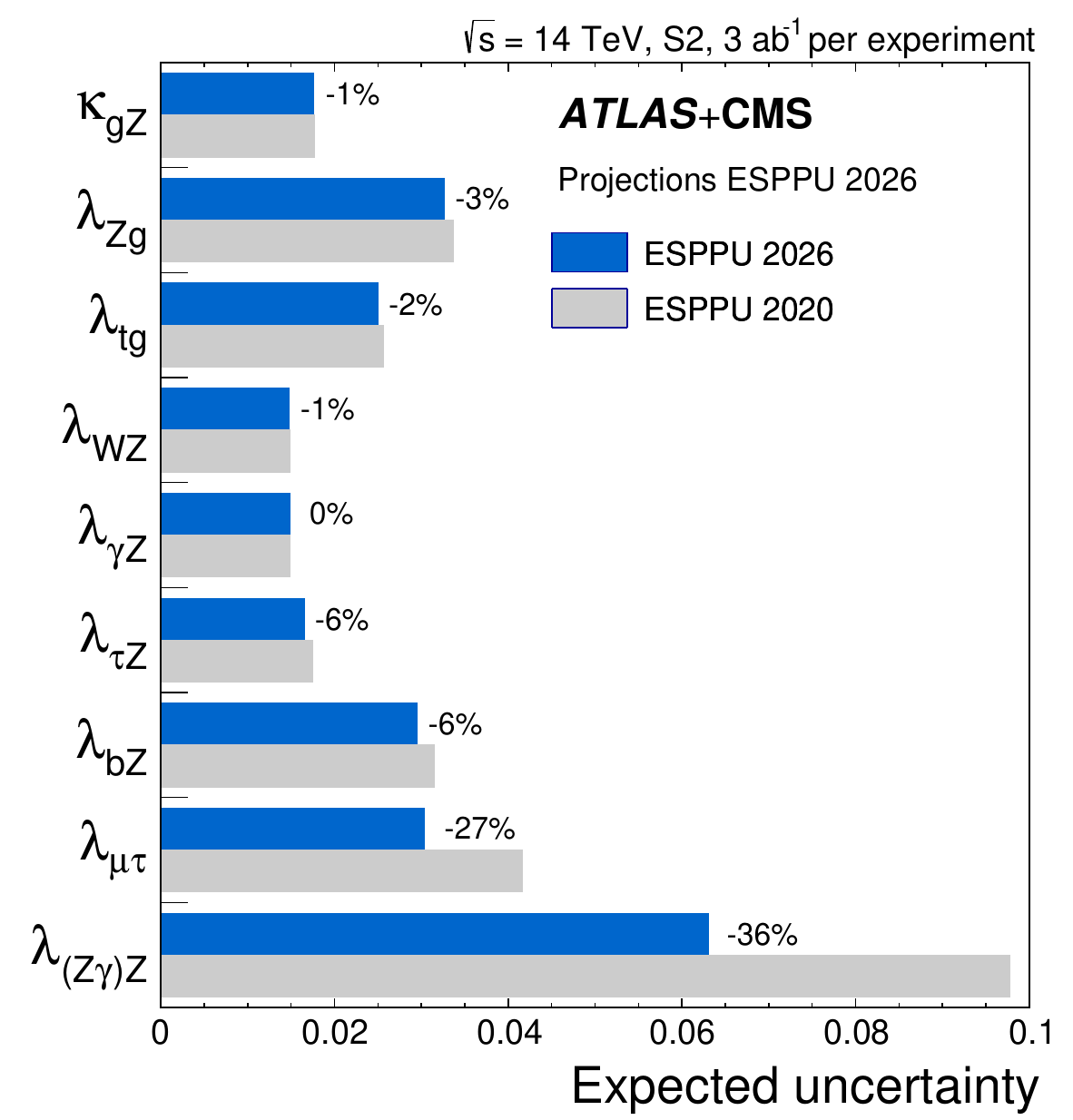}
          \caption{
           ESPPU 2026 projections of coupling modifier uncertainties (left) and their ratios (right), compared to the previous HL-LHC projections~\cite{ATLAS:2019mfr}. The shown percentages represent the relative difference between the two projections. The precision of $\kappa_Z$ is slightly lower due to the refined treatment of the $ZH$ theory uncertainty.
          }
    \label{fig:kappa_H_imp}
\end{figure}

\begin{figure}[htb]
    \centering
        \includegraphics[width=0.46\textwidth]{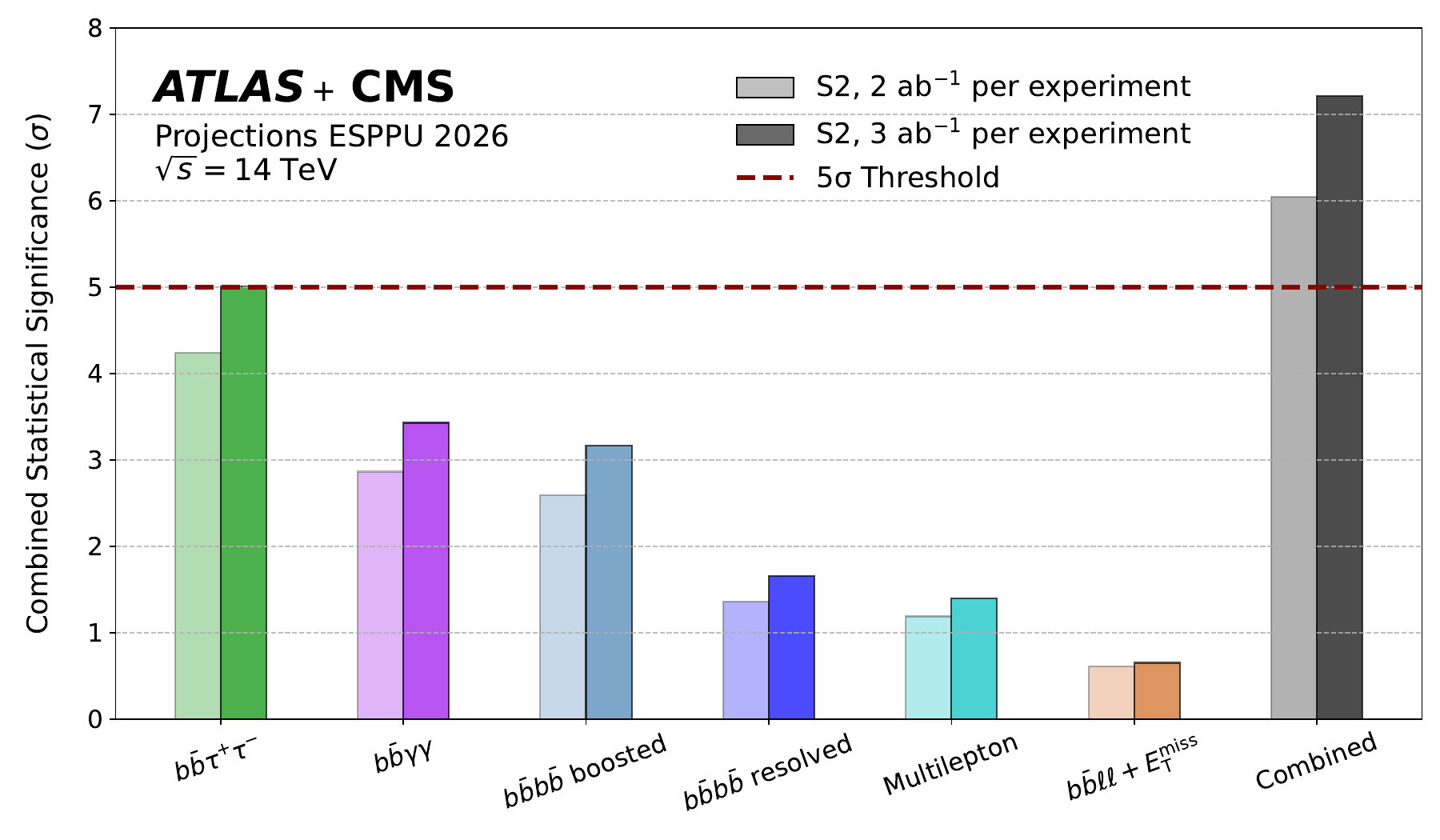}
        \includegraphics[width=0.46\textwidth]{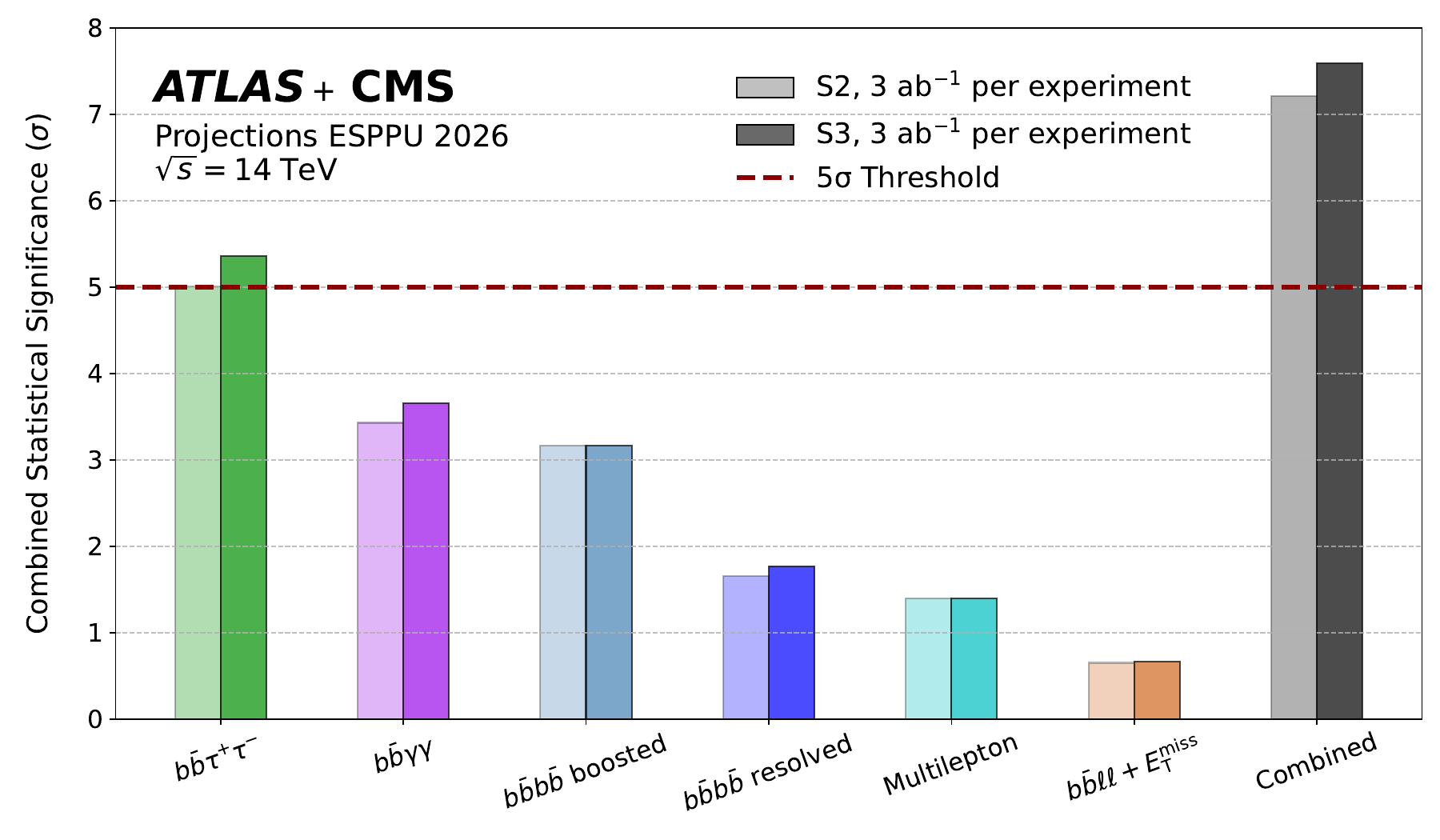}
    \caption{$HH$ significances per final state channel, combined between ATLAS and CMS,  in a comparison between integrated luminosities for the \Stwo systematics scenario (left) and  in a comparison between systematics scenarios at $3\;\textrm{ab}^{-1}$ (right).}
    \label{fig:HHmain_figure_app_comb}
\end{figure}

\begin{figure}[htb]
    \includegraphics[width=.49\linewidth]{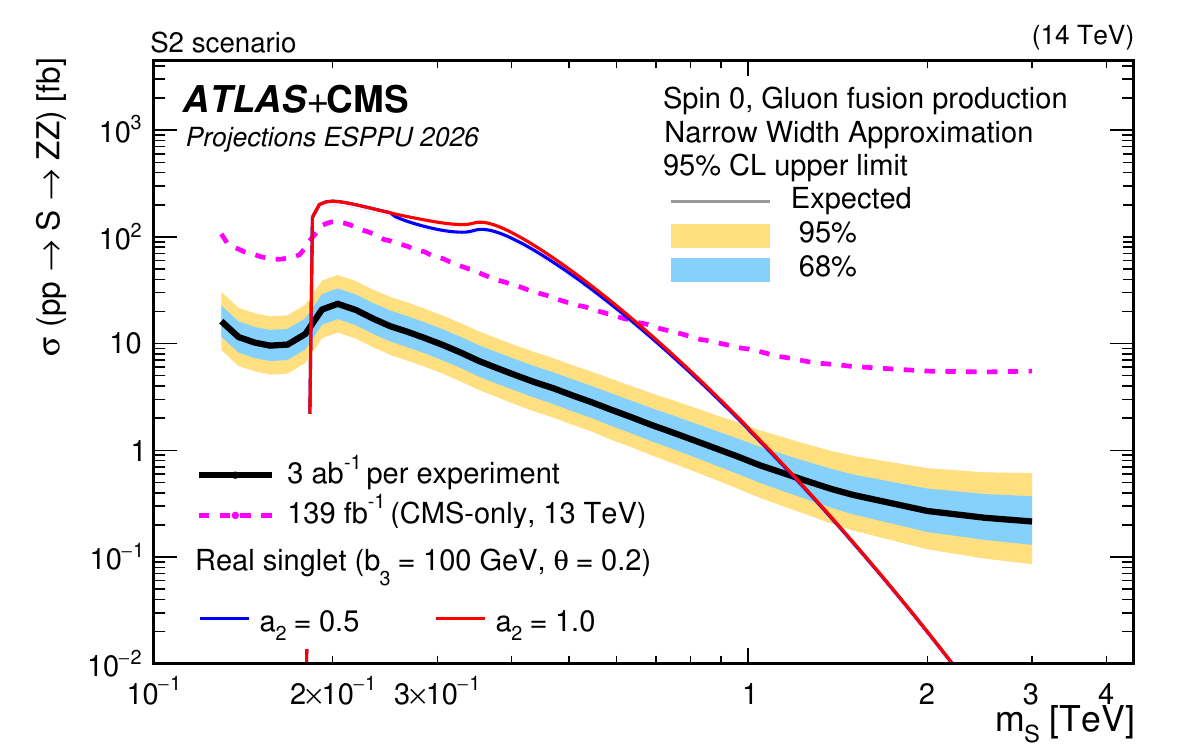}  
    \includegraphics[width=.43\linewidth,trim=2mm 5mm 2mm 1mm,clip]{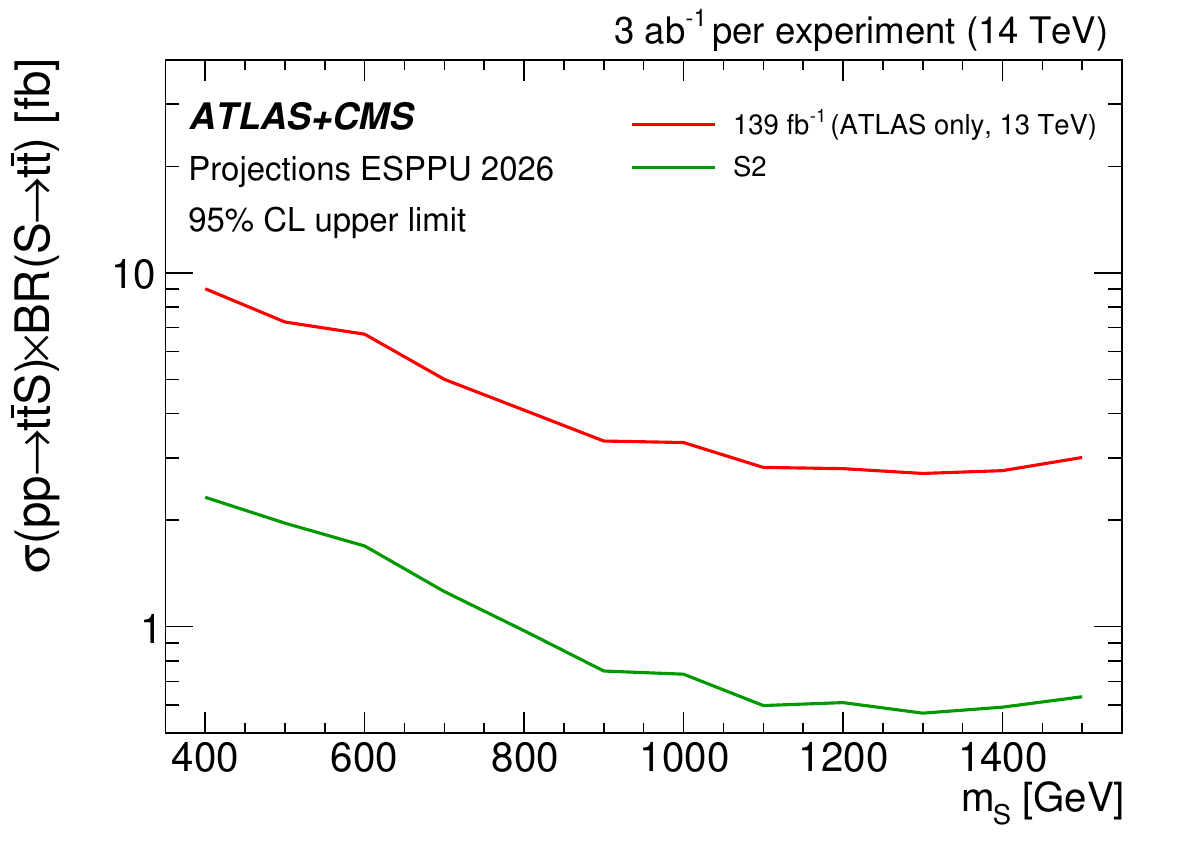}
    \caption{Expected 95\% CL upper limits on the $\sigma(pp \to S \to ZZ)$ (left) and $\sigma(pp \to S t \bar t \to t \bar t t \bar t)$ (right) cross sections as a function of the scalar mass $m_S$, produced via gluon fusion using the narrow width approximation. The projection is derived assuming 3 \abinv\ per experiment for the \Stwo\ scenario at 14~TeV.\label{fig:Resonant_limitsapp}}
\end{figure}

\begin{figure}[htb]
    \centering
    \includegraphics[width=0.48\textwidth]{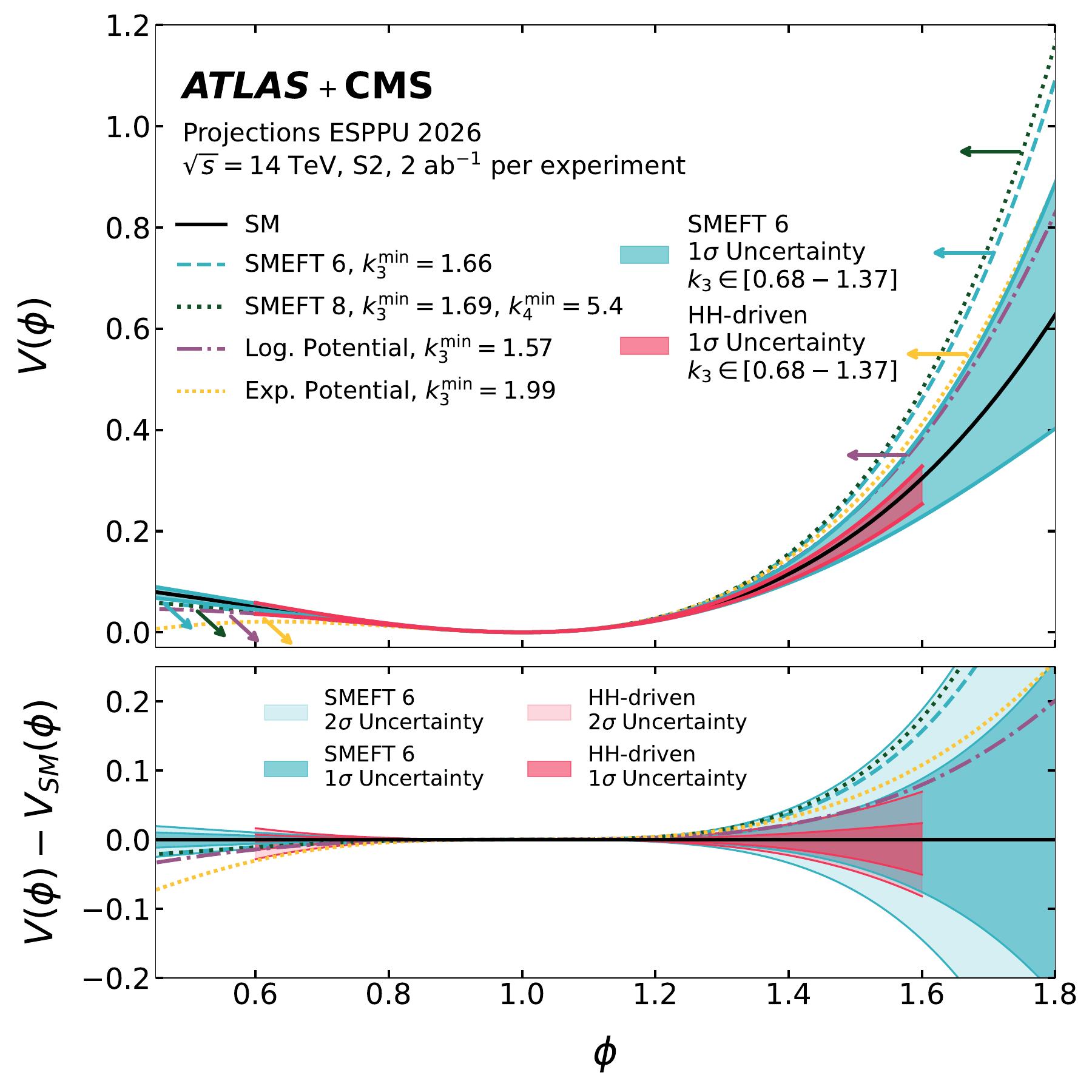}
    \includegraphics[width=0.48\linewidth]{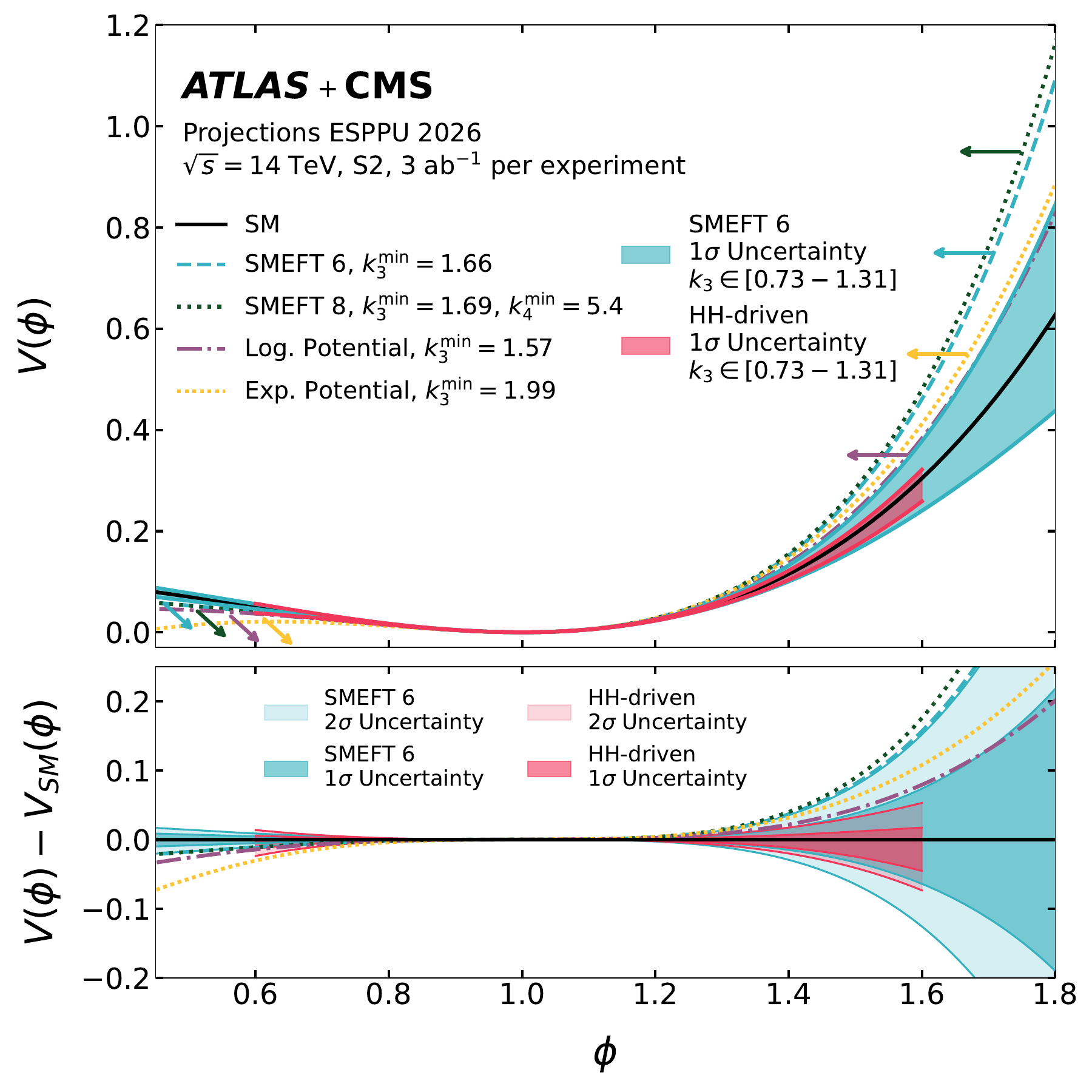}\\
    \hspace{2.5mm}
    \includegraphics[width=0.475\linewidth]{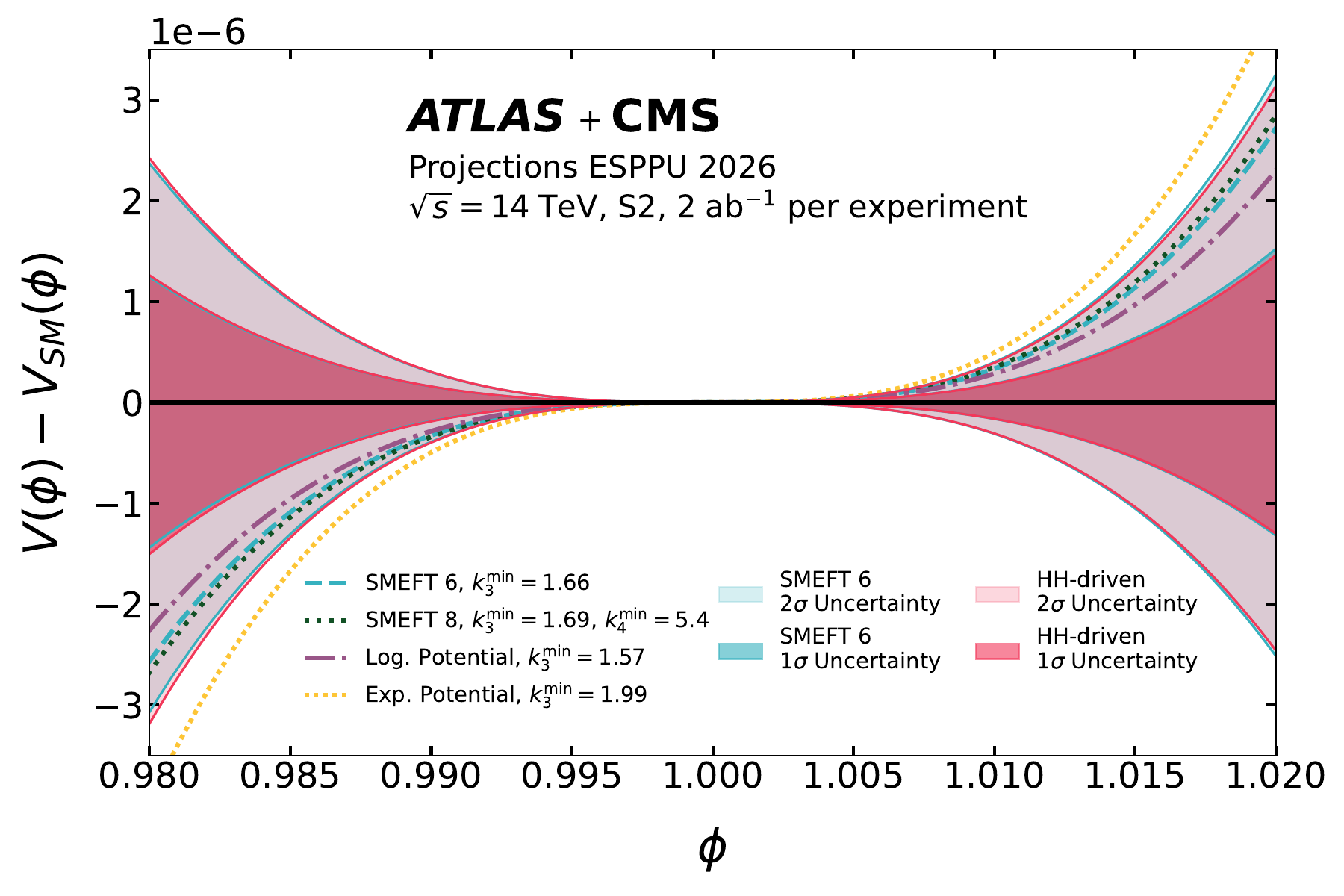}
    \hspace{0.5mm}
    \includegraphics[width=0.475\linewidth]{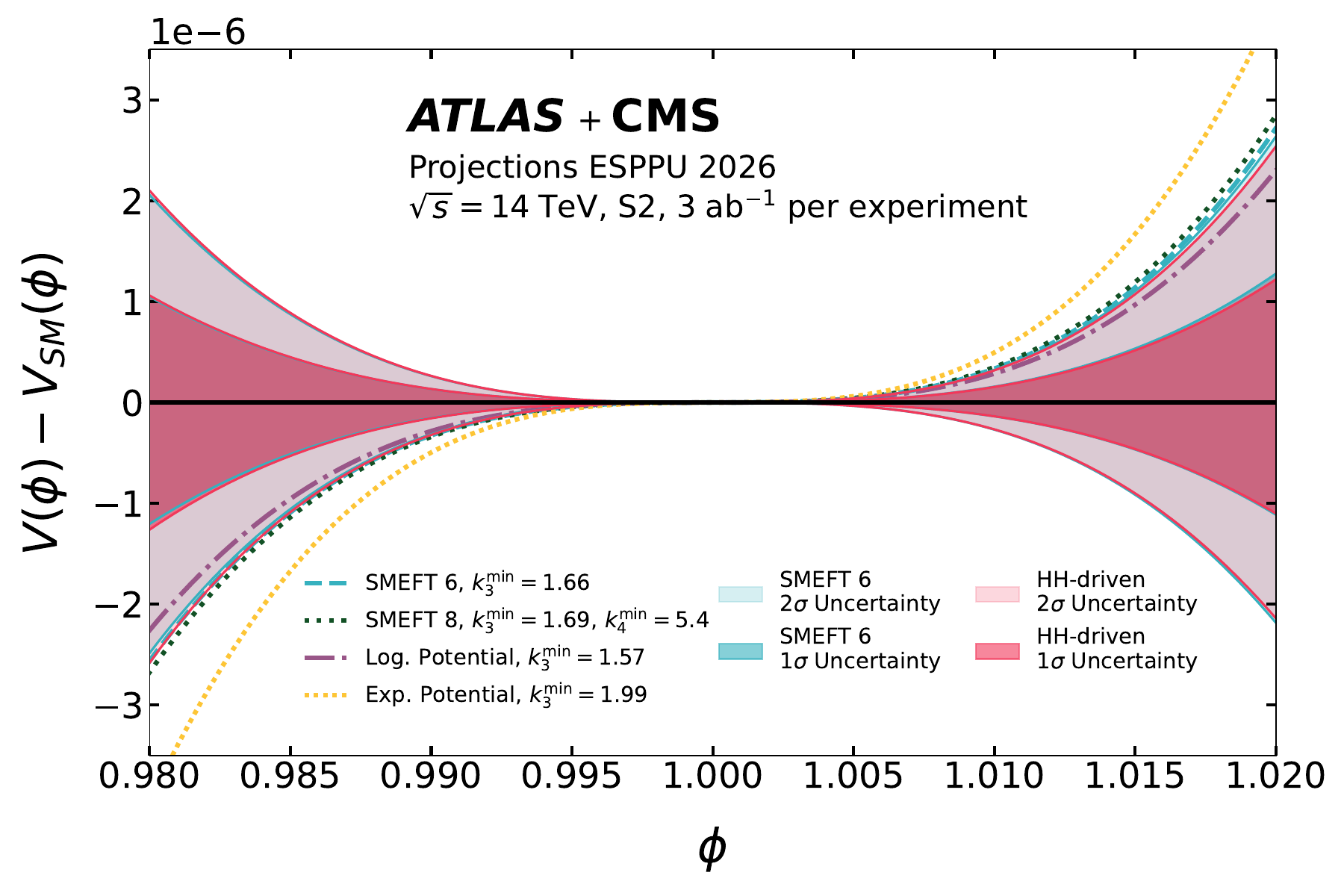}
    \caption{BEH potentials in various models which predict first-order phase transition~\cite{ADD}. The models are compared with the SM BEH potential. Two approaches (SMEFT~6 and HH-driven) are used to show the expected uncertainties on the Higgs self-coupling achieved by combining ATLAS and CMS at 2\abinv (top-left) and  at 3\abinv (top-right) in the \Stwo scenario, in a wide range of the BEH field value.
    The bottom panels show the difference between the potential $V(\phi)$ and its SM expectation $V_{SM}(\phi)$. Here, the 68\% and 95\% CL uncertainty bands on the shape of $V(\phi)$ are shown, for the HH-driven and SMEFT~6 potentials (see text). The bottom plots show the zoom into the $V(\phi)-V_{SM}(\phi)$ difference around the minimum of $V(\phi)$, corresponding to the validity range of the HH-driven band.
    \label{fig:Potentials_withExclusionAt2000_S2}}
\end{figure}

\begin{figure}[htb]
    \centering
           \includegraphics[width=0.48\linewidth]{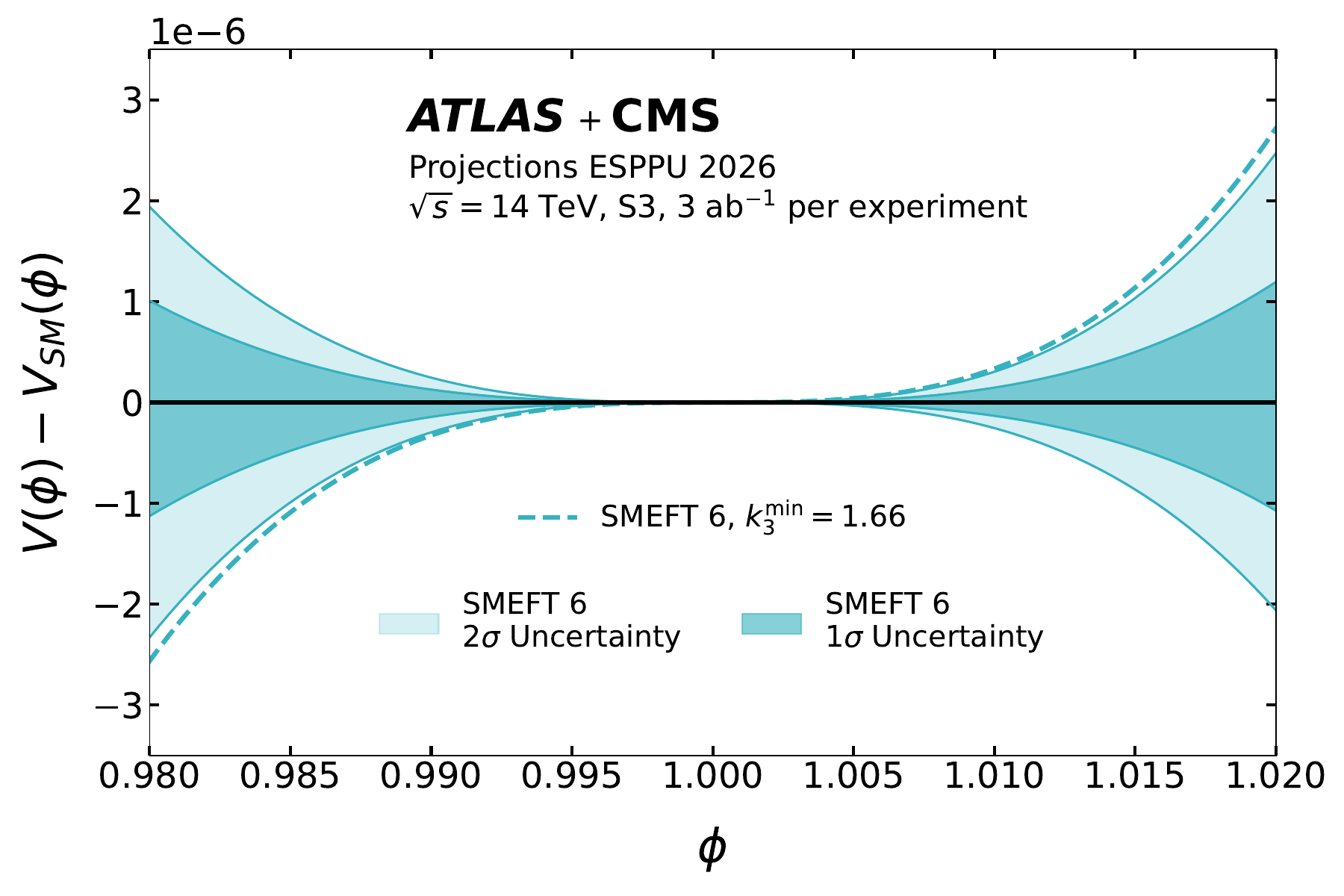}
           \includegraphics[width=0.48\linewidth]{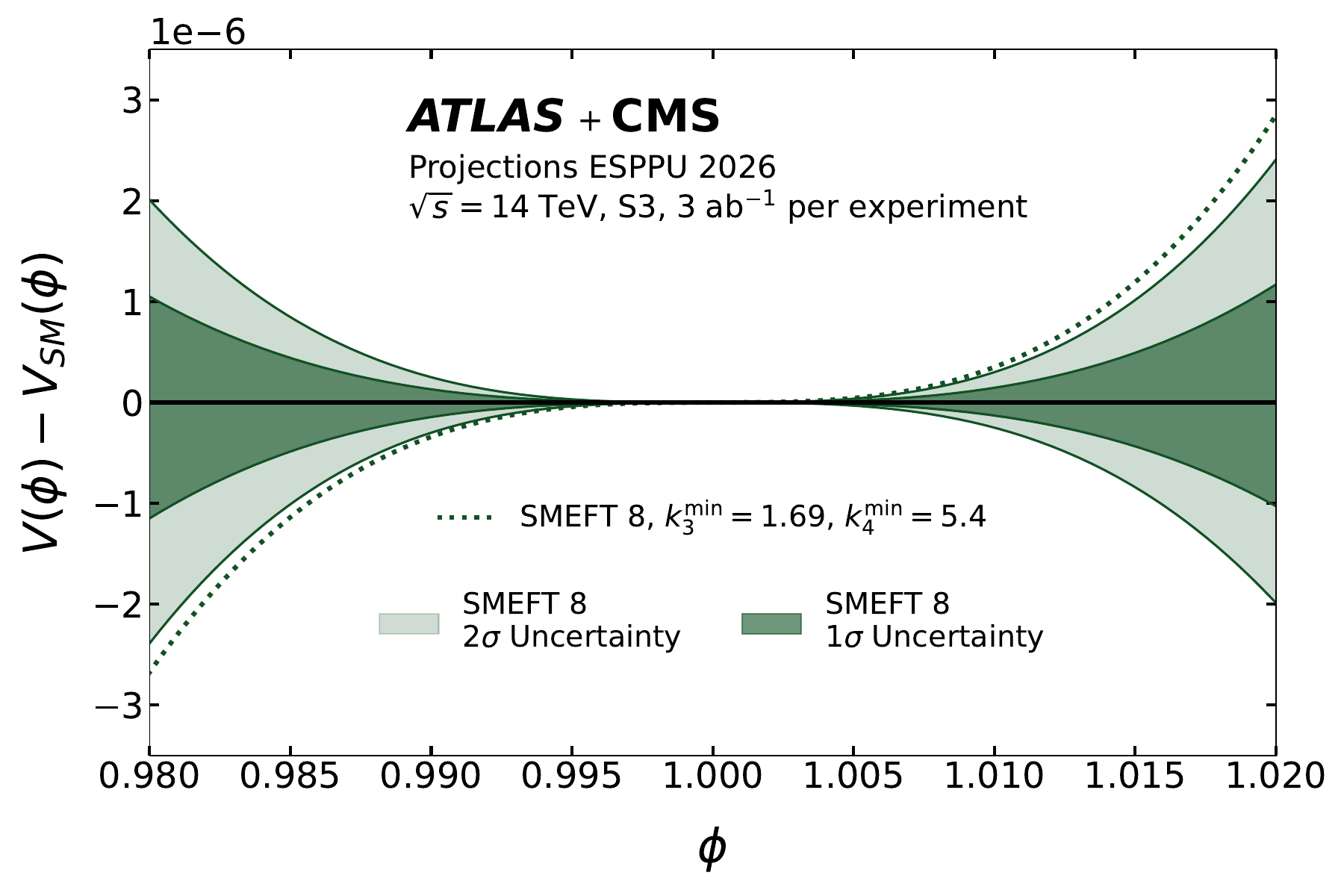}
\\
         \includegraphics[width=0.48\linewidth]{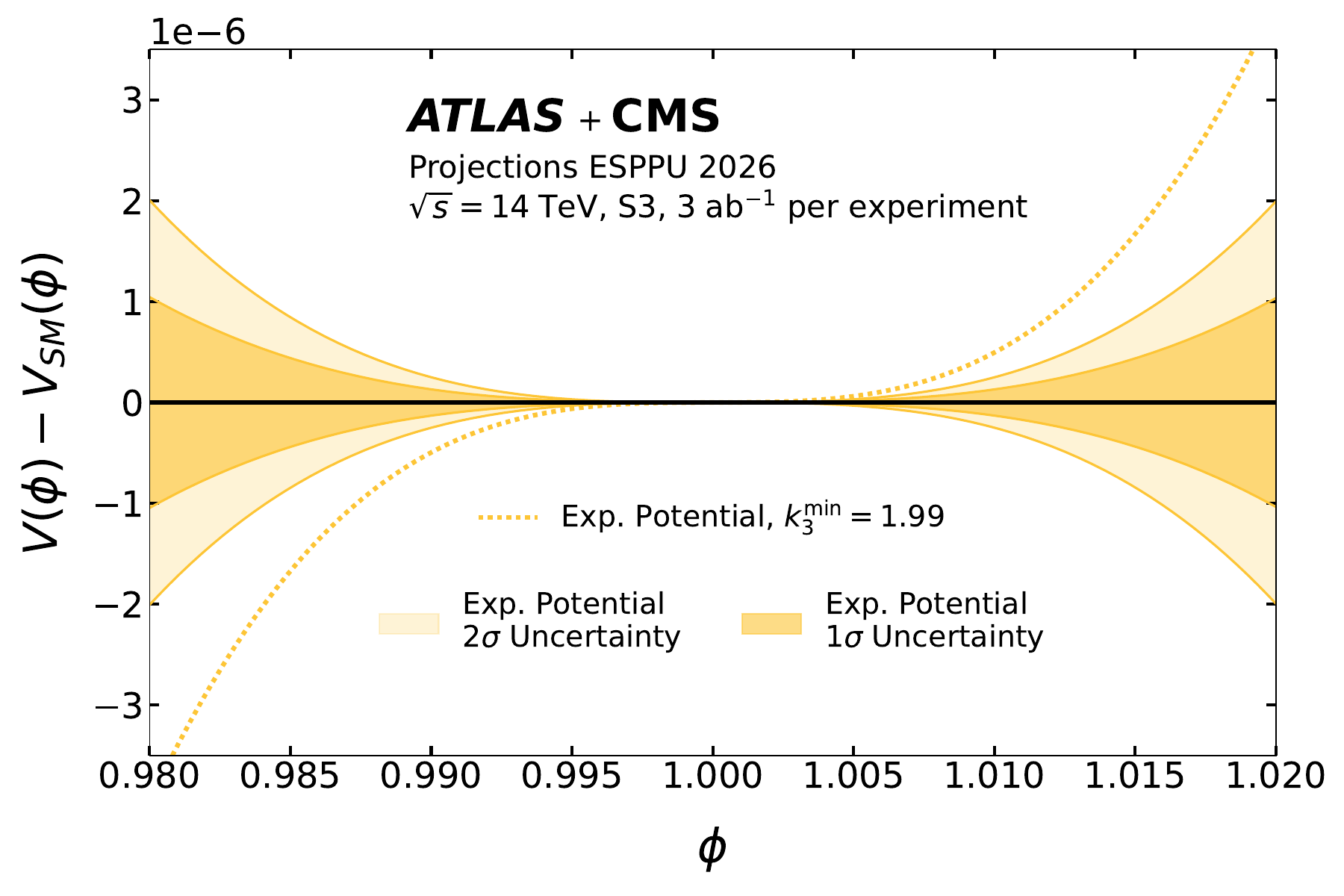} 
        \includegraphics[width=0.48\linewidth]
        {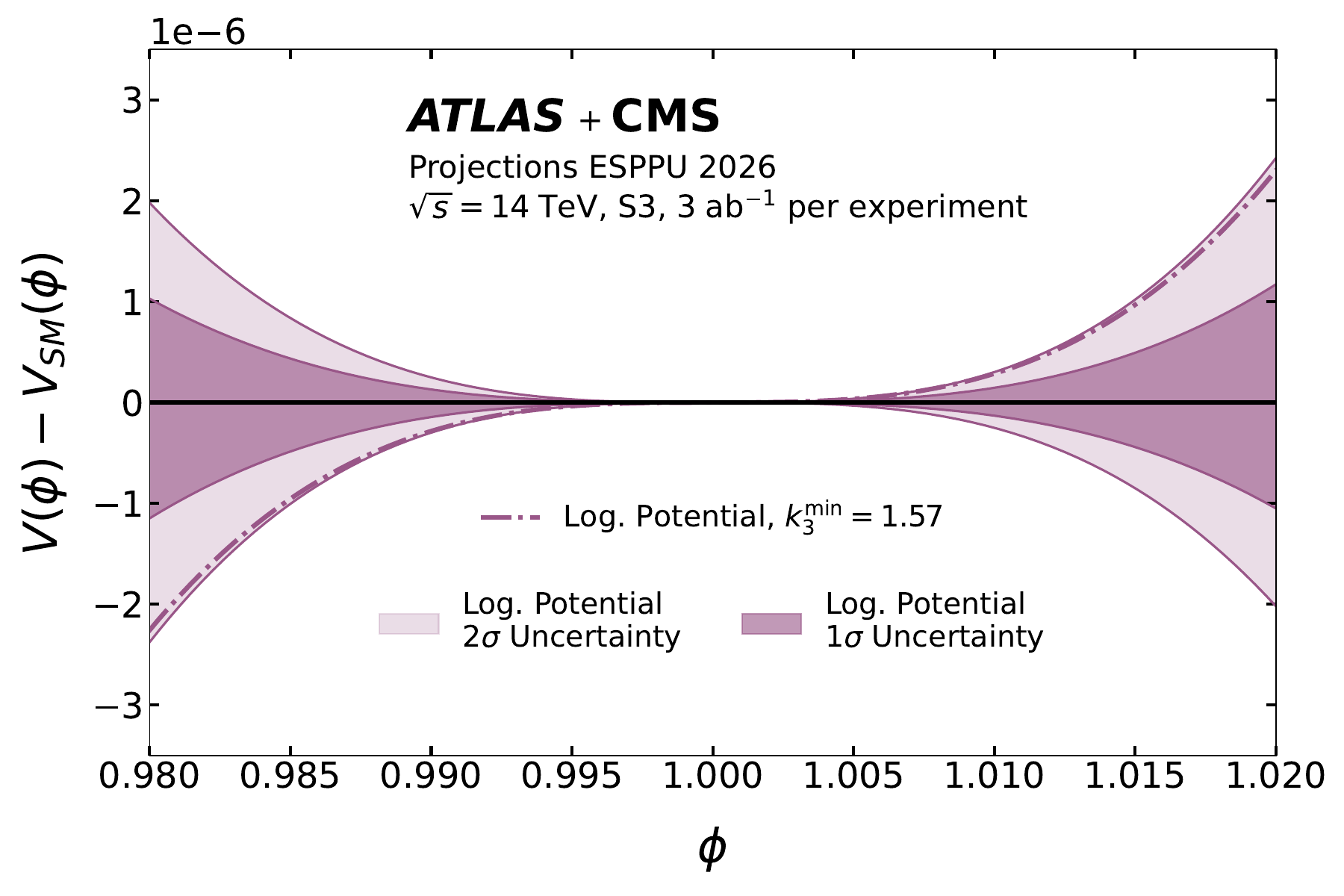}
     \caption{The difference between the BEH potential $V(\phi)$ and its SM expectation $V_{SM}(\phi)$ in four scenarios:
     SMEFT~6 (top left), SMEFT~8 (top right), exponential (bottom left), and logarithmic (bottom right)  potentials. The 68\% and 95\% uncertainty bands from the ATLAS+CMS combination is compared to the boundary of the regions for which each model predicts a strong first-order phase transition. The $\phi$ range  corresponds to the validity range of the HH-driven band.
     The projection is derived in \Sthree scenario assuming
      3\abinv per experiment.\label{fig:Potentials_withExclusionAt2000_S3_zoomOnly}}
\end{figure}

\begin{figure}[!htb]
    \centering
    \begin{subfigure}[b]{0.40\textwidth}
        \includegraphics[width=\textwidth]{figures/Combination_M300_b3=0_b4=0.25.pdf}
    \end{subfigure}
        \begin{subfigure}[b]{0.40\textwidth}
        \includegraphics[width=\textwidth]{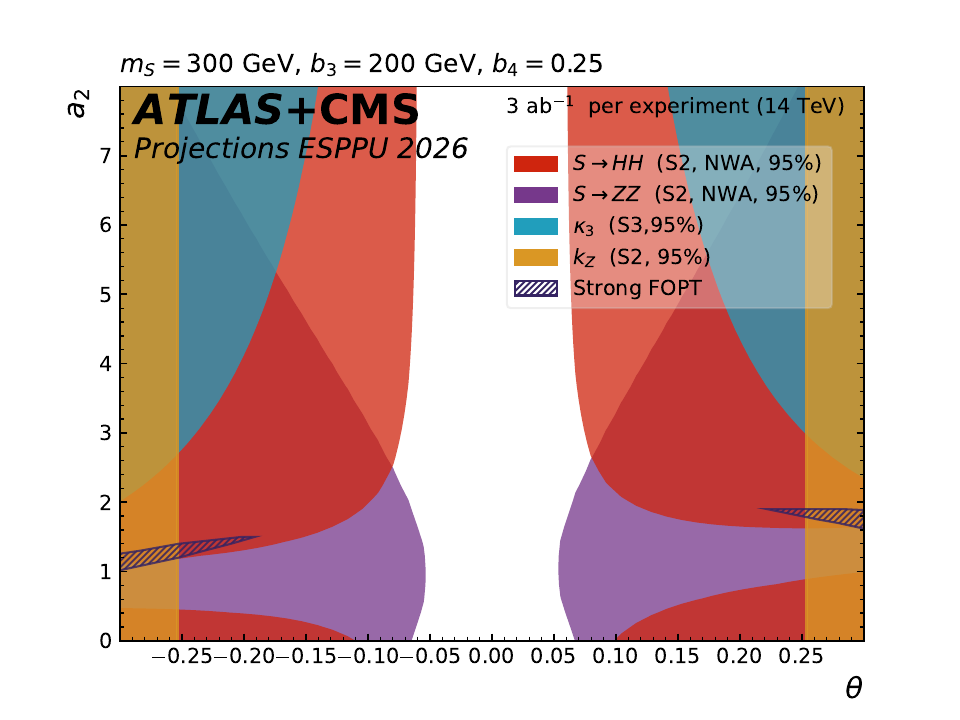}
    \end{subfigure}
        \begin{subfigure}[b]{0.40\textwidth}
        \includegraphics[width=\textwidth]{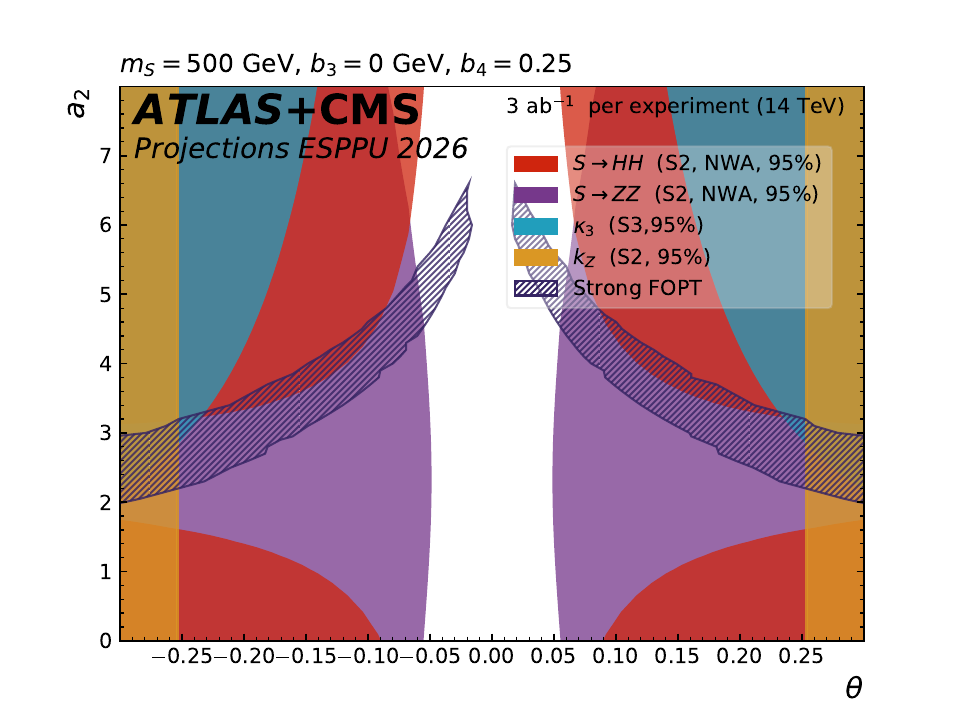}
    \end{subfigure}
        \begin{subfigure}[b]{0.40\textwidth}
        \includegraphics[width=\textwidth]{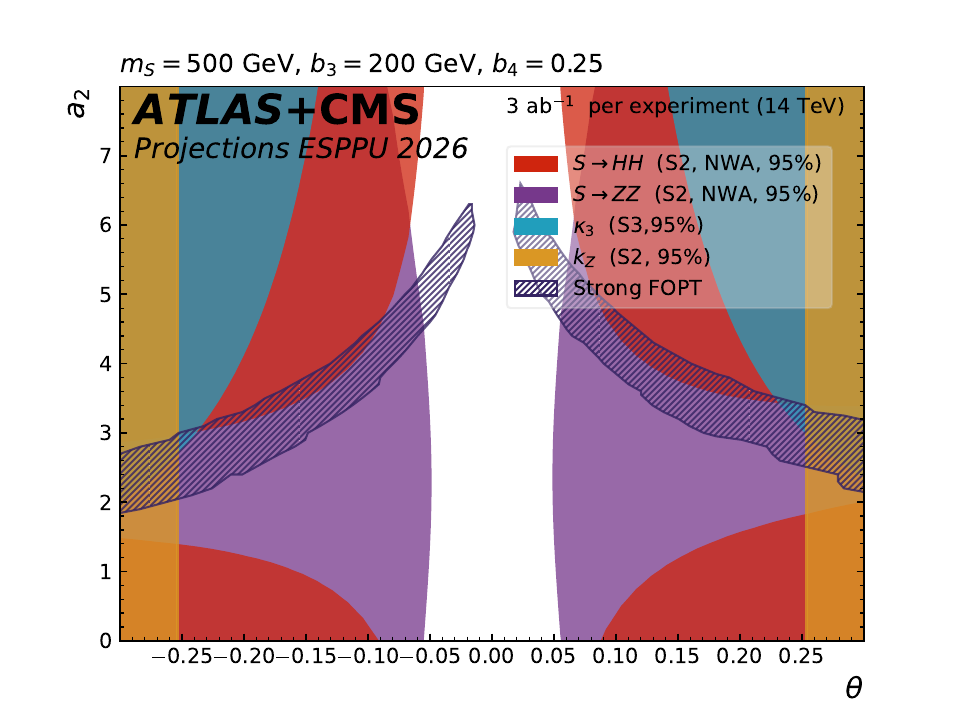}
    \end{subfigure}
           \begin{subfigure}[b]{0.40\textwidth}
        \includegraphics[width=\textwidth]{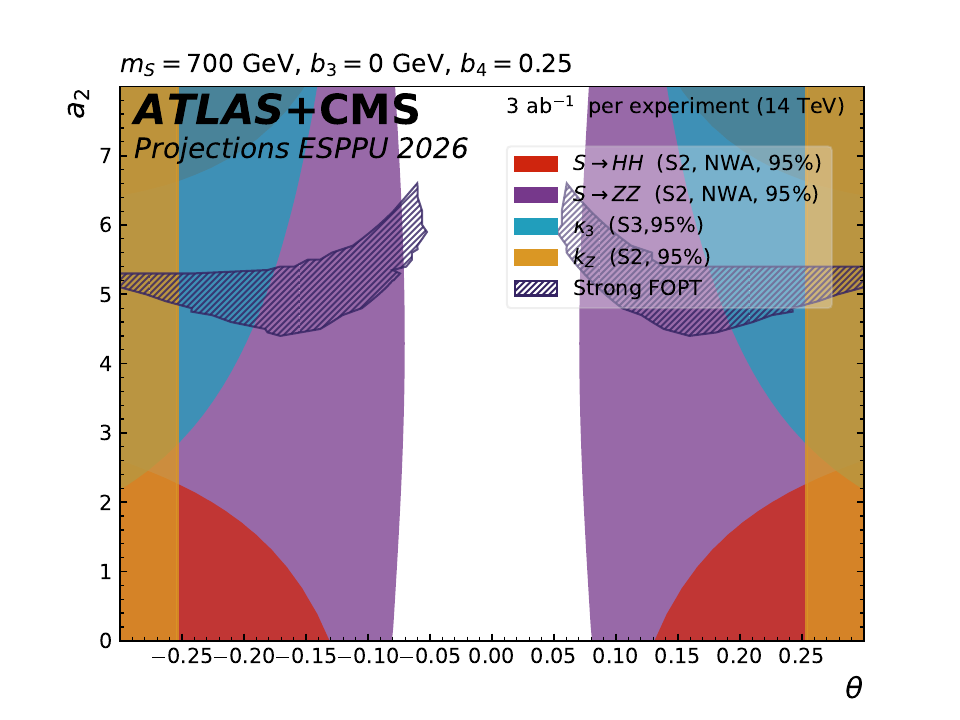}
    \end{subfigure}
        \begin{subfigure}[b]{0.40\textwidth}
        \includegraphics[width=\textwidth]{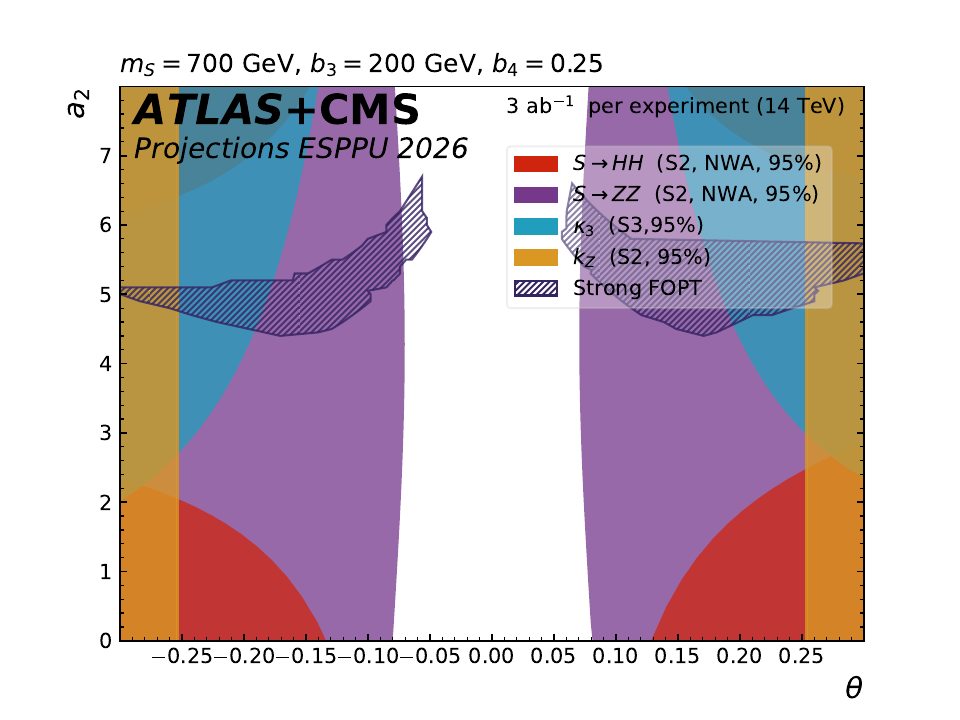}
    \end{subfigure}
           \begin{subfigure}[b]{0.40\textwidth}
        \includegraphics[width=\textwidth]{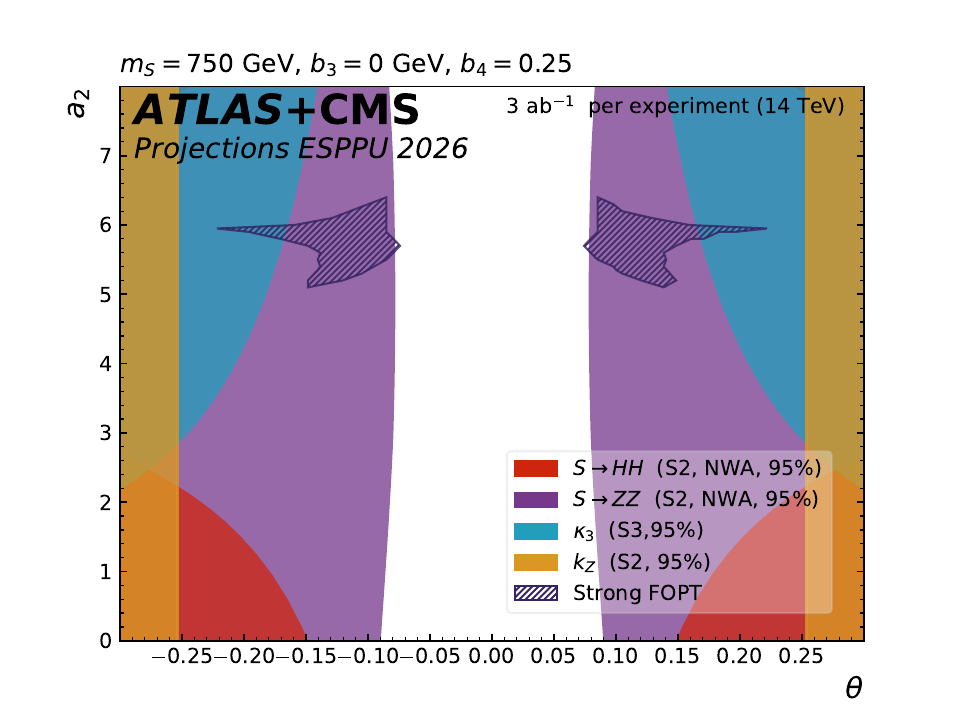}
    \end{subfigure}
        \begin{subfigure}[b]{0.40\textwidth}
        \includegraphics[width=\textwidth]{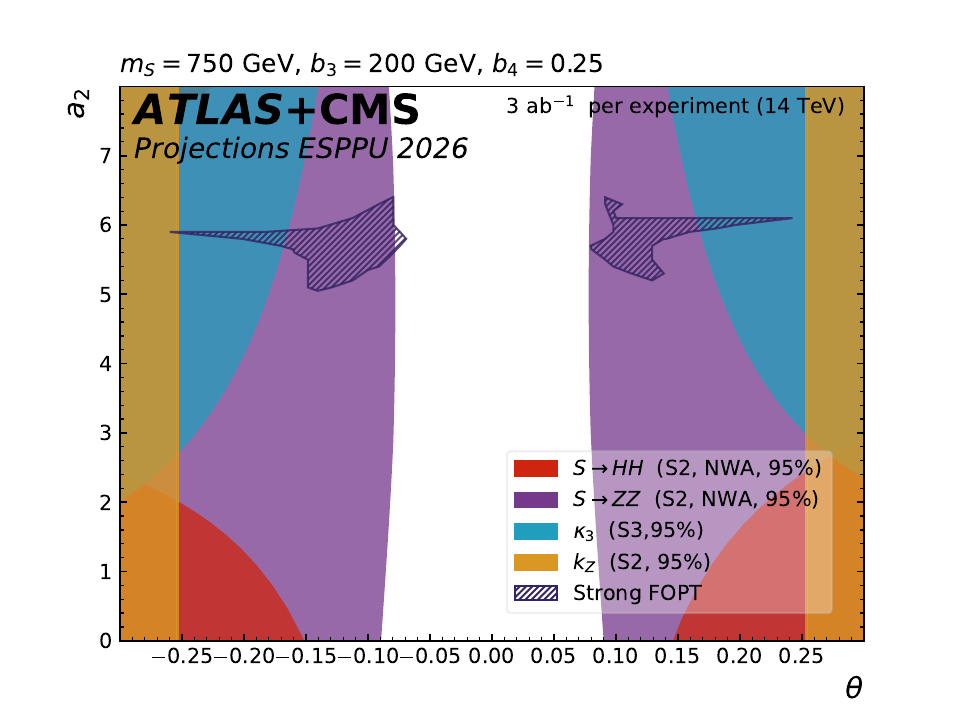}
    \end{subfigure}
    \caption{The dark blue hatched contours show the regions of the $a_2$ versus $\theta$ parameter space in the scalar singlet model where a strong first-order phase transition is possible. The other contours show the 95\% CL exclusion in this plane from the searches for resonant $S \to HH(ZZ)$ decays, from the $H$ coupling to $Z$ and from \kl constraints. The different plots correspond to representative parameter choices in the model.}
    \label{fig:Singletmasses}
\end{figure}

\begin{figure}[!htb]
    \centering
    \begin{subfigure}[b]{0.40\textwidth}
        \includegraphics[width=\textwidth]{figures/first_order_phase_transition_slice300_0.pdf}
    \end{subfigure}
        \begin{subfigure}[b]{0.40\textwidth}
        \includegraphics[width=\textwidth]{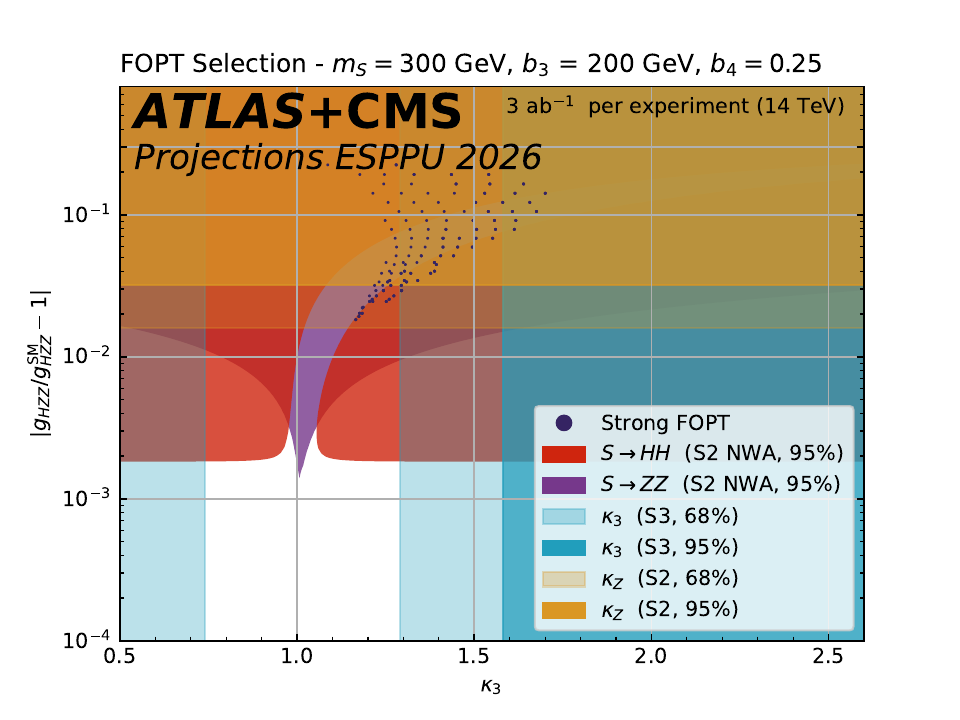}
    \end{subfigure}
        \begin{subfigure}[b]{0.40\textwidth}
        \includegraphics[width=\textwidth]{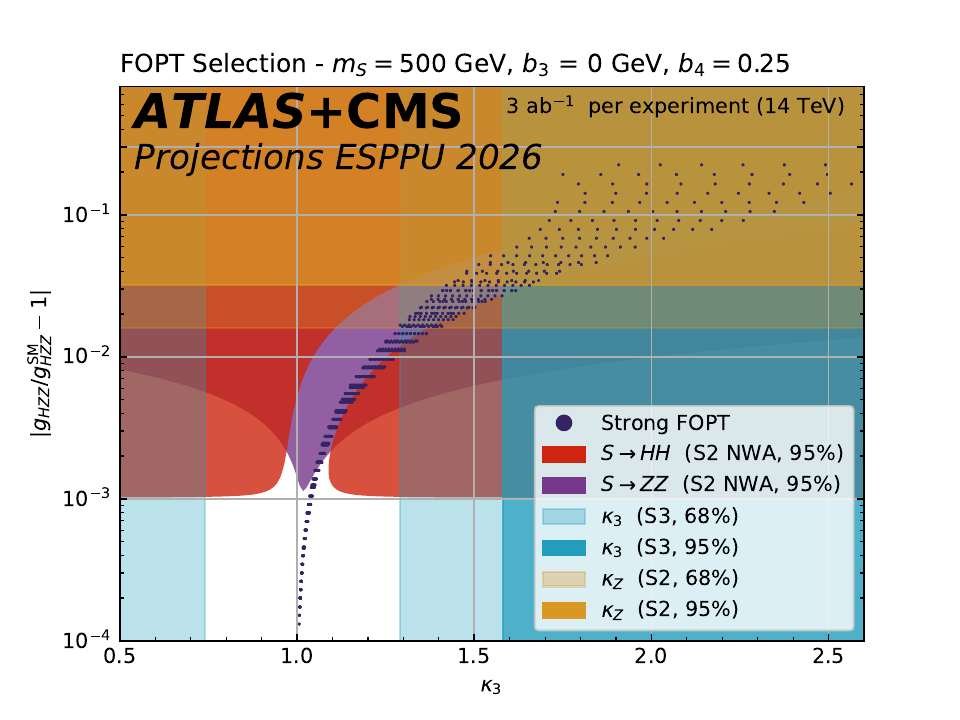}
    \end{subfigure}
        \begin{subfigure}[b]{0.40\textwidth}
        \includegraphics[width=\textwidth]{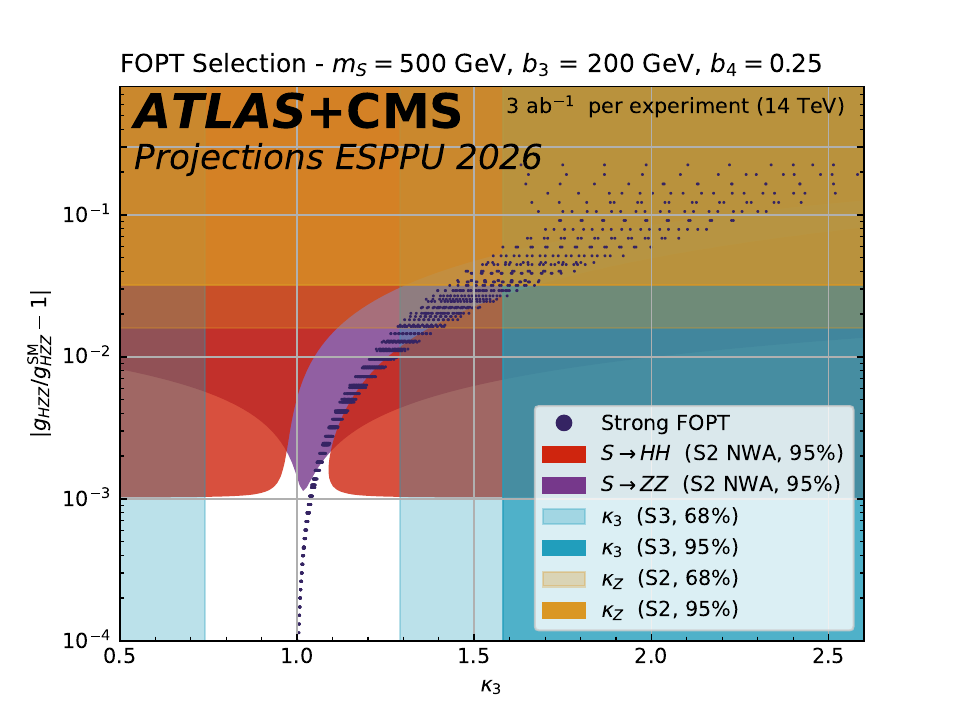}
    \end{subfigure}
           \begin{subfigure}[b]{0.40\textwidth}
        \includegraphics[width=\textwidth]{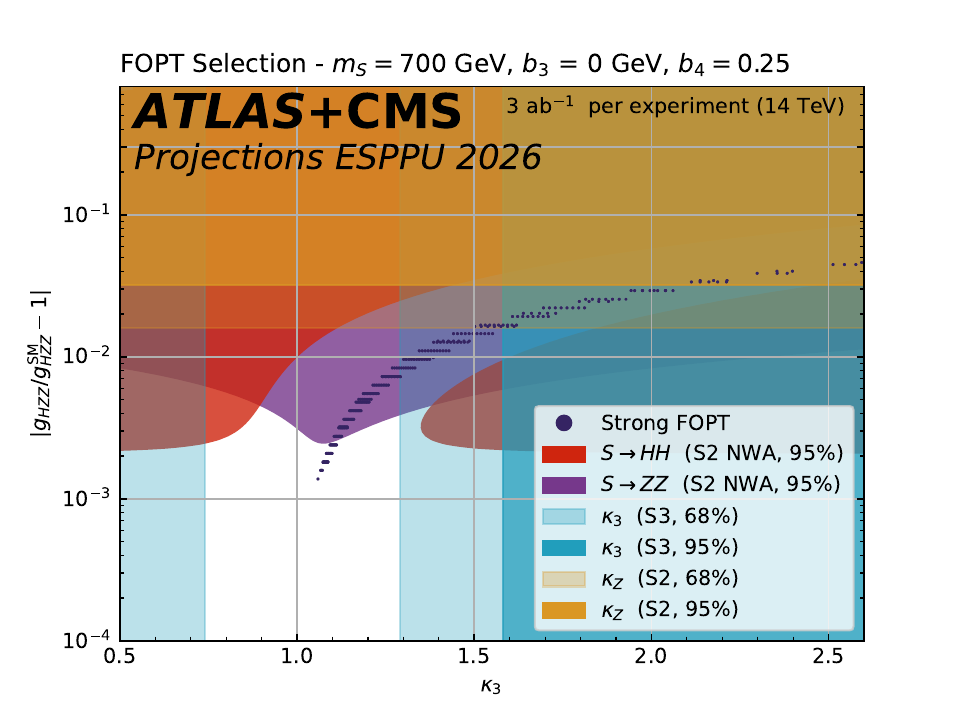}
    \end{subfigure}
        \begin{subfigure}[b]{0.40\textwidth}
        \includegraphics[width=\textwidth]{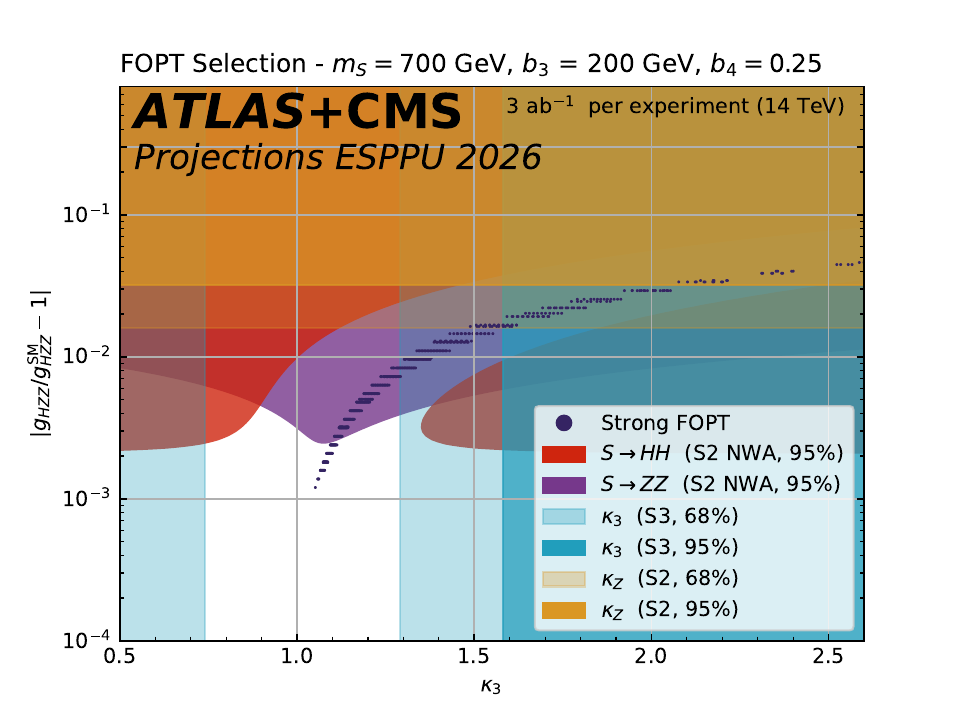}
    \end{subfigure}
           \begin{subfigure}[b]{0.40\textwidth}
        \includegraphics[width=\textwidth]{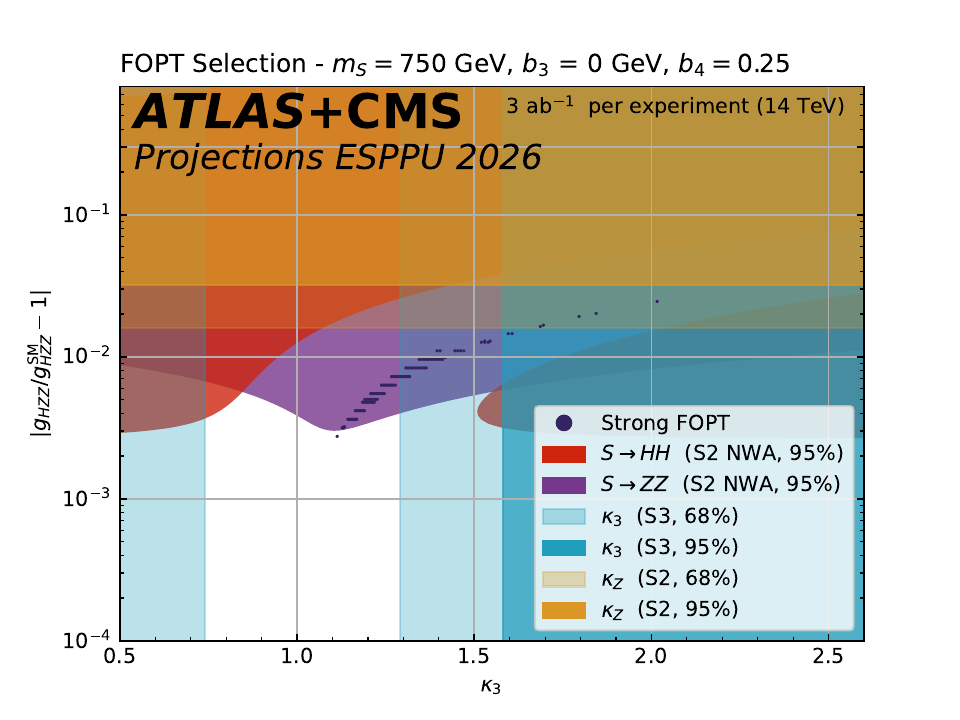}
    \end{subfigure}
        \begin{subfigure}[b]{0.40\textwidth}
        \includegraphics[width=\textwidth]{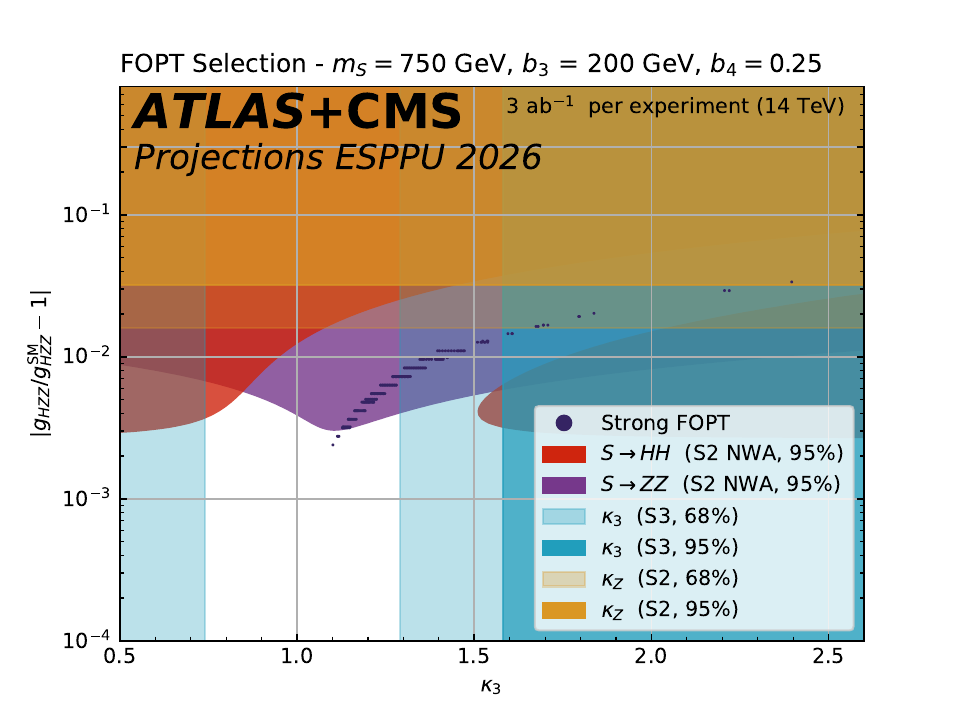}
    \end{subfigure}
    \caption{The dark blue hatched contours show the parameter space in the scalar singlet model where a strong first-order phase transition is possible in the plane  of the Higgs coupling to the Z  versus \kl.
    95\% CL exclusion contours from the searches for resonant $S \to HH(ZZ)$ decays and from \kl and $k_Z$ are overlaid. The plots show the exclusion in different $M_S$ versus $b_3$ slices of the parameter space when $b_4=0.25$}
    \label{fig:Singletmassesprojections}
\end{figure}

\begin{figure}[!htb]
  \centering
    \includegraphics[width=0.48\textwidth,trim=10mm 8mm 3mm 5mm,clip]
       {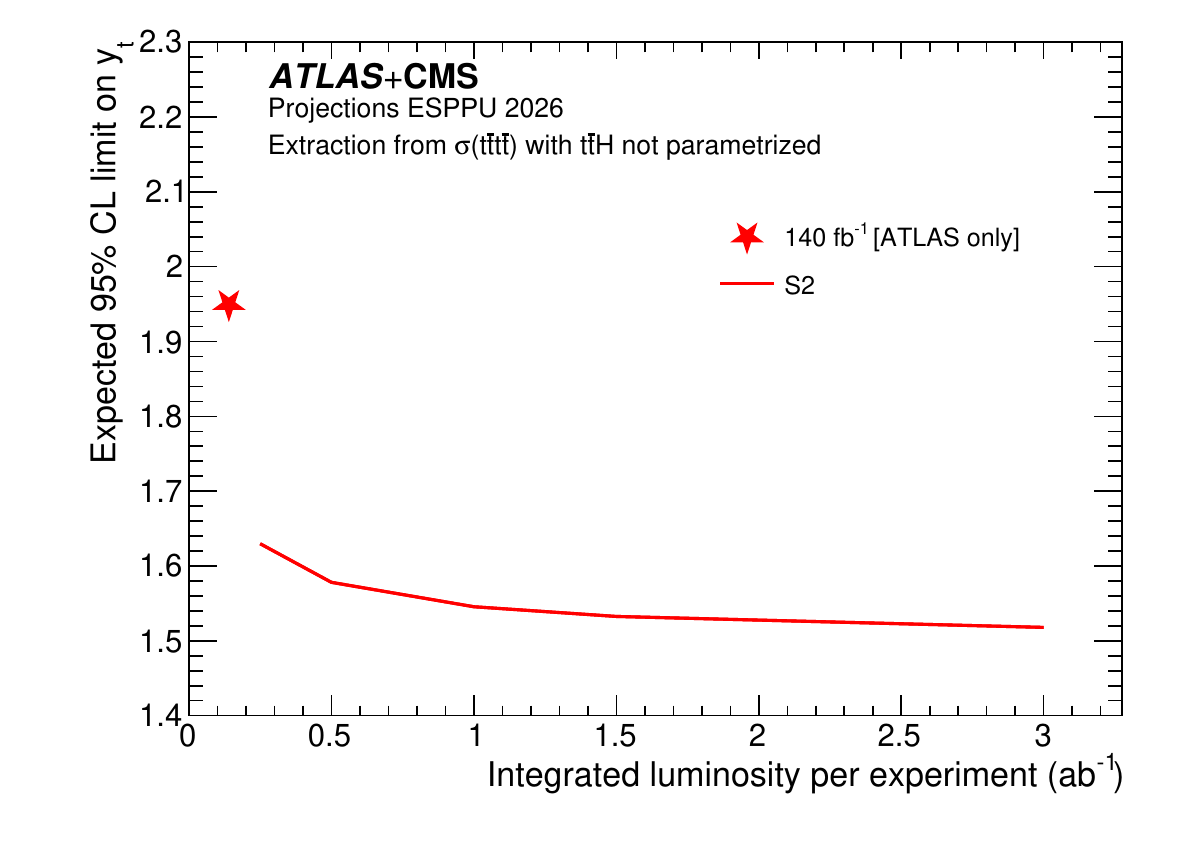}
    \includegraphics[width=0.48\textwidth,trim=10mm 8mm 3mm 5mm,clip]
       {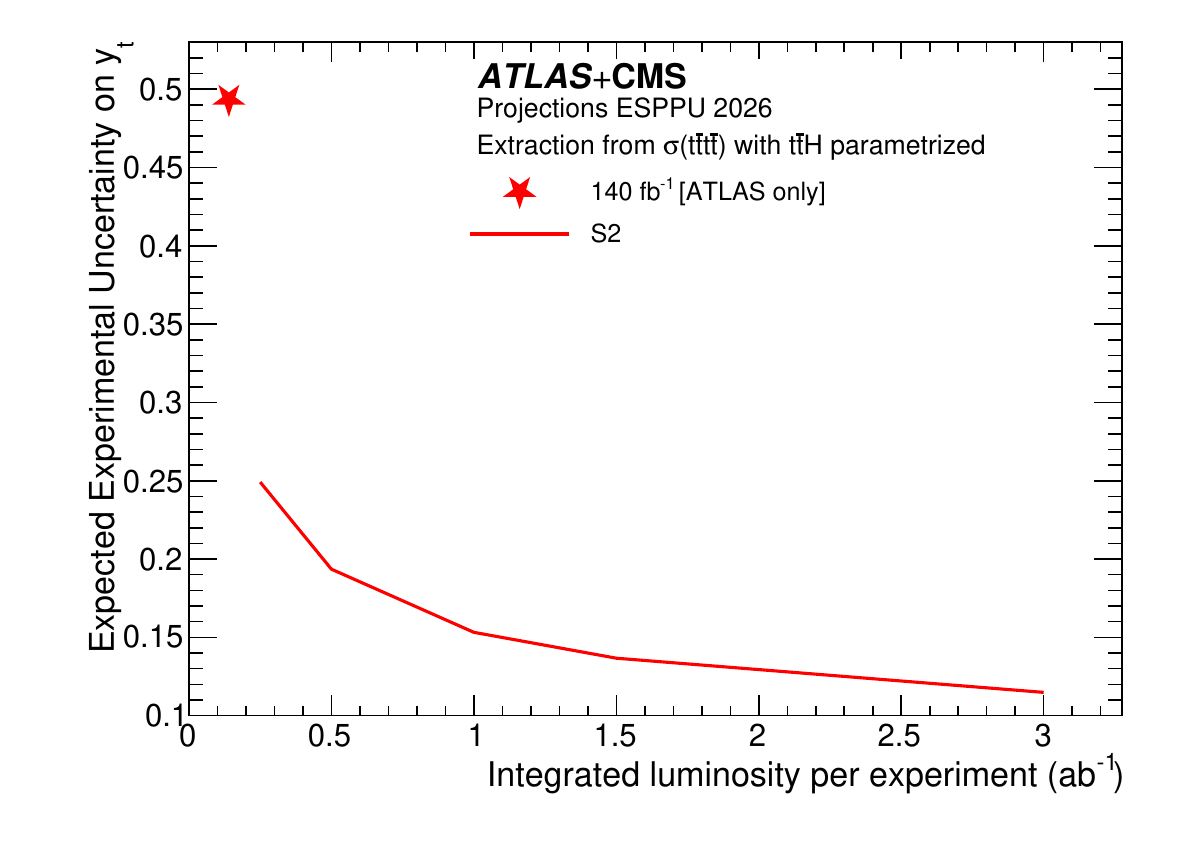}
  \caption{ Left: Expected uncertainty on $y_t$ as a function of the integrated luminosity, in the \Stwo scenario, obtained when the $t\bar{t}H$ contribution is freely floating.
  Right: Expected experimental uncertainty on $y_{t}$  as a function of the integrated luminosity, in the \Stwo scenario, obtained when the \ttH events are parametrized as a function of $\kappa_{t}$.
  }
  \label{fig:top_SMtttt_yukawa}
\end{figure}

\begin{figure}[htb]
  \centering
  \includegraphics[width=0.46\textwidth,trim=0mm 0mm 15mm 10mm]{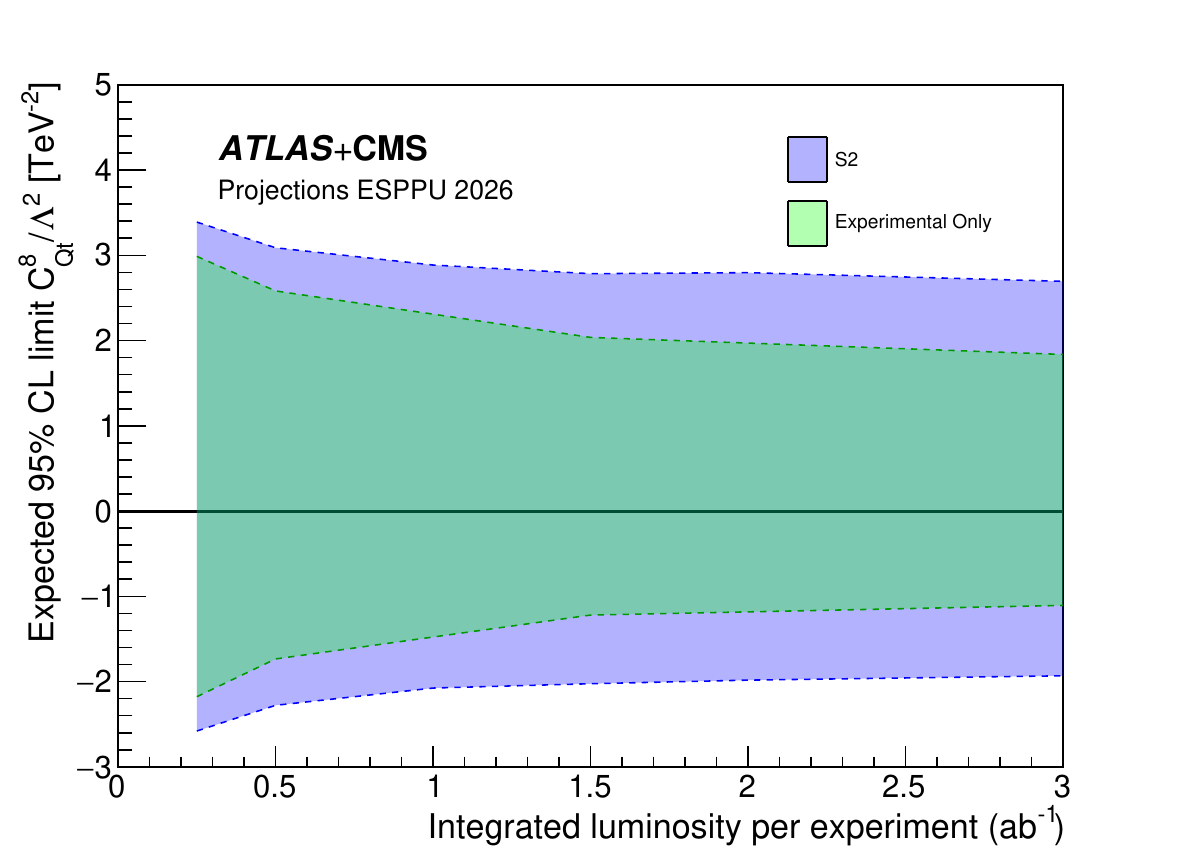}
  \caption{Evolution of the expected 95\% CL bound on $C_i/\Lambda^2$ as a function of integrated luminosity for the SMEFT operator $\mathcal{O}^{8}_{Qt}$ obtained from the ATLAS+CMS combination of the inclusive $t\bar{t}t\bar{t}$ production cross section measurements in the \Stwo scenario. The impact of the experimental and total uncertainties are shown separately.\label{fig:top_topX_HL-LHC}}
\end{figure}

\begin{table}[htbp!]
      \centering     
     \begin{tabular}{c| c|| c c}
    \toprule
           Operators & ATLAS only  & \multicolumn {2}{c}{ATLAS+CMS}  \\
            & 140~fb$^{-1}$ & 2~ab$^{-1}$ \Stwo & 3~ab$^{-1}$ \Stwo \\
    \midrule
           $\mathcal{O}^{1}_{QQ}$ & $[-1.9,~2.5]$ & $[-1.1,~1.7]$ & $[-1.1,~1.7]$ \\
           $\mathcal{O}^{1}_{Qt}$ & $[-2.0,~1.6]$ & $[-1.4,~0.9]$ & $[-1.4,~0.9]$\\
           $\mathcal{O}^{1}_{tt}$ & $[-0.9,~1.1]$ & $[-0.5,~0.8]$ & $[-0.5,~0.8]$ \\
           $\mathcal{O}^{8}_{Qt}$ & $[-3.4,~4.1]$ & $[-2.0,~2.8]$ & $[-1.9,~2.7]$ \\
    \bottomrule	   
      \end{tabular}
      \caption{Expected 95\% CL intervals on EFT coupling parameters, derived setting the new physics scale $\Lambda = 1$~TeV and assuming a single non-zero EFT parameter at a time, in the context of deviations induced in the \tttt process.}
      \label{tab:EFTresults}
  \end{table}

\clearpage

\section{Theoretical interpretation supporting material}
\label{app:theory}

The Higgs potential in the SM in the unbroken phase is given by
\begin{equation}
    V^{\rm SM}(\Phi) = -\mu^2 \Phi^\dagger \Phi + \lambda (\Phi^\dagger \Phi)^2\, 
    \label{eq:SMphi} \, ,
\end{equation}
with the two parameters $\mu>0$ and $\lambda>0$, and with $\Phi$ a complex scalar doublet.

Upon electroweak symmetry breaking, $\mu$ and $\lambda$  can be traded for two physical parameters, $v=1/\sqrt{\sqrt{2} G_F}$ and $m_H$, where  $G_F$ is the Fermi constant %extracted from muon decay 
and $m_H$ is the Higgs boson mass, constrained at high precision from direct mass measurements. Let us review the basic steps needed to rewrite the potential in the broken phase in terms of the Higgs boson field. First, one writes the Higgs doublet in terms of the physical Higgs scalar $H$
\begin{equation}
    \Phi = \left(0, \frac{H+v}{\sqrt{2}}\right)\,. 
    \label{eq:phi}
\end{equation}
with $v$ the value of $\Phi$ at its global minimum,
and substitute in Eq.~(\ref{eq:SMphi}). Second, the minima of the potential are found, by calculating the first derivative and setting it to zero (while the second derivative is positive). Third, impose that on the minimum $H=0$, which gives
\begin{equation}
    \mu = v \sqrt{\lambda}\,.
\end{equation}
Finally, substitute the value of $\mu$ back in the potential expressed in terms of $H$ and find the coefficients of the various powers of $H$:
\begin{equation}
    V(H) = \frac12 (2 \lambda v^2) H^2 + \lambda v H^3 + \frac {\lambda}{4} H^4 + {\rm const} \,. 
    \label{eq:SMh1}
\end{equation}
Note that the constant term can be associated to the vacuum energy, yet it cannot be determined in particle physics experiments.  The first term gives the Higgs mass 
\begin{equation}
    m_H^2 = 2 \lambda v^2 \,,
    \label{eq:hmass}
\end{equation}
which allows to eliminate $\lambda$ in favour of $m_H$, giving the final form of the Higgs potential at low energy in the SM:
\begin{equation}
    V^{\rm SM}(H) = \frac12 m_H^2 H^2 + \frac{m_H^2}{2 v} H^3 + \frac{m_H^2}{8 v} H^4 \,. 
    \label{eq:SMh2_1}
\end{equation}
which depends only on $v$ and $m_H$. In particular, in the SM the strength of both the trilinear and quadrilinear interactions is completely determined by a measurement of $m_H$ and of $v$,  the latter being constrained from low-energy data ($G_F$).

A generic potential at low energy can be written as
\begin{equation}
    V^{\rm low}(H) = \frac12 m_H^2 H^2 + \lambda_3 v  H^3 + \frac{\lambda_4}{4} H^4 + \frac{\lambda_5}{v}  H^5 + \frac{\lambda_6}{v^2}   H^6 + \ldots
    \label{eq:SMh2_2}
\end{equation}
where we have defined all the $\lambda_i$ adimensional using $v$ as a normalization scale.  Using the expression above, one can see that the SM corresponds to:
\begin{equation}
   \lambda_3^{\rm SM} = \lambda_4^{\rm SM} = \lambda= \frac{m_H^2}{2 v^2} \qquad {\rm and} \qquad\lambda_i=0 \quad {\rm for} \quad i\ge 5 \,, 
\end{equation}
as expected. 

Instead of starting from Eq.~(\ref{eq:SMphi}) and impose the minimum condition, one can implement directly that the vacuum of the theory should be found for $\Phi=(0,v/\sqrt{2})$
\begin{equation}
   V^{\rm SM}(\Phi) =  \lambda \left(\Phi^\dagger \Phi- \frac{v^2}{2}\right)^2\,.
    \label{eq:SMH2_3}
\end{equation}
It is easy to see by inserting Eq.~(\ref{eq:phi}) in the above expression that one directly obtains Eq.~(\ref{eq:SMh1}) without having to trade any other parameter and that the constant term is automatically zero. 
So the formulation given by  Eq.~(\ref{eq:SMH2_3}) is more convenient as a starting point when one can write SM deformations in a way that the mass and the vacuum are not modified. 
This is the case for SMEFT deformations, as will we see below. For more general potentials, one instead has to start from scratch, i.e. from  Eq.~(\ref{eq:SMphi}), and apply the same minimization procedure outlined above. 

In order to simplify  the notation, let us choose to measure $v$ in terms of units of the top quark mass, $v=\sqrt{2}$ and use the fact that in the SM $\lambda\sim1/8$ to obtain
{
\begin{equation}
   V^{\rm SM}(\phi) =  \frac18 \left(\phi^2-1\right)^2  \,,
    \label{eq:SMH3}
\end{equation}
}
with
\begin{equation}
    \phi = \frac{H}{\sqrt{2}}+ 1\,,\, 
    m_H  = \frac{1}{\sqrt{2}}   \,. 
    \label{eq:hmass2_app}
\end{equation}

\subsection{EWSB potentials in BSM scenarios}
\label{eq:bsm}

We consider now EWSB potentials that may arise in realistic BSM scenarios, where all new particles are heavy and above some cutoff mass scale $\Lambda$. 

There are two EFT approaches, the so-called linear one (SMEFT) and non-linear one (HEFT). 
In the case of the Higgs potential the HEFT one corresponds exactly to using $V^{\rm low}(H)$ as defined in Eq.~(\ref{eq:SMh2_2}), using arbitrary $\lambda_i$.

The SMEFT case is more interesting. In this case one can write the potential 
\begin{equation}
   V^{\rm SMEFT}(\Phi) =  \lambda \left(\Phi^\dagger \Phi- \frac{v^2}{2}\right)^2 + \sum^\infty_{i=6,8,10,\ldots} \frac{c_i}{\Lambda^{i-4}} \left(\Phi^\dagger \Phi- \frac{v^2}{2}\right)^{i/2}\,, 
    \label{eq:SMEFT}
\end{equation}
which including terms only up to dim-8 gives:
\begin{equation}
   V^{\rm SMEFT}_{8}(\Phi) =  \lambda \left(\Phi^\dagger \Phi- \frac{v^2}{2}\right)^2 
   +  \frac{c_6}{\Lambda^{2}} \left(\Phi^\dagger \Phi- \frac{v^2}{2}\right)^{3}
   +  \frac{c_8}{\Lambda^{4}} \left(\Phi^\dagger \Phi- \frac{v^2}{2}\right)^{4} \label{eq:SMEFT8}\,,
\end{equation}
in terms of two Wilson coefficients $c_6$ and $c_8$.

Substituting Eq.~(\ref{eq:phi}) and mapping it to $V^{\rm low}(H)$ directly gives all the $\lambda_i$:
\begin{eqnarray}
  \lambda_3 &=&  \frac{m_H^2}{2v^2} + \frac{c_6 v^2}{\Lambda^2} = \lambda  \left(1 + c_6 \frac{2 v^4}{m_H^2\Lambda^2}\right)\equiv \lambda (1+\bar c_6)\,,\\
  \lambda_4 &=&  \frac{m_H^2}{2v^2} + \frac{6 c_6 v^2}{\Lambda^2} 
  + \frac{4 c_8 v^4}{\Lambda^4} = \lambda 
  \left(1 +  6 c_6 \frac{2 v^4}{m_H^2\Lambda^2}
          +    c_8 \frac{8 v^6}{m_H^2\Lambda^4} \right) \equiv 
          \lambda (1 + 6 \bar c_6 + \bar c_8)\,, \\
        &&\ldots\\
        \lambda_8 &=&   \frac{c_8 v^4}{16 \Lambda^4} = \lambda  \frac{1}{64}\bar c_8\,, 
     \\
     \lambda_i &=&  0 \qquad i\ge 9\,, 
\end{eqnarray}
where we have redefined $\bar{c}_6$ and $\bar{c}_8$ to absorb the dependence on the heavy scale $\Lambda$, to which low-energy measurements are not directly sensitive. 
Note that the redefined $\bar{c}_6$ and $\bar{c}_8$ coefficients are still dimensionless, with $\bar{c}_6 \gg \bar{c}_8$ if $\Lambda \gg v$.  We note a few interesting features that come from the parametrization that we used. First, dim-8 effects starts influencing  interactions from $i=4$ onwards. 
In other words the parametrization is such that by measuring the trilinear coupling one gets information on $c_6$ only, while the quartic interaction remains unconstrained.
Once that is measured, one would get information on $c_8$ from a measurement of the quadrilinear. 
Note, however, that $c_6$ enters also in the higher point interactions with fixed coefficients and in particular in the four point interaction. So in the SMEFT up to dim-6, trilinear and quadrilinear couplings are not independent.
Starting at dim-8 they become independent, while all non-zero higher-point interactions depend on them.  
In other words, a plot of ($\lambda_3,\lambda_4$) is just a straight line if dim-6 effects only are included, while it becomes a family of curves if dim-8 effects are also considered. 

In all cases, however, the potential is different from the HEFT where only $\lambda_3$ and $\lambda_4$ are modified independently and all higher order terms neglected. 
Vice-versa, it is also clear that from the SMEFT point of view, assuming that $\lambda_3$ is the only coupling modified by new physics is inconsistent (or highly unnatural). 
Hence measurements of $\lambda_3$ and $\lambda_4$, when available, allow to discriminate between SMEFT and HEFT scenarios.

To simplify the notation,  let us measure $v$ in terms of units of the top quark mass, $v=\sqrt{2}$ and use the fact that $\lambda\sim1/8$ and set $\Lambda=1$ (in unit of top quark mass) to get
\begin{equation}
   V^{\rm SMEFT}_{8}(\phi) =  \frac18 \left(\phi^2- 1\right)^2 
   +  c_6 \left(\phi^2- 1\right)^{3}
   + c_8  \left(\phi^2- 1 \right)^{4} \label{eq:SMEFT8simp_app1}\,,
\end{equation}
giving
\begin{eqnarray}
  k_3 &=& \lambda_3/\lambda = \left(1 + 16 c_6\right)\equiv  (1+\bar c_6)\,,\\
  k_4 &=& \lambda_4/\lambda =
  \left(1 +  96 c_6 + 128   c_8 \right) \equiv 
         (1 + 6 \bar c_6 + \bar c_8)\,.
\end{eqnarray}
We therefore obtain
\begin{equation}
V^{\rm SMEFT}_{6}(\phi) = { \frac18 \left(\phi^2- 1\right)^2} 
   +  \frac{k_3-1}{16} \left(\phi^2- 1\right)^{3}
\label{eq:SMEFT8simp_app2}\,,
\end{equation}
and 
\begin{equation}
V^{\rm SMEFT}_{8}(\phi) =  {\frac18 \left(\phi^2- 1\right)^2} 
   +  \frac{k_3-1}{16} \left(\phi^2- 1\right)^{3}
   +  \frac{k_4-6 k_3+5}{128} \left(\phi^2- 1 \right)^{4} \label{eq:SMEFT8simp_app3}\,.
\end{equation}

We now consider a modification of the low-energy Higgs potential in the SM by a small term with a logarithmic dependence on the scalar doublet inner product $\Phi^\dagger \Phi$ (as required if SU$_L$(2) breaking is linearly realized as in the case of the SM).
The inspiration comes from the Coleman-Weinberg potential obtained by including the loop effects in the Lagrangian, and is often presented as a valid alternative to the SM EWSB potential which is consistent with single Higgs production measurements.

In this case we need to apply the procedure we used at the beginning, Eq.~(\ref{eq:SMphi}). Following the notation of Ref.~\cite{Reichert:2017puo}, we define this potential as

\begin{equation}
    V^{\rm CW}(\Phi) = -\mu^2 \Phi^\dagger \Phi + \lambda (\Phi^\dagger \Phi)^2 - c_{\rm CW} (\Phi^\dagger \Phi) \Lambda^2 \log \frac{\Phi^\dagger \Phi}{2\Lambda^2}\,,
    \label{eq:CW}
\end{equation}
such that the only difference with the SM in the extra logarithmic term.

\newcommand{\hov}{\frac{h}{v}}

By following the same procedure as for the SM potential, we obtain ($c_{\rm CW} = \frac38 (k_3-1) $)
{ 
\begin{eqnarray}
 V^{\rm CW}(\phi)=
{\frac18 \left(\phi^2- 1\right)^2} 
 + \frac38 (k_3-1) 
 \left[
 (\frac12 \phi^4-1)-  \phi^2 \log \phi^2
 \right]
 \label{eq:log}\,. 
\end{eqnarray}
}

Finally, we consider a modification of the SM Higgs potential by a small term with an exponential depending on the $\Phi^\dagger \Phi$ inner product. 
The inspiration for this model comes from including instanton effects. 
Again, we need to apply the procedure we used at the beginning, Eq.~(\ref{eq:SMphi}). Following the notation of Ref.~\cite{Reichert:2017puo}, we define this exponential potential as follows
\begin{equation}
    V^{\rm I}(\Phi) = -\mu^2 \Phi^\dagger \Phi + \lambda (\Phi^\dagger \Phi)^2 \pm (\Phi^\dagger \Phi)^2 e^{-\frac{2}{c_I} \frac{\Lambda ^2}{\Phi^\dagger \Phi}} \,,
    \label{eq:instanton}
\end{equation}
where the label ``I'' stands for instanton. 
The extra exponential term can be either positive or negative, and depends on the ratio $c_I/\Lambda^2$. 

By following the same procedure as above 
\begin{eqnarray}
V^I (\phi)={\frac18 \left(\phi^2- 1\right)^2} \pm \phi ^4 e^{-\frac{2}{{c_I} \phi ^2}}+\frac{e^{-2/{c_I}} \pm \left(-({c_I} ({c_I}+2)+2) \phi ^4+2 ({c_I}+2) \phi
   ^2-2\right)}{{c_I}^2}
   \label{eq:exp}
\end{eqnarray}
where the $c_I$ for the two possible signs are to  be expressed as function of $k_3$, by inverting numerically the expressions above.

\clearpage 

\printbibliography

\end{document}